\documentclass[preprint]{./aastex}

\newcommand{\rp}{$R_{50}$}
\newcommand{\re}{$R_{\rm eff}$}

\newcommand{\xhap}{$x(\alpha_{50})$}
\newcommand{\xhae}{$x(\alpha_{\rm eff})$}
\newcommand{\xhahbp}{$x(\alpha\beta_{50})$}
\newcommand{\xhahbe}{$x(\alpha\beta_{\rm eff})$}
\newcommand{\xnhap}{$x({\rm N}2_{50})$}
\newcommand{\xnhae}{$x({\rm N}2_{\rm eff})$}
\newcommand{\xonp}{$x({\rm O}3{\rm N}2_{50})$}
\newcommand{\xone}{$x({\rm O}3{\rm N}2_{\rm eff})$}

\newcommand{\xohp}{$x({\rm OH}_{50})$}
\newcommand{\xohe}{$x({\rm OH}_{\rm eff})$}

\slugcomment{Not to appear in Nonlearned J., 45.}

\shorttitle{Aperture effects on abundance determinations}
\shortauthors{Iglesias-P\'{a}ramo et al.}

\begin{document}

\title{Aperture effects on the oxygen abundance determinations from CALIFA data}

\author{J. Iglesias-P\'{a}ramo\altaffilmark{1,2}, J.M. V\'{\i}lchez\altaffilmark{1}, F.F. Rosales-Ortega\altaffilmark{3}, S.F. S\'{a}nchez\altaffilmark{4}, S. Duarte Puertas\altaffilmark{1}, V. Petropoulou\altaffilmark{5}, A. Gil de Paz\altaffilmark{6}, L. Galbany\altaffilmark{7,8}, M. Moll\'{a}\altaffilmark{9}, C. Catal\'{a}n-Torrecilla\altaffilmark{6}, A. Castillo Morales\altaffilmark{6}, D. Mast\altaffilmark{10,11}, B. Husemann\altaffilmark{12}, R. Garc\'{\i}a-Benito\altaffilmark{1}, M.A. Mendoza\altaffilmark{1}, C. Kehrig\altaffilmark{1}, E. P\'{e}rez-Montero\altaffilmark{1}, P. Papaderos\altaffilmark{13}, J.M. Gomes\altaffilmark{13}, C.J. Walcher\altaffilmark{14}, R.M. Gonz\'{a}lez Delgado\altaffilmark{1}, R.A. Marino\altaffilmark{15,6}, \'A. R. L\'opez-S\'anchez\altaffilmark{16,17}, B. Ziegler\altaffilmark{18}, H. Flores\altaffilmark{19} \and J. Alves\altaffilmark{18}}
\email{jiglesia@iaa.es}
\altaffiltext{1}{Instituto de Astrof\'{\i}sica de Andaluc\'{\i}a - CSIC, Glorieta de la Astronom\'{\i}a s.n., 18008 Granada, Spain}
\altaffiltext{2}{Estaci\'{o}n Experimental de Zonas \'{A}ridas - CSIC, Ctra. de Sacramento s.n., La Ca\~{n}ada, Almer\'{\i}a, Spain}
\altaffiltext{3}{Instituto Nacional de Astrof\'{\i}sica, \'{O}ptica y Electr\'{o}nica, Luis E. Erro 1, 72840 Tonantzintla, Puebla, Mexico}
\altaffiltext{4}{Instituto de Astronom\'{\i}a, Universidad Nacional Aut\'{o}noma de M\'{e}jico, A.P. 70-264, 04510 M\'{e}xico, D.F., Mexico}
\altaffiltext{5}{INAF-OA Brera, via Brera 28, 20121 Milano, Italy}
\altaffiltext{6}{Departamento de Astrof\'{\i}sica y CC. de la Atm\'{o}sfera, Universidad Complutense de Madrid, E-28040, Madrid, Spain}
\altaffiltext{7}{Millennium Institute of Astrophysics, Chile}
\altaffiltext{8}{Departamento de Astronom\'{\i}a, Universidad de Chile, Camino El Observatorio 1515, Las Condes, Santiago, Chile}
\altaffiltext{9}{Departamento de Investigaci\'{o}n B\'{a}sica, CIEMAT, Avda. Complutense 40, 28040 Madrid, Spain}
\altaffiltext{10}{Observatorio Astron\'omico, Laprida 854, X5000BGR, C\'ordoba, Argentina}
\altaffiltext{11}{Consejo de Investigaciones Cient\'{\i}ficas y T\'ecnicas de la Rep\'ublica Argentina, Avda. Rivadavia 1917, C1033AAJ, CABA, Argentina}
\altaffiltext{12}{European Southern Observatory (ESO),  Karl-Schwarzschild-Str.2, D-85748 Garching b. M\"{u}nchen, Germany}
\altaffiltext{13}{Instituto de Astrof\'{\i}sica e Ci\^{e}ncias do Espa\c co, Universidade do Porto, CAUP, Rua das Estrelas, P-4150-762 Porto, Portugal}
\altaffiltext{14}{Leibniz-Institude f\"{u}r Astrophysik Potsdam (AIP), An der Sternwarte 16, D-14482 Potsdam, Germany}
\altaffiltext{15}{Department of Physics, Institute for Astronomy, ETH Z\"{u}rich, CH-8093 Z\"{u}rich, Switzerland}
\altaffiltext{16}{Australian Astronomical Observatory, PO Box 915, North Ryde, NSW 1670, Australia}
\altaffiltext{17}{Department of Physics and Astronomy, Macquarie University, NSW 2109, Australia}
\altaffiltext{18}{University of Vienna, T\"{u}rkenschanzstrasse 17, 1180 Vienna, Austria}
\altaffiltext{19}{GEPI, Observatoire de Paris, CNRS, 5 Place Jules Janssen, Meudon F-92195, France}

\begin{abstract}
This paper aims at providing aperture corrections for emission lines in a sample of spiral galaxies from the Calar Alto Legacy Integral Field Area Survey (CALIFA) database.
In particular, we explore the behavior of the $\log$([\ion{O}{3}]~$\lambda$5007/H$\beta$)/([\ion{N}{2}]~$\lambda$6583/H$\alpha$) (O3N2) and $\log$[\ion{N}{2}]~$\lambda$6583/H$\alpha$ (N2) flux ratios 
since they are closely connected to different empirical calibrations of the oxygen abundances in star forming galaxies.

We compute median growth curves of H$\alpha$, H$\alpha$/H$\beta$, O3N2 and N2 up to 2.5\rp~and 1.5 disk~\re.
These distances cover most of the optical spatial extent of the CALIFA galaxies.
The growth curves simulate the effect of observing galaxies through apertures of varying radii.
We split these growth curves by morphological types and stellar masses to check if there is any dependence on these properties.

The median growth curve of the H$\alpha$ flux shows a monotonous increase with radius with no strong dependence on galaxy inclination, morphological type and stellar mass.
The median growth curve of the H$\alpha$/H$\beta$ ratio monotonically decreases from the center towards larger radii, showing for small apertures a maximum value of $\approx 10\%$ larger than
the integrated one.
It does not show any dependence on inclination, morphological type and stellar mass.
The median growth curve of N2 shows a similar behavior, decreasing from the center towards larger radii. 
No strong dependence is seen with the inclination, morphological type and stellar mass.
Finally, the median growth curve of O3N2 increases monotonically with radius, and it does not show dependence with the inclination.
However, at small radii it shows systematically higher values for galaxies of earlier
morphological types and for high stellar mass galaxies.

Applying our aperture corrections to a sample of galaxies from the SDSS survey
at $0.02 \leq z \leq 0.3$ shows that the average difference between fiber-based and aperture
corrected oxygen abundances, for different galaxy stellar mass and redshift ranges, 
reaches typically to $\approx 11\%$, depending on the abundance calibration used. 
This average difference is found to be systematically biased, though still within 
the typical uncertainties of oxygen abundances derived from empirical calibrations. 
Caution must be exercised when using observations of galaxies for small radii 
(e.g. below 0.5 \re) given the high dispersion shown around the median growth curves. 
Thus, the application of these median aperture corrections to derive abundances 
for individual galaxies is not recommended when their fluxes come from radii much smaller than either \rp~ or \re.

\end{abstract}


\section{Introduction}

Bidimensional spectroscopy is gaining more and more importance as a powerful observational technique capable to provide new results on 
the properties of galaxies.
Many instruments (commonly known as Integral Field Spectrographs, IFS) have been developed in the last years to produce bidimensional spectroscopy, most of them based on arrays of fibers that collect light from the sky area of interest
and drive it through a dispersor.
The main limitation of these fiber-fed spectrograps is their limited field-of-view, which makes impossible the blind observation of large areas of the sky to observe large amounts of galaxies as it is the case of the large scale surveys like SDSS (York et al. 2000), 2dFGRS (Colless et al. 2001), VVDS (Le F\`{e}vre et al. 2005), z-COSMOS (Lilly et al. 2007) and DEEP/DEEP2 (Davis et al. 2003).

The best option for IFS surveys is thus to focus on selected individual objects, but this is very time consuming and prevents the study of large samples of galaxies.
So far, there are only a few surveys employing IFS, from which we highlight: SAURON (Bacon et al. 2001), ATLAS-3D (Cappellari et al. 2011), PINGS (Rosales-Ortega et al. 2010), VENGA (Blanc et al. 2010), VIXENS (Heiderman et al. 2011) and CALIFA (S\'{a}nchez et al. 2012a).

The CALIFA Survey (S\'{a}nchez et al. 2012a) is observing a statistically well-defined
sample of 600 galaxies in the local Universe with the Potsdam
Multi Aperture Spectrograph in the PPAK mode (Roth et al.
2005) at the 3.5m telescope at Calar Alto Observatory.
The survey benefits from the wide field-of-view of PPAK (about 1~arcmin$^{2}$) compared to similar instruments at other telescopes.
Thus, CALIFA galaxies are mapped over most of their optical spatial extent,
and so far it has allowed up to date complete bidimensional studies of galaxy properties like 
the star formation histories (Cid Fernandes et al. 2013, Gonz\'{a}lez-Delgado et al. 2014), 
the properties of the ionized gas in early-type galaxies (Kehrig et al. 2012, Papaderos et al. 2013, Singh et al. 2013, Gomes et al. 2015a), the properties of large samples of \ion{H}{2} regions (S\'{a}nchez et al. 2013,2014), 
or the effects of the spatial resolution at different redshifts (Mast et al. 2014) among others.

In addition to this, the CALIFA database allows the study of the biases introduced when galaxies are observed through small and size-limited apertures, which usually are single-fiber spectrographs.
Several studies have already notices the existence of such aperture effects in the properties of early and late-type galaxies (e.g. Kehrig et al. 2013, Gomes et al. 2015b).
Aperture effects on galaxy properties have been previously addressed by following different approaches (e.g. Hopkins et al. 2003, Brinchmann et al. 2004, Ellis et al. 2005, Kewley et al. 2005, Salim et al. 2007, Gerssen et al. 2012, Zahid et al. 2013).
On top of that, a recent study by Richards et al. (2015) based on part of the SAMI Galaxy Survey (Allen et al. 2015; Bryant et al. 2015; Sharp et al. 2015) concludes that biases in the estimation of the total instantaneous star formation rate of a galaxy arise when the aperture correction is built only from spectra of the nuclear region of galaxies.

A preliminary study of the aperture effects based on CALIFA data was presented in Iglesias-P\'{a}ramo et al. (2013, IP13) and was devoted to the H$\alpha$ and H$\beta$ emission line fluxes.
In the present work we extend their analysis using a much larger sample of galaxies which allows us to perform a more detailed study. 
We try to go beyond and focus on observational parameters related to the derivation of the oxygen abundances, like the (widely used in the literature) N2 and O3N2 parameters. 

This is a crucial point since spiral galaxies are known to show radial abundance gradients, which means that observing them through a reduced aperture must not necessarily provide a complete information on the spatial abundance distribution. 
Regarding this point, it has been reported in the literature 
that the oxygen abundance derived for the integrated fluxes 
of emission lines of spiral galaxies equals the corresponding 
abundance of their H{\sc ii} regions at a typical galactocentric 
distance of $0.4 \times R_{opt}$ (Pilyugin et al. 2004, Moustakas et al. 2006). 
This result linking the abundances of H{\sc ii} regions at a fixed 
galactocentric radius and the abundance obtained for the integrated 
emission line flux ratios from the whole disk of spiral galaxies, 
give some support to adopting a reference characteristic value 
for the abundance of a spiral galaxy. Nonetheless, we should bear in mind 
that the integrated emission of the spiral disks represents 
in fact a composite spectrum, including different H{\sc ii} regions 
(plus diffuse ionized ISM) with varying physical conditions and chemical compositions.  

Although there is a classical method to derive oxygen abundances of bright star forming regions based on atomic data and the fluxes of emision lines ([O{\sc ii}]$\lambda$3727\AA, [O{\sc iii}]$\lambda$4363\AA, [O{\sc iii}]$\lambda$5007\AA), known as the direct method, it cannot be used for more distant or intrinsically fainter galaxies since some of these lines are usually not detected.
For this reason, abundance calibrations based on theoretical or empirical methods based only on the brightest emission lines must be used.
Significant differences among the different calibrations commonly used in the literature (see Kewley \& Ellison 2008 for an extensive discussion on this topic) have been reported.
Moreover, in some cases different calibrations are used for a single observed emission line flux ratio.
For these two reasons care must be taken when comparing relations involving oxygen abundances derived from different methods since the differences could be due to the choice of the calibration rather than to the real abundance.

In this work we study the effect of the aperture on two observational quantities, namely N2 ($\log$[N{\sc ii}]$\lambda$6583\AA/H$\alpha$) and O3N2 ($\log$([O{\sc iii}]$\lambda$5007\AA/H$\beta$)/([N{\sc ii}]$\lambda$6583\AA/H$\alpha$)), that have been widely used in the literature to estimate oxygen abundanes of star formation galaxies (e.g. Pettini \& Pagel 2004, P\'{e}rez-Montero \& Contini 2009, Marino et al. 2013).
The applicability intervals of these indicators are $-2.5 < $N2$< -0.3$ and O3N2$< 2$.
Marino et al. (2013) found dispersions of $\sigma \approx 0.16$ and 0.18~dex when fitting temperature-based abundances to abundances derived from the N2 and O3N2 methods using the following calibrations:
\begin{equation}
12 + \log {\rm O/H} = 8.743 + 0.462 \times {\rm N2}
\end{equation}
and 
\begin{equation}
12 + \log {\rm O/H} = 8.533 - 0.214 \times {\rm O3N2}
\end{equation} 

Our study is performed for these two observational quantities, so the effect of the aperture on the abundances depends on the choice of the calibration based on any of these observed quantities. 
A further advantage of this larger sample, compared to the one used in IP13, is that it allows us to study the effect of parameters like inclination, morphological type and stellar masses, on the derived average aperture corrections.

The paper is organized as follows: 
Section~2 explains the selection of the sample.
The main results from our analysis are detailed in Section~3.
Section~4 contains a discussion on the implications of the aperture effects on the abundance determination and an example to illustrate this interesting point.
Finally, the conclusions of the paper are enumerated in Section~5.

\section{Sample selection}

The sample of galaxies used for this work has been selected from the CALIFA database updated to December 2013, reduced with the v1.4
version of the pipeline (Garc\'{\i}a-Benito et al. 2015).
Details on the instrumental setup and properties of the spectra can be found in S\'{a}nchez et al. (2012a), where the survey is presented.
We started with an initial sample of 402 galaxies, the ones observed up to December 2013, out of the 937 galaxies comprising the total CALIFA mother sample (Walcher et al. 2014). 
Then we removed the elliptical and lenticular galaxies (E, S0, S0a) in order to keep only spiral galaxies. 
As we are interested in covering the galaxy disks as much as possible within the CALIFA aperture ($\approx 36''$ radius), we have to choose a proper scale related to the spatial extent of the galaxies. 
For this, two possibilities, related to different structural component of the galaxies, arise as the most interesting:
\begin{itemize}
\item The Petrosian radius in the SDSS-$r'$ band within a circular aperture containing 50\% of the total Petrosian flux (SDSS $petroR50\_r$, and hereafter \rp): this scale takes into account the stellar emission from both the bulge and the disk inside this circular aperture. 
Thus, this spatial scale is sensitive to light coming from stars of different ages, and has the great advantage of its availability for a huge number of galaxies, in particular the whole CALIFA sample.
\item The effective radius encompasing 50\% of the light coming from the disk component (hereafter \re): this scale is computed after a morphological decomposition bulge/disk of the galaxy SDSS g'-band surface brightness profile, removing the light coming from the bulge and assuming that the disk has an exponential profile of the form $I(r) = I_{0} e^{-1.678 r/R_{\rm eff}}$. 
Details about the procedure can be found in S\'{a}nchez et al. (2014).
\end{itemize}

\begin{figure}
\plottwo{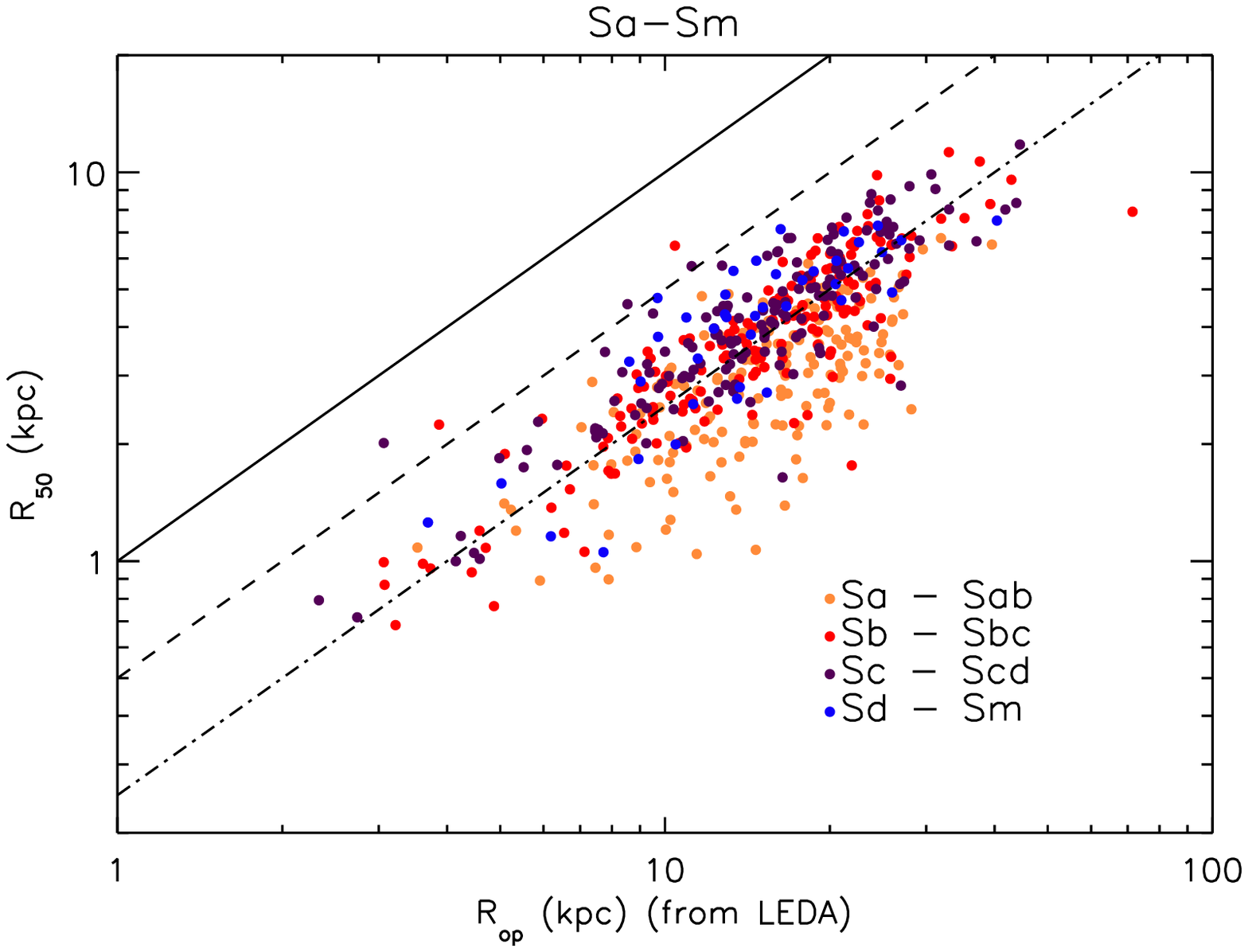}{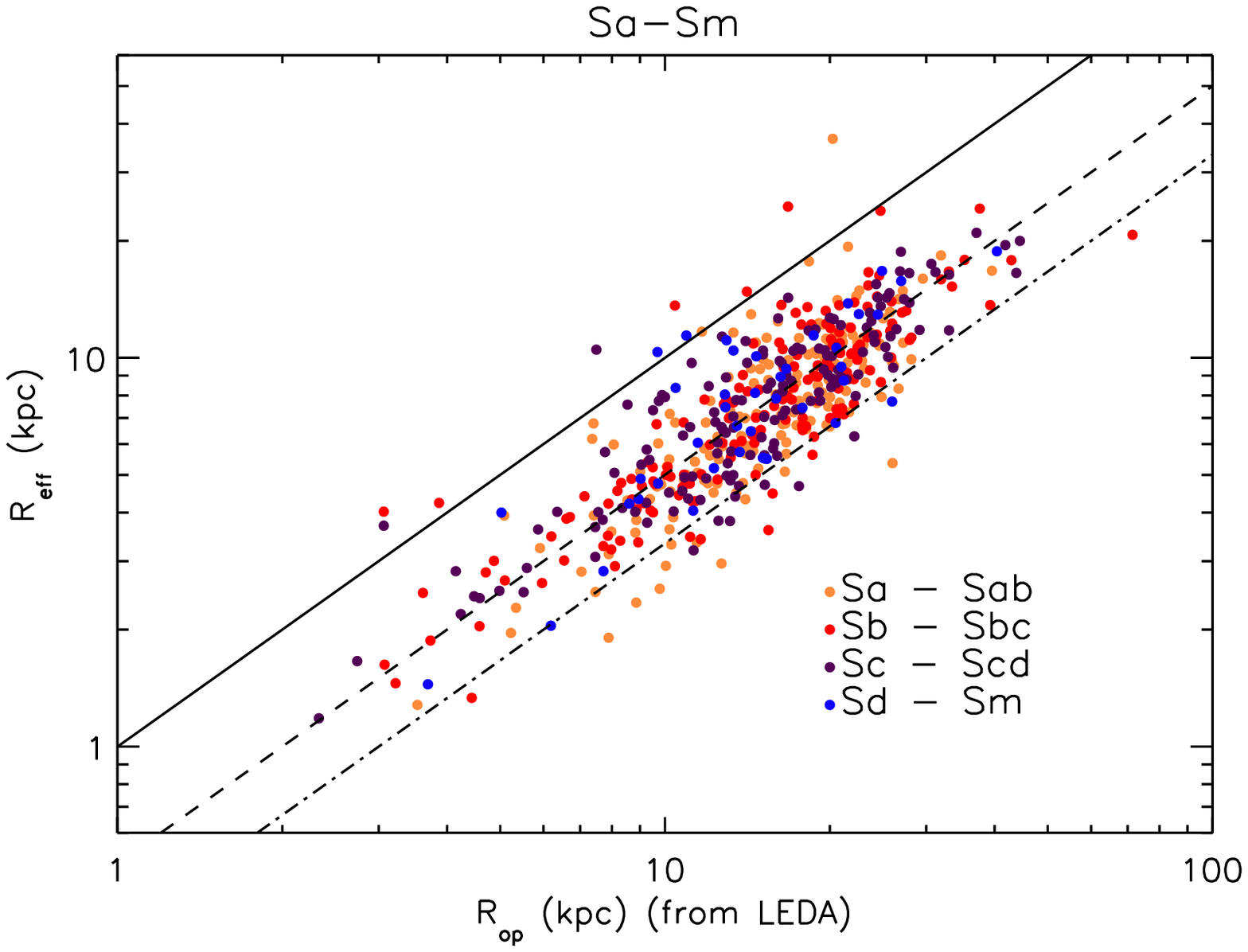}
\caption{{\bf Left:} $R_{\rm op}$ vs. \rp~for the spiral galaxies of the CALIFA sample.
Solid, dashed and dot-dashed lines correspond to $R_{\rm op} = $~\rp, $R_{\rm op} = 2\times$~\rp~ and $R_{\rm op} = 4\times$~\rp~ respectively.
{\bf Right:} $R_{\rm op}$ vs. \re~for the spiral galaxies of the CALIFA sample.
Solid, dashed and dot-dashed lines correspond to $R_{\rm op} = $~\re, $R_{\rm op} = 2\times$~\re~ and $R_{\rm op} = 3\times$~\re~ respectively.\label{compa_rad}}
\end{figure}

Figure~\ref{compa_rad} shows the comparison of \rp~and \re~with the optical radii\footnote{From LEDA database (Makarov et al. 2014)}, which correspond to the major semi-axis of the elliptical aperture at $\mu_{B} = 25$~mag~arcsec$^{-2}$ isophote (hereafter $R_{\rm op}$) of the CALIFA spirals.
As the figure shows, for most spiral galaxies 2\rp~$\leq R_{\rm op} \leq$ 4\rp, with the exception of the Sa-Sab galaxies for most of which $R_{\rm op}$ seems to be lower than 4\rp. 
This differential behavior is likely due to the fact that Sa-Sab galaxies present promiment bulges that result in reduced values of \rp.
In addition, it is shown that \re~$\leq R_{\rm op} \leq$ 3\re, and this relation holds in the same way for all morphological types.
In this case, all the morphological types behave similarly because \re~ is estimated by using only the light from the disk.

In the subsequent analysis we will present the results of the growth curves (as a function of \rp~and \re) of several observable fluxes or flux ratios based on a selected set of emission lines.
They will be used as indicators of the bias induced when estimating galaxy properties 
from fluxes measured within reduced apertures instead of taking the integrated values of these fluxes.
Hereafter we will refer to the median growth curves of the H$\alpha$ flux, and the H$\alpha$/H$\beta$, N2 and O3N2 ratios as a function of \rp~ as \xhap, \xhahbp, \xnhap~ and \xonp~ respectively.
These curves represent the value of the parameter measured within a circular aperture of a given radius normalized to the value of the parameter measured within 36'', which is the radius of the largest circular aperture considered.
This way, the value of all parameters within a circular aperture of radius 36'' will be taken as the integrated value of this property.
In the case of N2 and O3N2, the growth curves are logarithmic.
Thus, 

\xhap$_{r}$ = $f$(H$\alpha$)$_{r}$/$f$(H$\alpha$)$_{36''}$, 

\xhahbp$_{r}$ = ($f$(H$\alpha$)$_{r}$/$f$(H$\beta$)$_{r}$)/($f$(H$\alpha$)$_{36''}$/$f$(H$\beta$)$_{36''}$), 

\xnhap$_{r}$ = $\log$[$f$([N{\sc ii}]6583)/$f$(H$\alpha$)]$_{r}$$-$$\log$[$f$([N{\sc ii}]6583)/$f$(H$\alpha$)]$_{36''}$, 

and

\xonp$_{r}$ = $\log$[($f$([O{\sc iii}]5007)/$f$(H$\beta$))/($f$([N{\sc ii}]6583)/$f$(H$\alpha$))]$_{r}$$-$

$-$$\log$[($f$([O{\sc iii}]5007)/$f$(H$\beta$))/($f$([N{\sc ii}]6583)/$f$(H$\alpha$))]$_{36''}$.

Correspondingly, we will refer to the growth curves of the H$\alpha$ flux, and the H$\alpha$/H$\beta$, N2 and O3N2 ratios as a function of \re~ as \xhae, \xhahbe, \xnhae~ and \xone~ respectively.

The procedure followed to produce the growth curves is similar to the one described in IP13.
For each galaxy, unidimensional spectra are constructed by adding all the pixels within circular apertures of radii varying from 3'' to 36'', in steps of 3''.
In order to obtain the pure emission line spectra, we use a single stellar population (SSP) fitting to remove the contribution of the underlying continuum of the stellar population. 
We apply a linear combination of two SSP synthesis models of Vazdekis et al. (2010) based on the MILES stellar library (S\'{a}nchez-Bl\'{a}zquez et al. 2006) and a Kroupa IMF (Kroupa 2001). 
The ages of the SSPs used range from 0.10 to 0.79 Gyr for the case of the young stellar population and from 2.00 to 14.13 Gyr for the old stellar population. 
Five different metallicities are considered for each age ([M/H] values equal to 0.00, 0.20, -0.40, -0.71 and -1.31 dex offset from the solar value).
Once the underlying stellar continuum is removed, the fluxes of the emission lines were obtained from gaussian fits to the residual spectra.
The procedure followed to get the fits is as follows: first we fit the triplet H$\alpha$+[N{\sc ii}] assuming a common recession velocity and FWHM for the three lines, and the usual line ratio $f$(6548)/$f$(6583)=0.333.
No broad components in H$\alpha$ were considered because we have removed AGNs from our samples and thus only narrow emission lines are expected. 
Then we produced individual fits for H$\beta$ and [O{\sc iii}] leaving free the recession velocity and the FWHM but using as first guess the values obtained for H$\alpha$+[N{\sc ii}].
After that, for each galaxy we have 12-element vectors containing the fluxes of the emission lines contained within each of the circular apertures. 
The median growth curves as a function of \rp~or\re~where then produced by combining the properly interpolated vectors.

In addition to this, given that we focus on star-forming galaxies, for the subsequent analysis we remove those CALIFA galaxies classified as AGN, according to the prescriptions of Best et al. (2012), Kewley et al. (2001, 2006) and Cid Fernandes et al. (2011), and to the NED\footnote{\tt {http://ned.ipac.caltech.edu/}} database.
In order to keep only galaxies with good quality data we will remove those galaxies for which $f$(H$\alpha$)~$> 0$ and $\sigma$(H$\alpha$)/$f$(H$\alpha$)~$\leq 0.333$ in all the circular apertures.
We note that the results of this study are the same if we apply a 2-$\sigma$ cut in signal-to-noise, resulting in a significantly larger sample, instead of the 3-$\sigma$ applied in what follows.

\begin{figure}
\plottwo{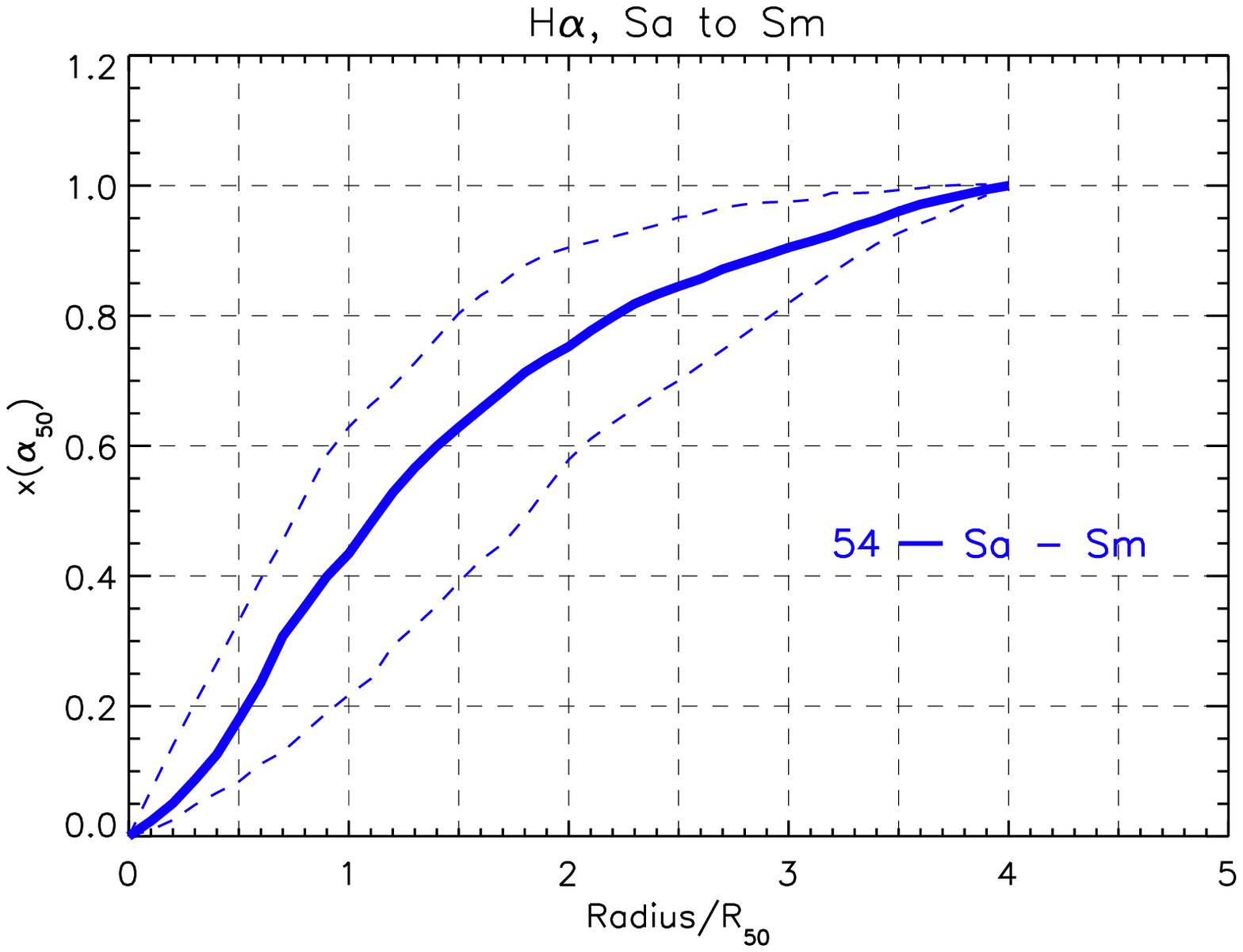}{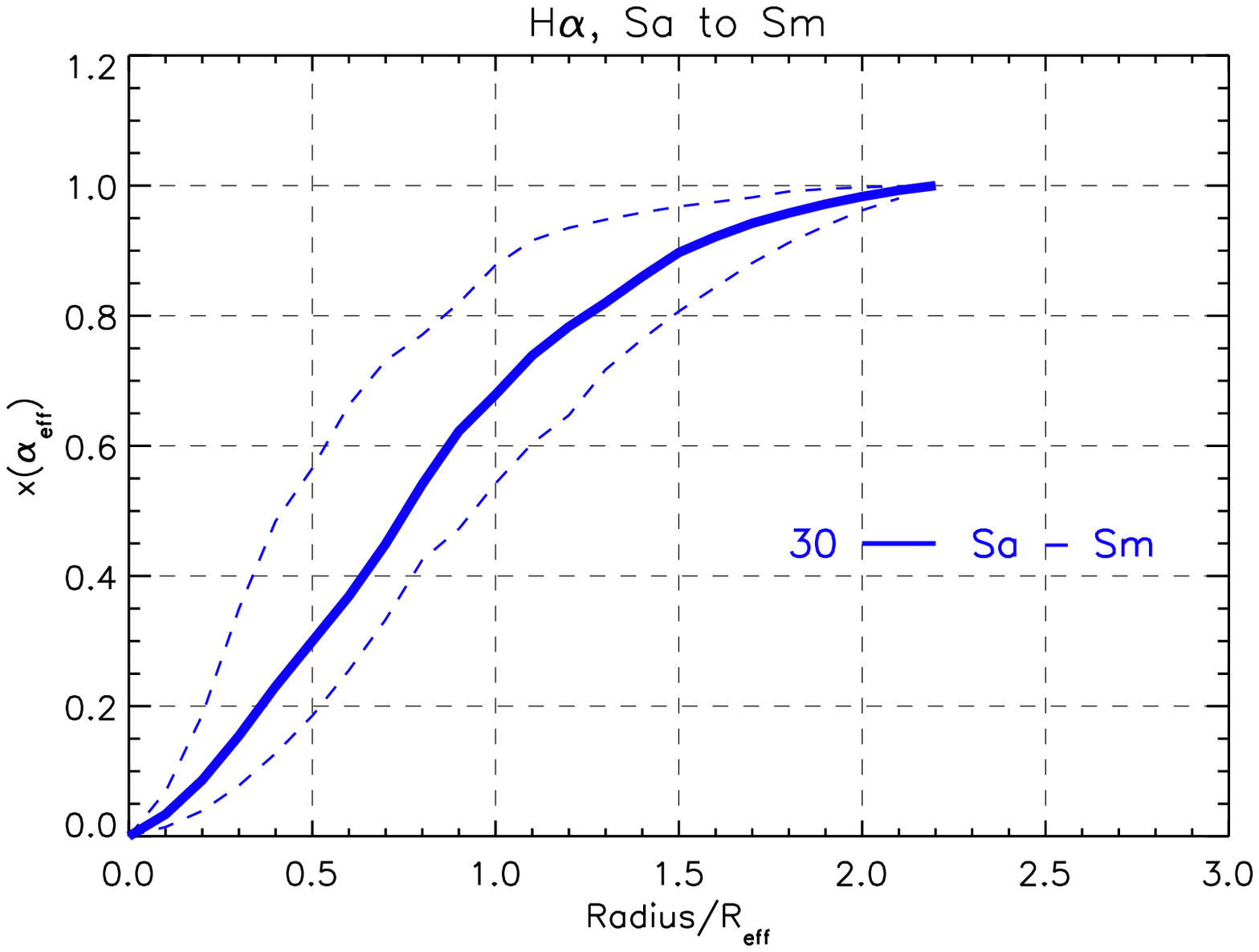}
\caption{{\bf Left:} \xhap~ for CALIFA spiral galaxies with \rp~$< 9$'' normalized to 4\rp.
Solid lines correspond to median values. 
Dashed lines contain 68.2\% of the values.
{\bf Right:} \xhae~ for CALIFA spiral galaxies with \re~$< 16.4$'' normalized to 2.2\re.\label{ha_40}}
\end{figure}

As a first step we investigate the distribution of the H$\alpha$ emission in the CALIFA galaxies.
Figure~\ref{ha_40} shows \xhap~ and \xhae~ for the spirals covered in the CALIFA aperture up to 4\rp~and 2.2\re~respectively.
These limits were imposed with the compromise of covering as much as possible the disks of the CALIFA galaxies, but keeping samples with reasonable numbers of galaxies.
In both cases, \xhap~ and \xhae~ seem to saturate at 4\rp~and 2.2\re~respectively, suggesting that the bulk of the H$\alpha$ emission of spiral galaxies is contained within these apertures.
But the numbers of galaxies available for this test are very low for a detailed study of the aperture effects including the role of the stellar mass and morphological type.
For this reason, we chose two less conservative limits, namely 2.5\rp~and 1.5\re, that contain on average $85$\%\footnote{This value is slightly lower than the one, 90\%, reported in IP13. Given that the current sample is larger, we are more confident about the value reported in this paper.} and $90$\% of the total H$\alpha$ emission in spiral galaxies (see figure~\ref{ha_40}).

After imposing this last condition, we end up with 
165 galaxies covered up to 2.5\rp~and 133 galaxies covered up to 1.5\re, which is the case for galaxies with \rp~ $\leq$ 14.4\arcsec\ and \re~ $\leq$ 24\arcsec, respectively\footnote{Further restrictions must be imposed to the subsamples when analyzing the H$\alpha$/H$\beta$, N2 and O3N2 growth curves, namely $f$(H$\beta$)~$> 0$, $\sigma$(H$\beta$)/$f$(H$\beta$)~$\leq 0.333$, $f$([\ion{N}{2}])~$> 0$, $\sigma$([\ion{N}{2}])/$f$([\ion{N}{2}])~$\leq 0.333$, $f$([\ion{O}{3}])~$> 0$, and $\sigma$([\ion{O}{3}])/$f$([\ion{O}{3}])~$\leq 0.333$, in all circular apertures. These restrictions shorten $S_{50}$ to 159, 106 and 104 galaxies, and $S_{\rm eff}$ to 129, 92 and 90 galaxies respectively.}. 
These two subsamples ($S_{50}$ and $S_{\rm eff}$) will be the basis of our statistical study of aperture effects, taking into account that $S_{50}$ galaxies will be studied up to $\approx 2.5$\rp, which contains on average $85$\% of their H$\alpha$ emission, and that $S_{\rm eff}$ galaxies will be be studied up to $\approx 1.5$\re, which contains on average $90$\% of their H$\alpha$ emission.
Table~\ref{samples} shows the number of galaxies of each subsample after imposing each of the filters previously mentioned.
Also, Table~\ref{samples_prop} contains some basic properties of the two subsamples.

\section{Results}

\subsection{Corrections for fixed angular apertures}

\subsubsection{H$\alpha$ growth curves}

We start discussing the behavior of the growth curves of \xhap~ and \xhae~ of our galaxies.
For this we first compare \xhap~ of IP13 with that obtained in this work.
As in IP13, we produce median growth curves and a confidence interval equivalent to 1$\sigma$, consistent in that one containing 68.2\% of the individual growth curves at each radius.
Table~\ref{compa} lists the values of \xhap~ of IP13 and the one produced with the present sample.
As it can be seen from Table~\ref{compa}, both curves are consistent within 1$\sigma$ along the probed range.
The increased number of galaxies in our current sample will make possible a more detailed analysis of the aperture corrections split by morphological types or stellar masses.

\begin{figure}
\plottwo{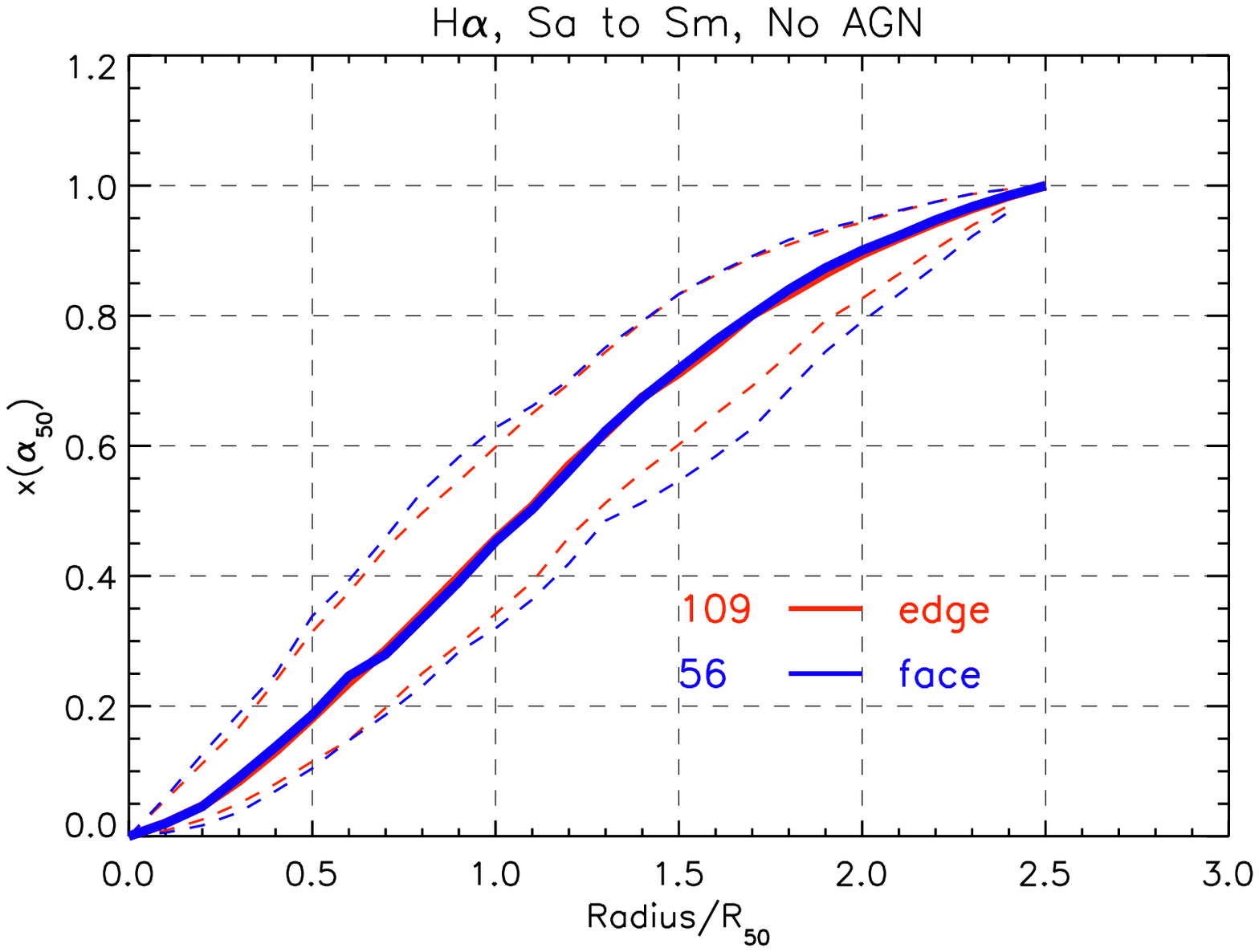}{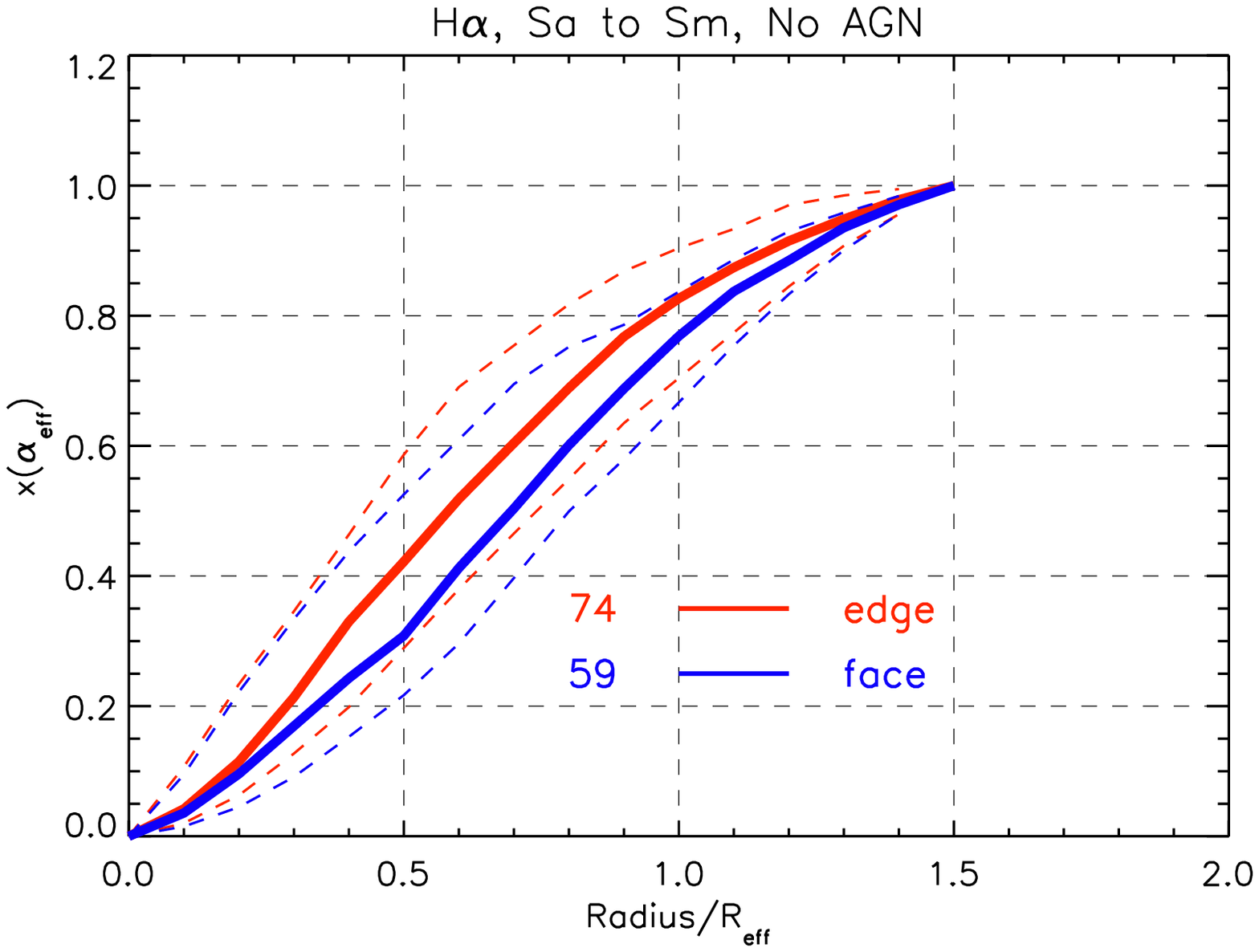}
\caption{\xhap~(left) and \xhae~(right) for galaxies with different inclination.
Solid lines correspond to median values. Dashed lines contain 68.2\% of the distribution.
Face (blue) and edge (red) galaxies have $b/a > 0.4$ and $b/a \leq 0.4$ respectively. 
The numbers within the plot box indicate the sizes of the samples.\label{ha_incli}}
\end{figure}

\begin{figure}
\plottwo{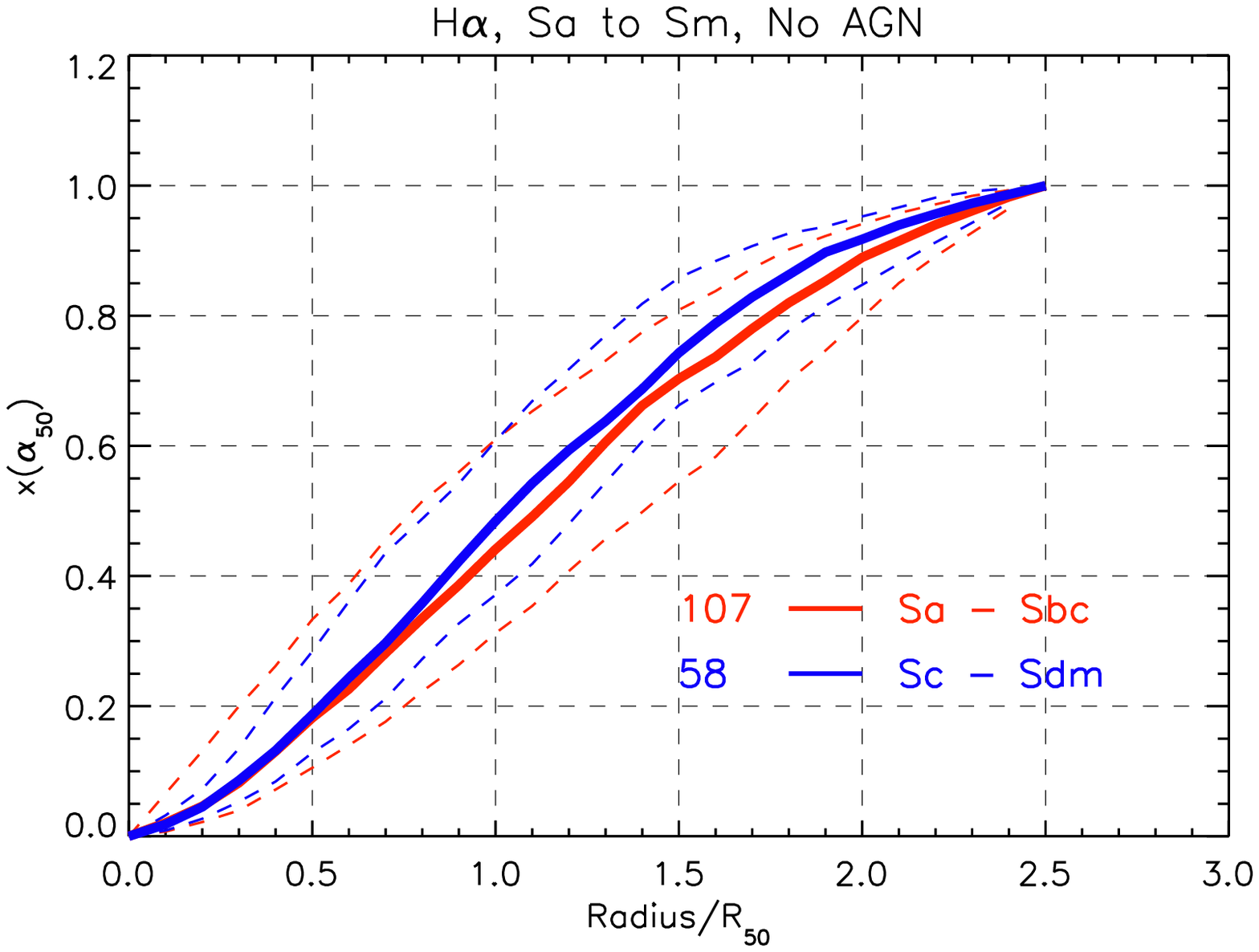}{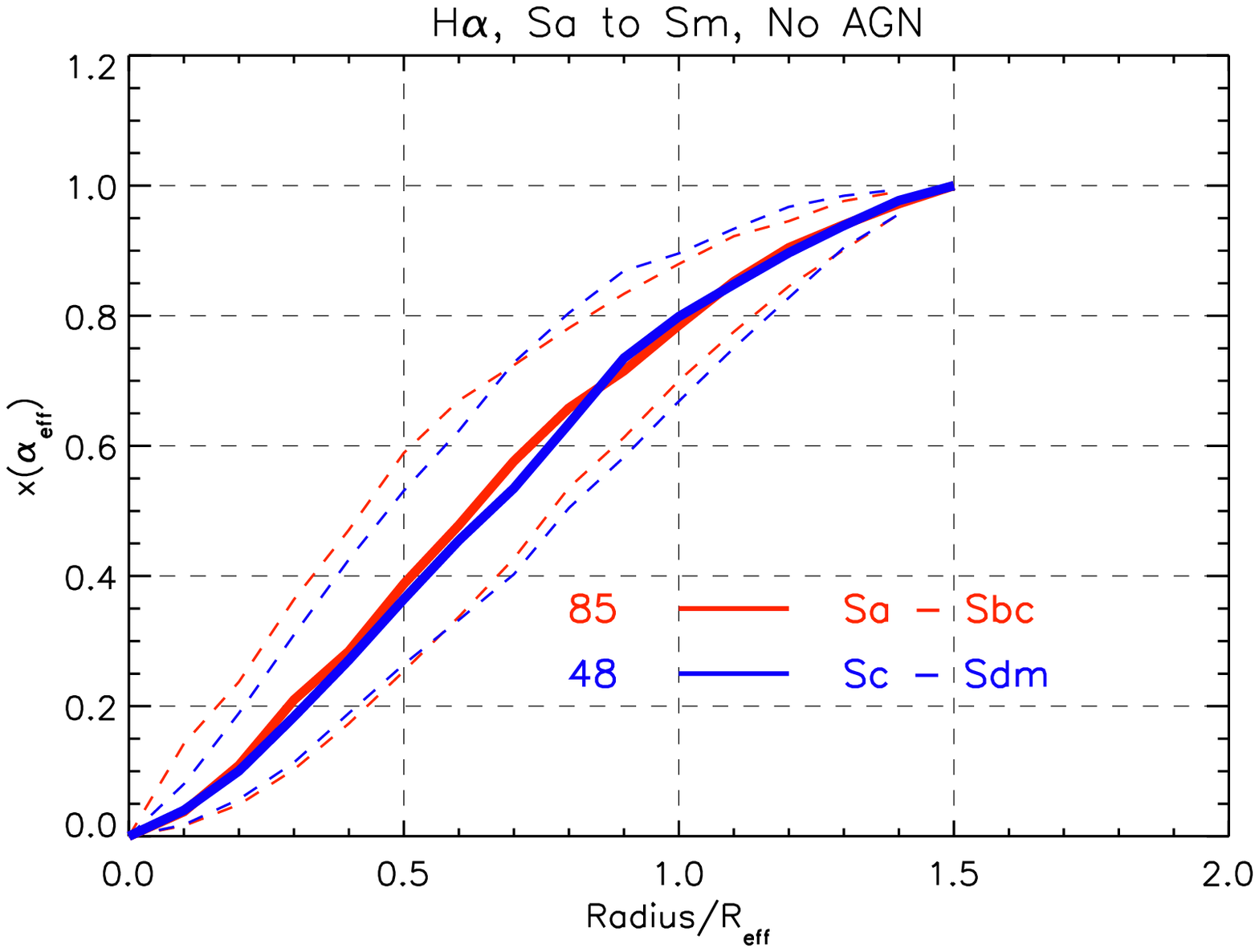}
\caption{\xhap~(left) and \xhae~(right) for galaxies with different morphological types.
Solid lines correspond to median values. Dashed lines contain 68.2\% of the distribution.
Sa-Sbc and Sc-Sm galaxies are represented in blue and red respectively.
The numbers within the plot box indicate the sizes of the samples.\label{ha_type}}
\end{figure}

\begin{figure}
\plottwo{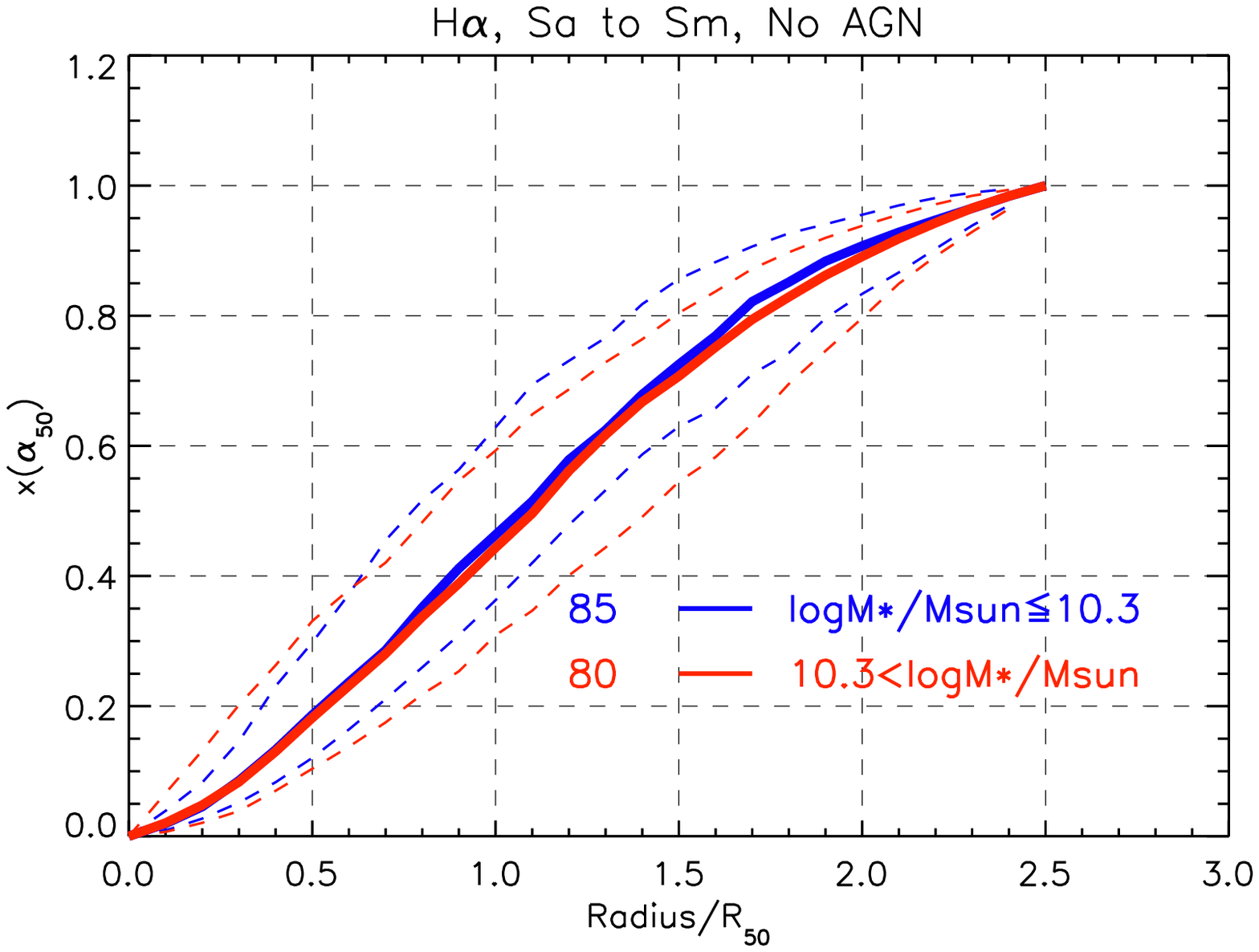}{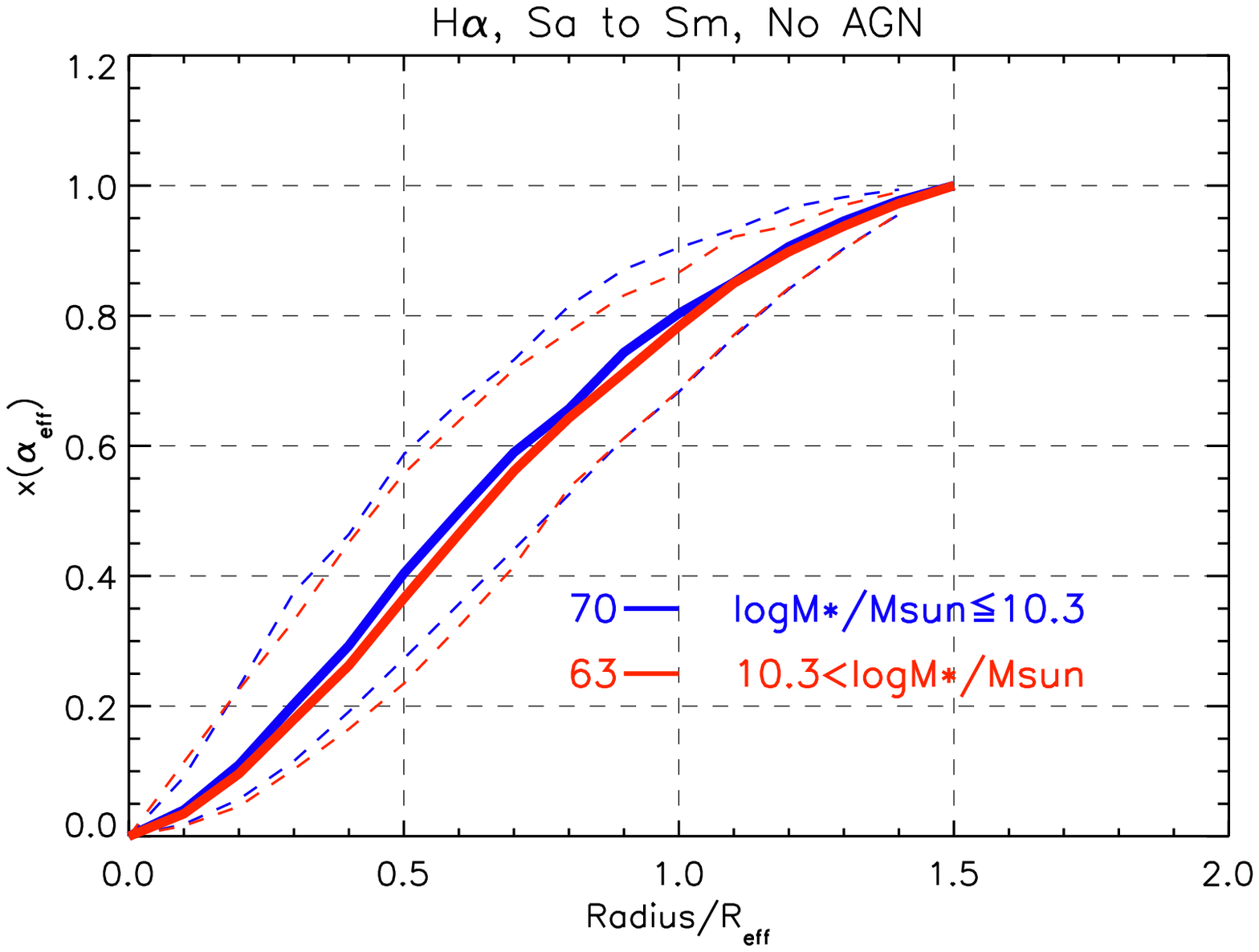}
\caption{\xhap~(left) and \xhae~(right) for galaxies with different stellar masses.
Solid lines correspond to median values. Dashed lines contain 68.2\% of the distribution.
Galaxies with $\log M^{*}/M_{\odot} \leq 10.3$ and $\log M^{*}/M_{\odot} > 10.3$ are represented in blue and red respectively.
The numbers within the plot box indicate the sizes of the samples.\label{ha_mass}}
\end{figure}

Figures~\ref{ha_incli} to \ref{ha_mass} show the H$\alpha$ growth curves split by galaxy inclination, morphological type and stellar mass.
Inclination is defined as the ratio of SDSS $isob\_r/isoa\_r$ and we consider face-on galaxies those with inclination larger than 0.4, the rest being edge-on galaxies.
Concerning the morphological types, two bins were defined as follows: Sa to Sbc, and Sc to Sdm. 
This splits our sample into early spirals and late spirals.
Finally, the stellar mass was also split in two ranges, separated at $\log M^{*}/M_{\odot} = 10.3$. 
This limit was imposed since after removing E-S0 galaxies and AGNs, it splits the sample in two subsamples of similar numbers of galaxies.
As it is clear from the figures, the H$\alpha$ growth curves do not seem to show any dependence on inclination, morphological type and stellar mass.
The only exception, for which a slight dependence is seen, is \xhae~when different inclinations are taken into account. 
In this case, the growth curves for edge-on and face-on galaxies look different, although still consistent with each other within the 1$\sigma$ limits shown in the figure.

\subsubsection{H$\alpha$/H$\beta$ growth curves}

\begin{figure}
\plottwo{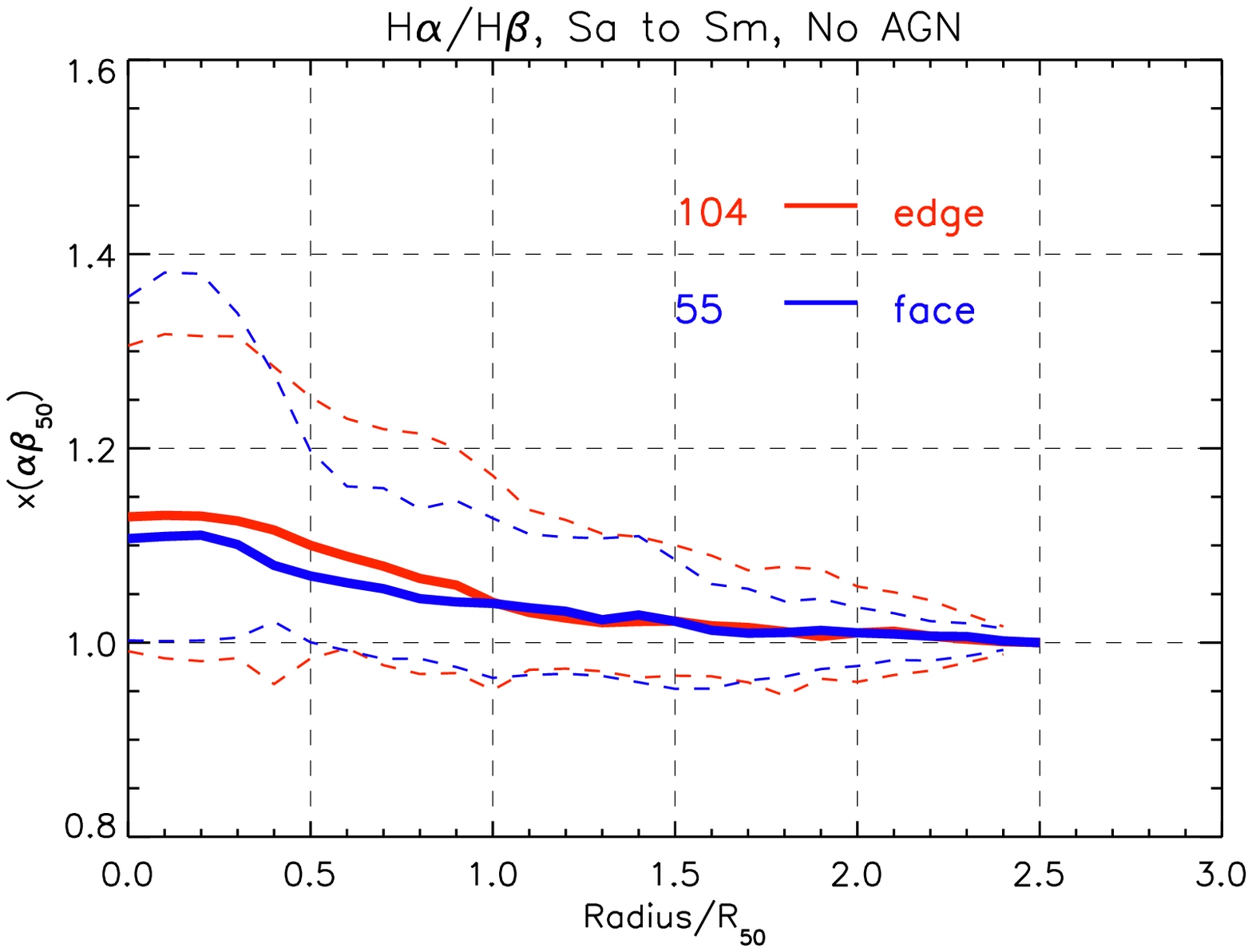}{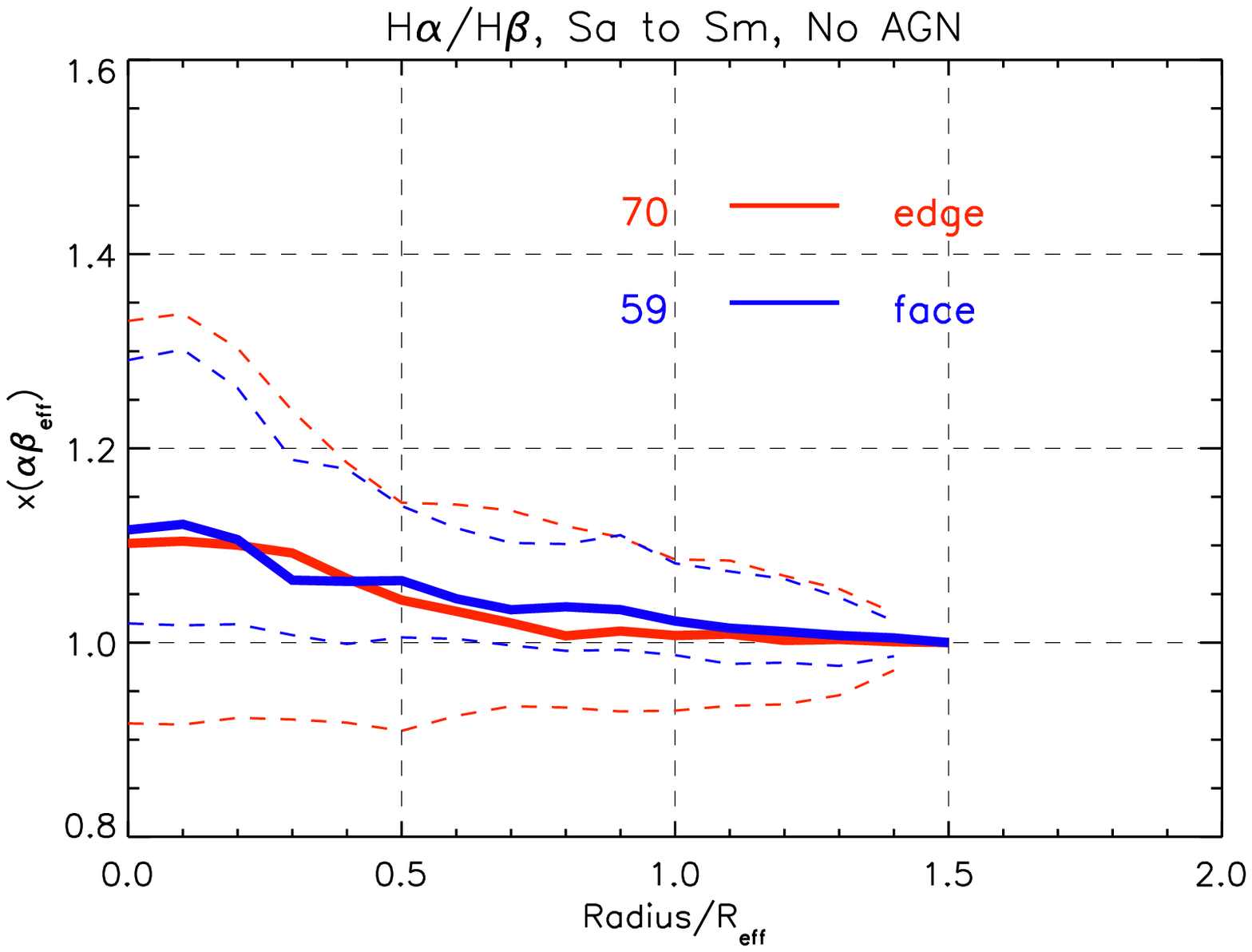}
\caption{\xhahbp~(left) and \xhahbe~(right) for galaxies with different inclination.
Solid lines correspond to median values. Dashed lines contain 68.2\% of the distribution.
Face (blue) and edge (red) galaxies have $b/a > 0.4$ and $b/a \leq 0.4$ respectively. 
The numbers within the plot box indicate the sizes of the samples.\label{hahb_incli}}
\end{figure}

\begin{figure}
\plottwo{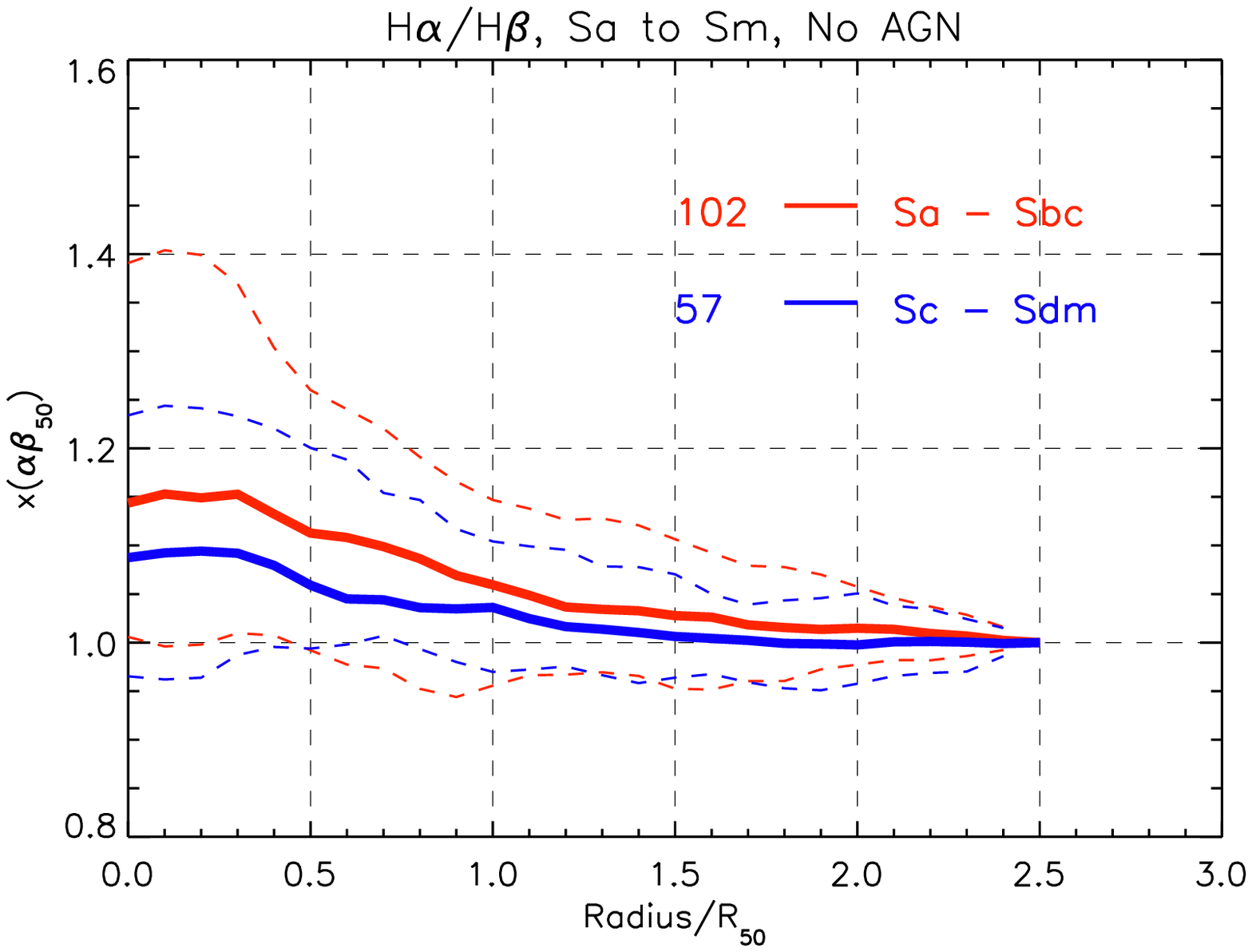}{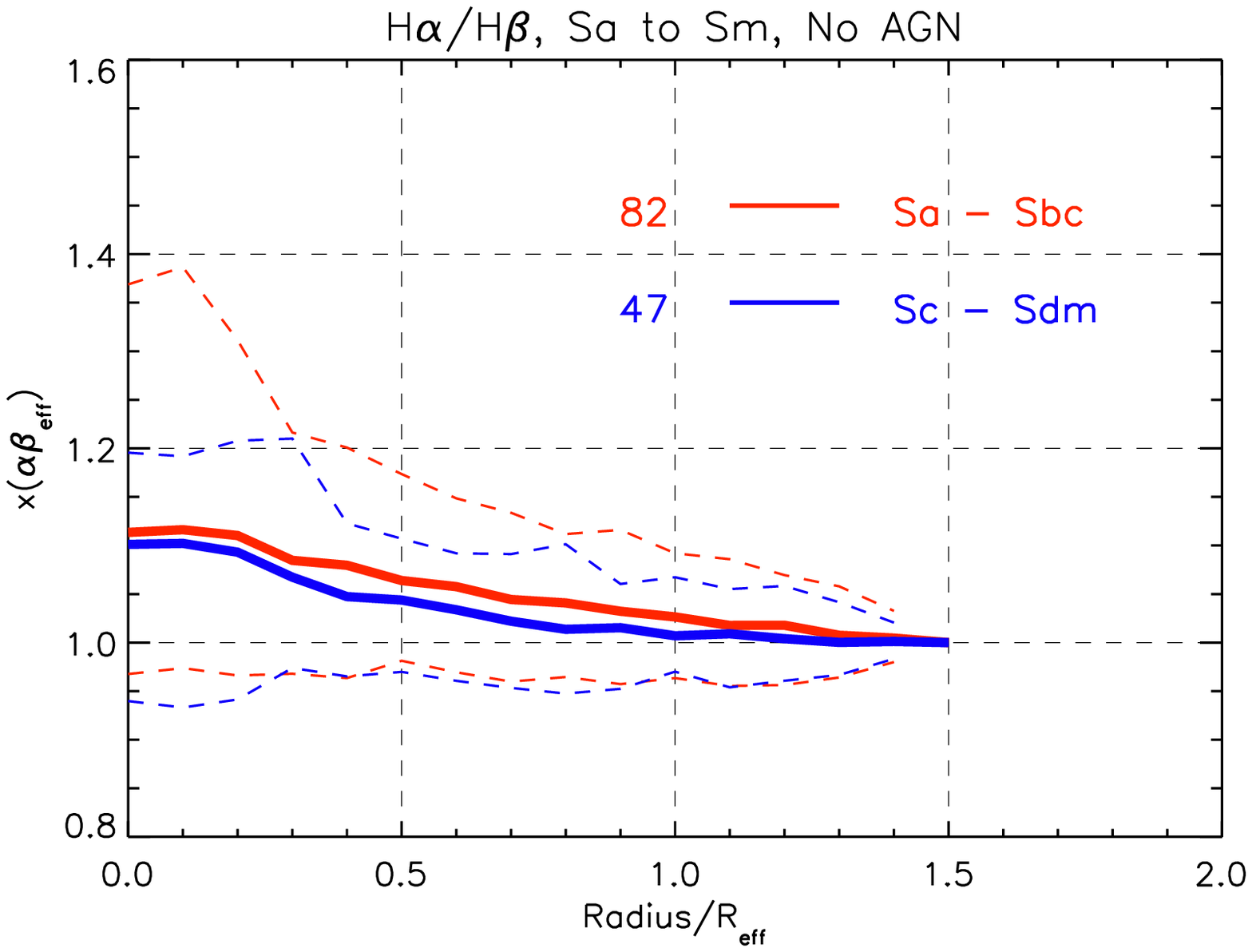}
\caption{\xhahbp~(left) and \xhahbe~(right) for galaxies with different morphological types.
Solid lines correspond to median values. Dashed lines contain 68.2\% of the distribution.
Sa-Sbc and Sc-Sm galaxies are represented in blue and red respectively.
The numbers within the plot box indicate the sizes of the samples.\label{hahb_type}}
\end{figure}

\begin{figure}
\plottwo{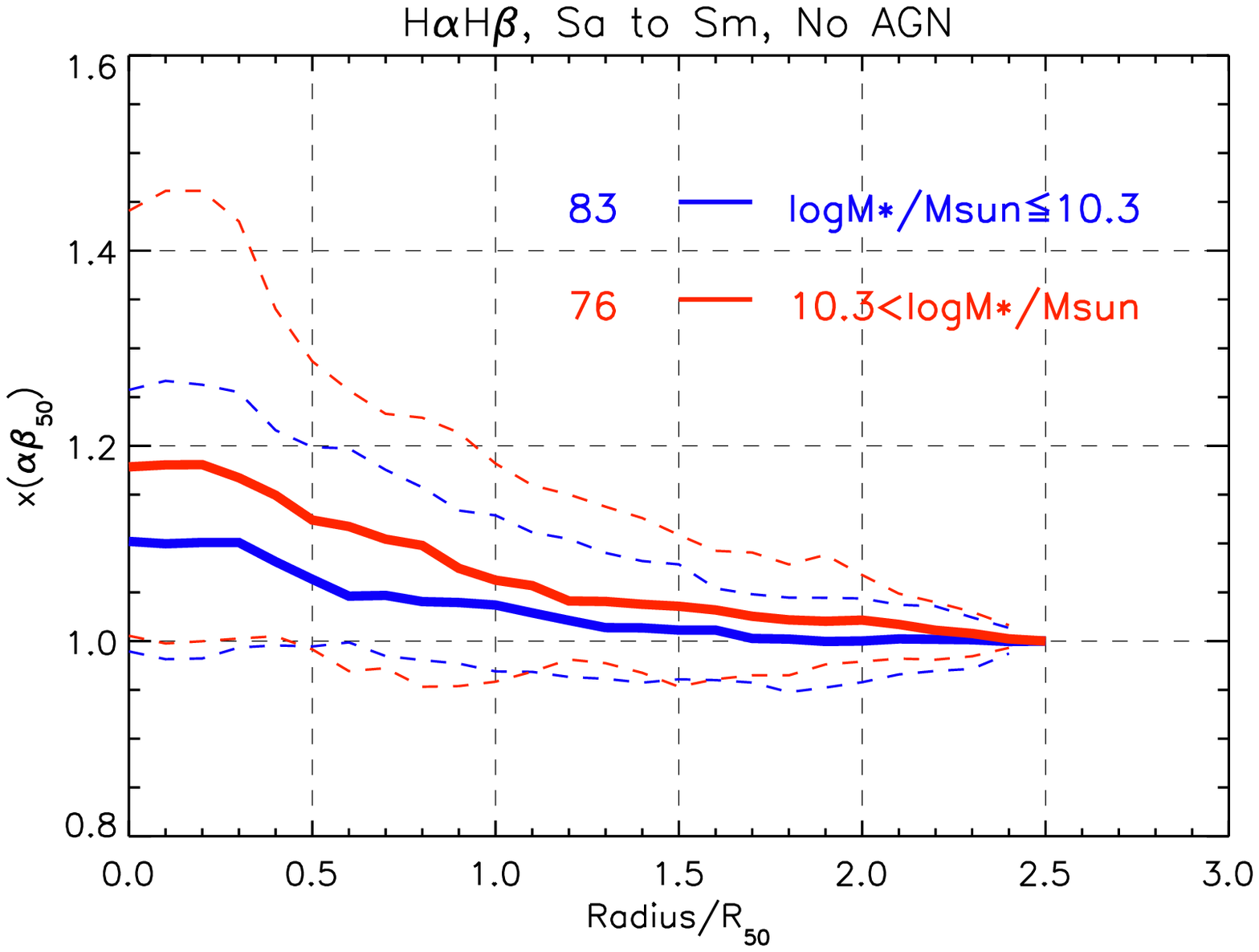}{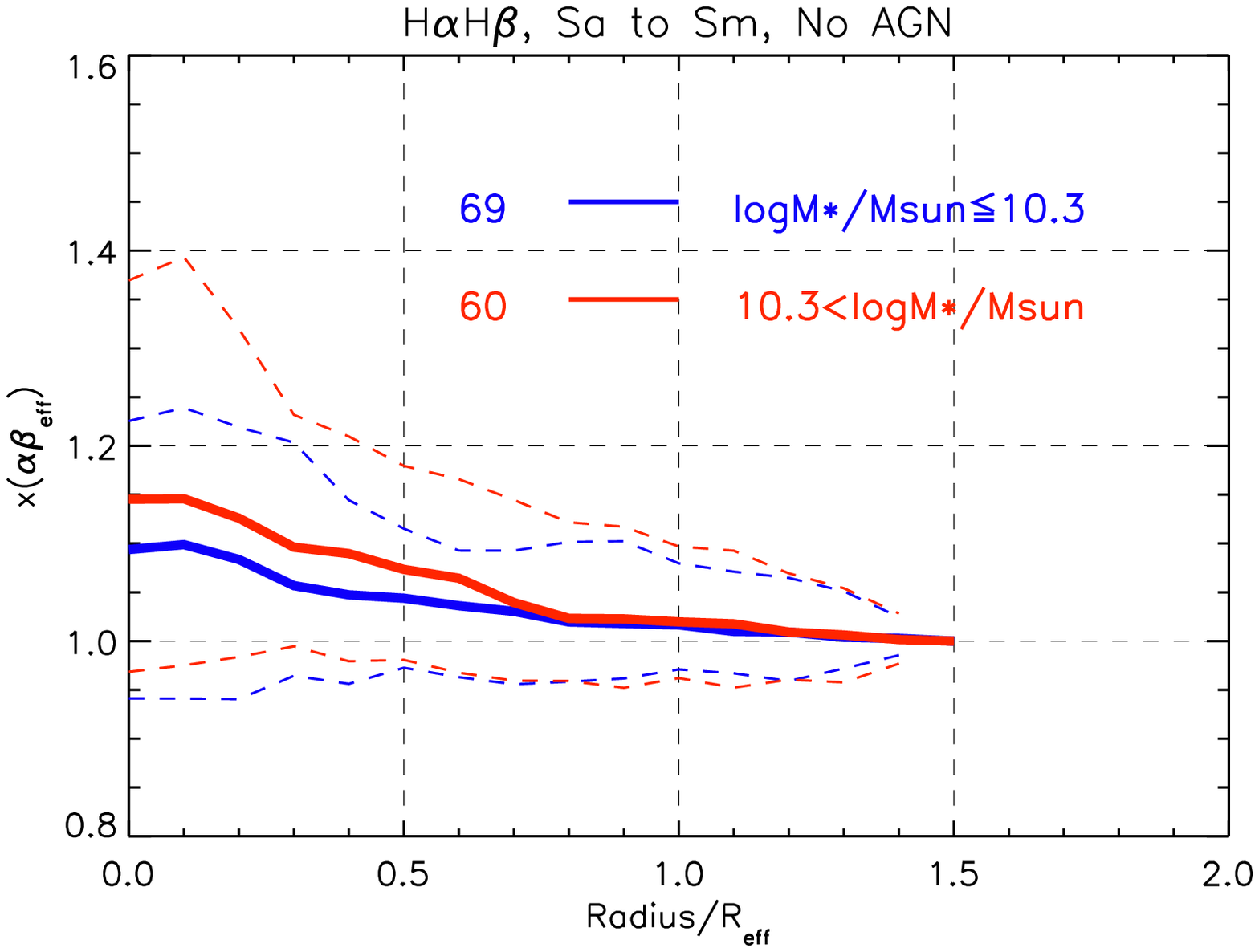}
\caption{\xhahbp~(left) and \xhahbe~(right) for galaxies with different stellar masses.
Solid lines correspond to median values. Dashed lines contain 68.2\% of the distribution.
Galaxies with $\log M^{*}/M_{\odot} \leq 10.3$ and $\log M^{*}/M_{\odot} > 10.3$ are represented in blue and red respectively.
The numbers within the plot box indicate the sizes of the samples.\label{hahb_mass}}
\end{figure}

Figure~\ref{hahb_incli} shows \xhahbp~ and~ \xhahbe~ for different inclinations.
Both curves are almost coincident irrespective of the inclination of the galaxies.
It is remarkable that both \xhahbp~and \xhahbe~show a very mild decline from the central regions to the outskirts, with a maximum value of $\approx 1.1$, which corresponds to an increase of 0.20~dex in $c$(H$\beta$) or 0.43~mag in $A_{V}$ at the center of the galaxies.
This means that observing galaxies through a small aperture results in values of the extinction larger than those obtained with the integrated values of the H$\alpha$ and H$\beta$ fluxes.
Figure~\ref{hahb_type} shows \xhahbp~and \xhahbe~ for different morphological types.
\xhahbp~is slightly higher for Sa-Sbc than for Sc-Sm galaxies, reaching values of $\approx 1.15$ at the center of the galaxies in the first case.
This difference is not observed for \xhahbe.
Figure~\ref{hahb_mass} shows \xhahbp~and \xhahbe~ for different stellar masses.
As in the previous case, \xhahbp~is slightly higher for galaxies with $\log M^{*}/M_{\odot} > 10.3$ than for galaxies with $\log M^{*}/M_{\odot} \leq 10.3$ for small apertures, but this difference almost disappears in \xhahbe.
However, the differences observed in the median growth curves for galaxies of different morphological types and stellar masses are always lower than the dispersions around these median growth curves, which as the figures show, increase as the radii of the apertures decrease.

This slight increase in the H$\alpha$/H$\beta$ growth curve observed for small apertures with respect
to the integrated values can be compared with the variation in this Balmer line 
ratio as a function of electron temperature. 
Assuming case B recombination and low-density limit, a substantial change, 
from 20000~K to 5000~K, in electron temperature should be 
present to see a $10\%$ change in H$\alpha$/H$\beta$.
This change in electronic temperature is not observed in typical H{\sc ii} regions in the disks of spirals (P\'{e}rez-Montero \& Contini 2009). For this reason, a radial gradient on the dust content of star forming regions, 
likely related to the abundance gradient measured in spiral galaxies,
would explain this aperture effect.

\subsubsection{N2 growth curves}

\begin{figure}
\plottwo{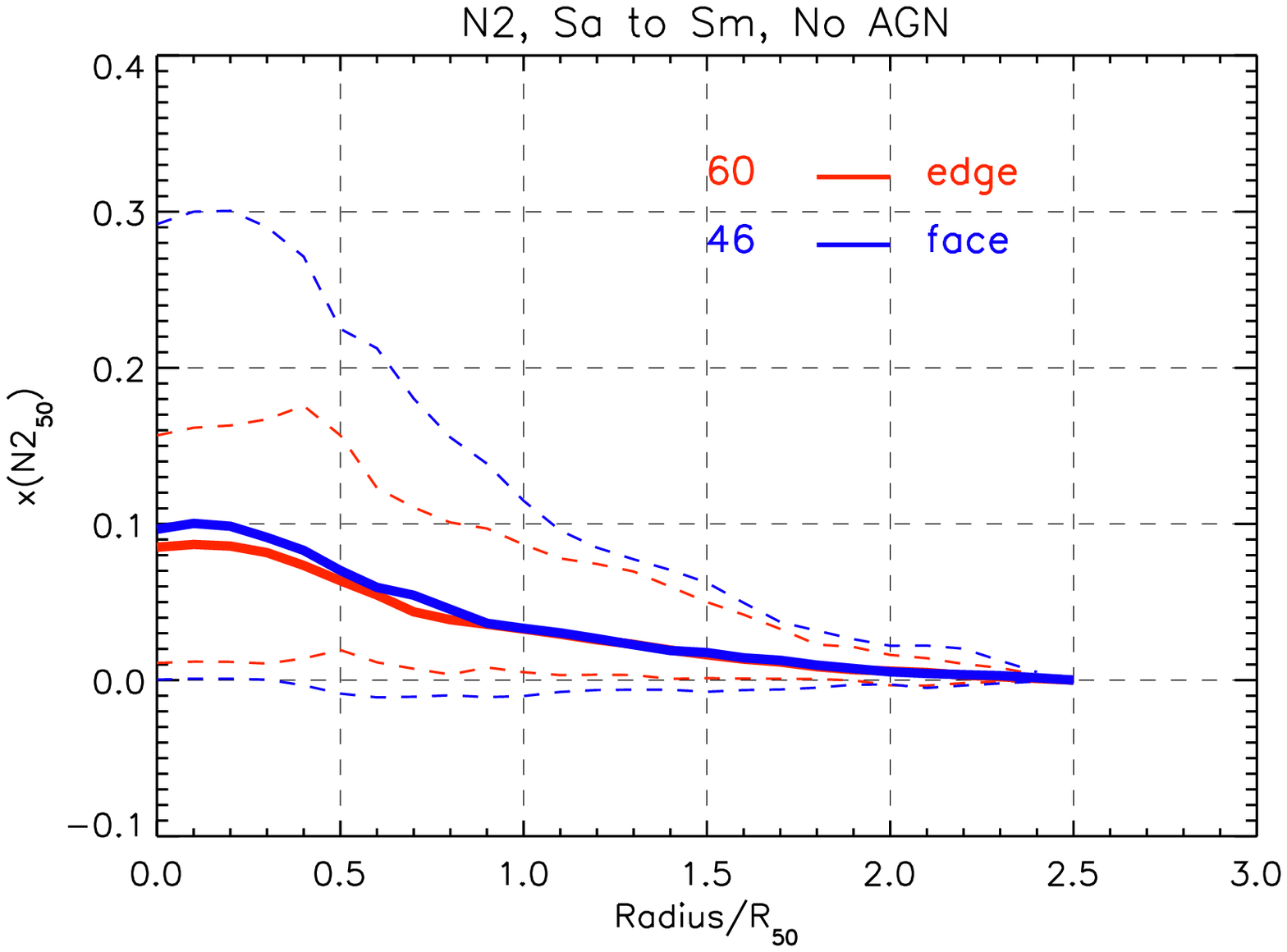}{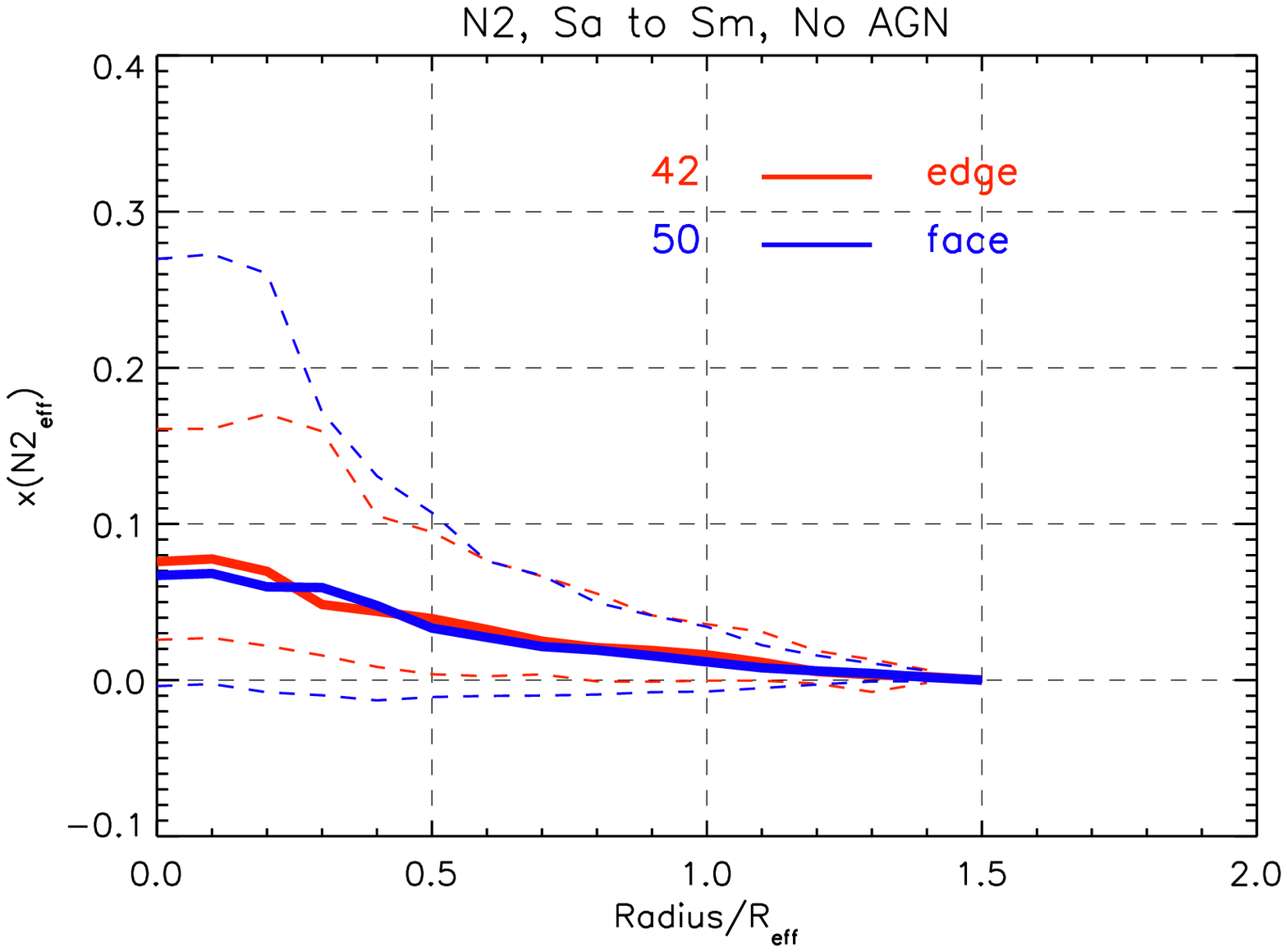}
\caption{\xnhap~(left) and \xnhae~(right) for galaxies with different inclination.
Solid lines correspond to median values. Dashed lines contain 68.2\% of the distribution.
Face (blue) and edge (red) galaxies have $b/a > 0.4$ and $b/a \leq 0.4$ respectively. 
The numbers within the plot box indicate the sizes of the samples.\label{n2ha_incli}}
\end{figure}

\begin{figure}
\plottwo{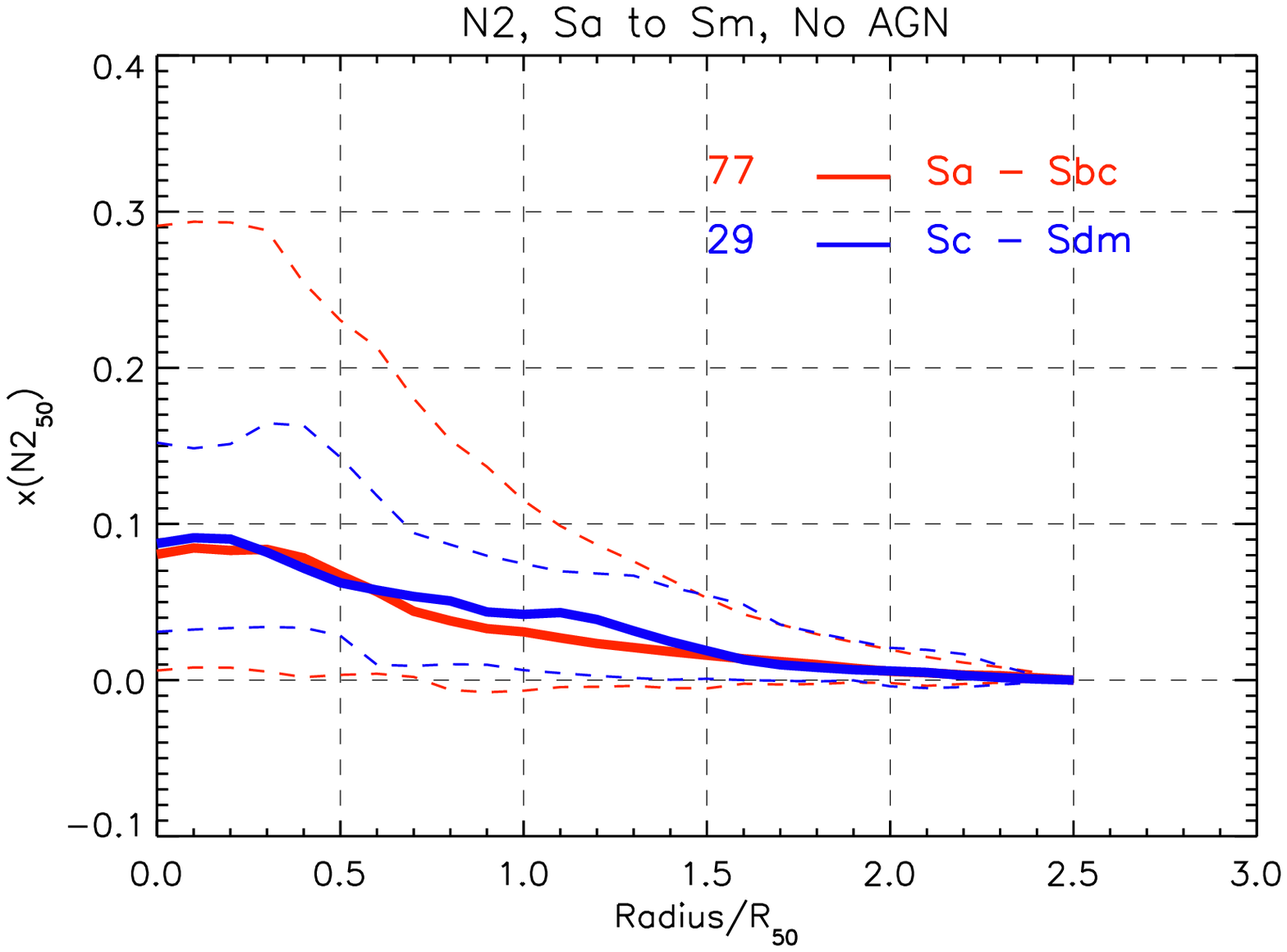}{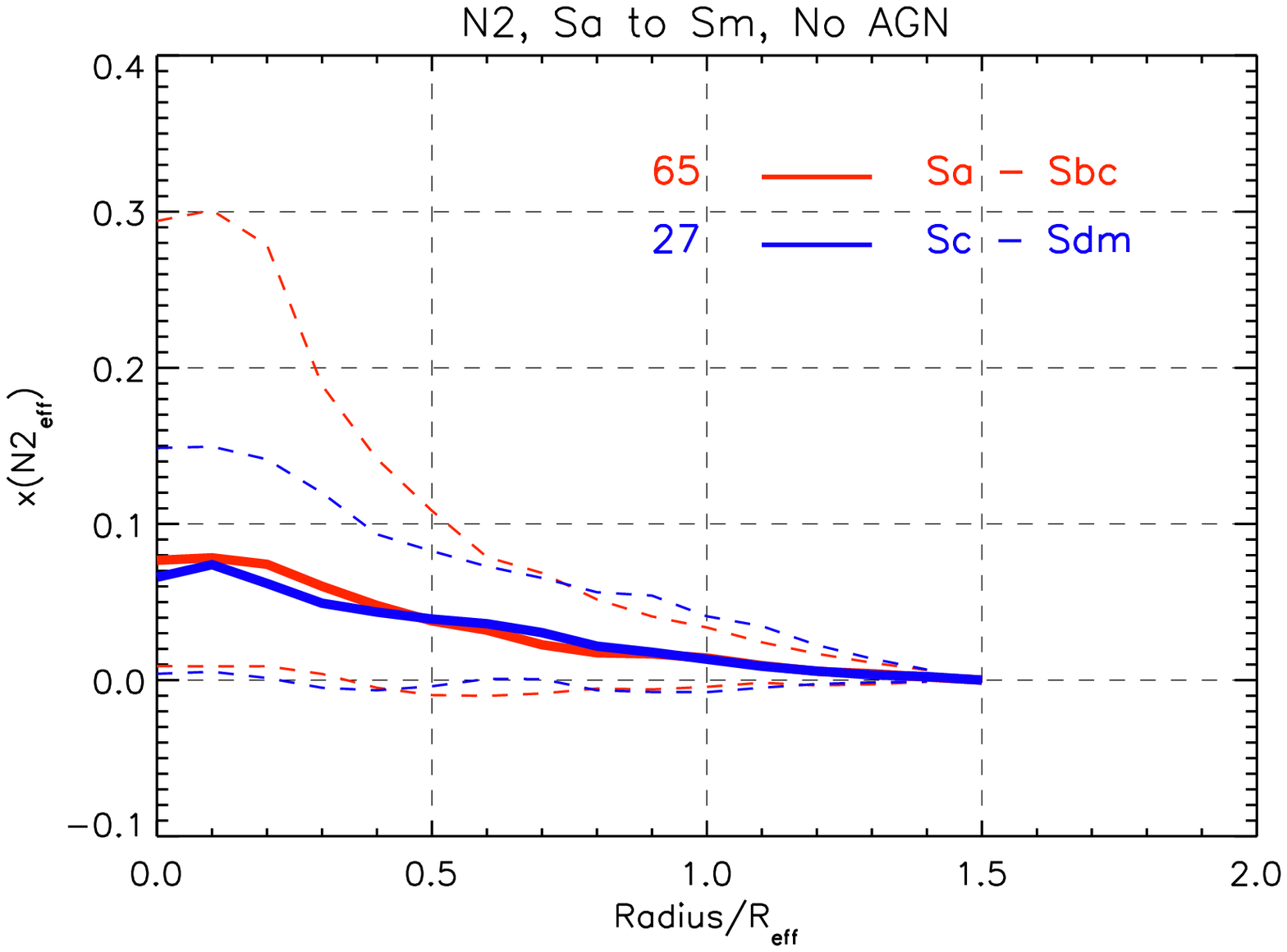}
\caption{\xnhap~(left) and \xnhae~(right) for galaxies with different morphological types.
Solid lines correspond to median values. Dashed lines contain 68.2\% of the distribution.
Sa-Sbc and Sc-Sm galaxies are represented in blue and red respectively.
The numbers within the plot box indicate the sizes of the samples.\label{n2ha_type}}
\end{figure}

\begin{figure}
\plottwo{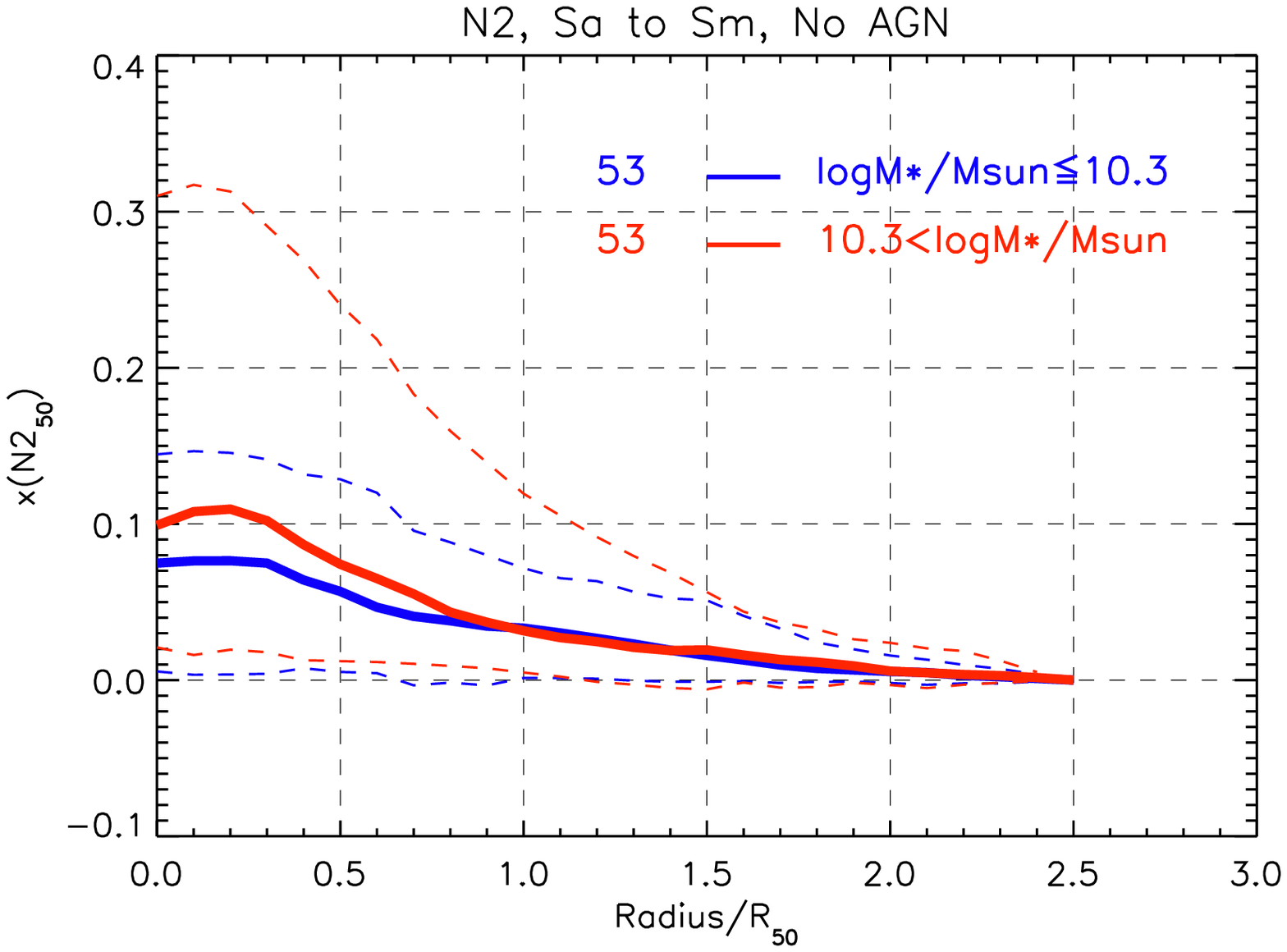}{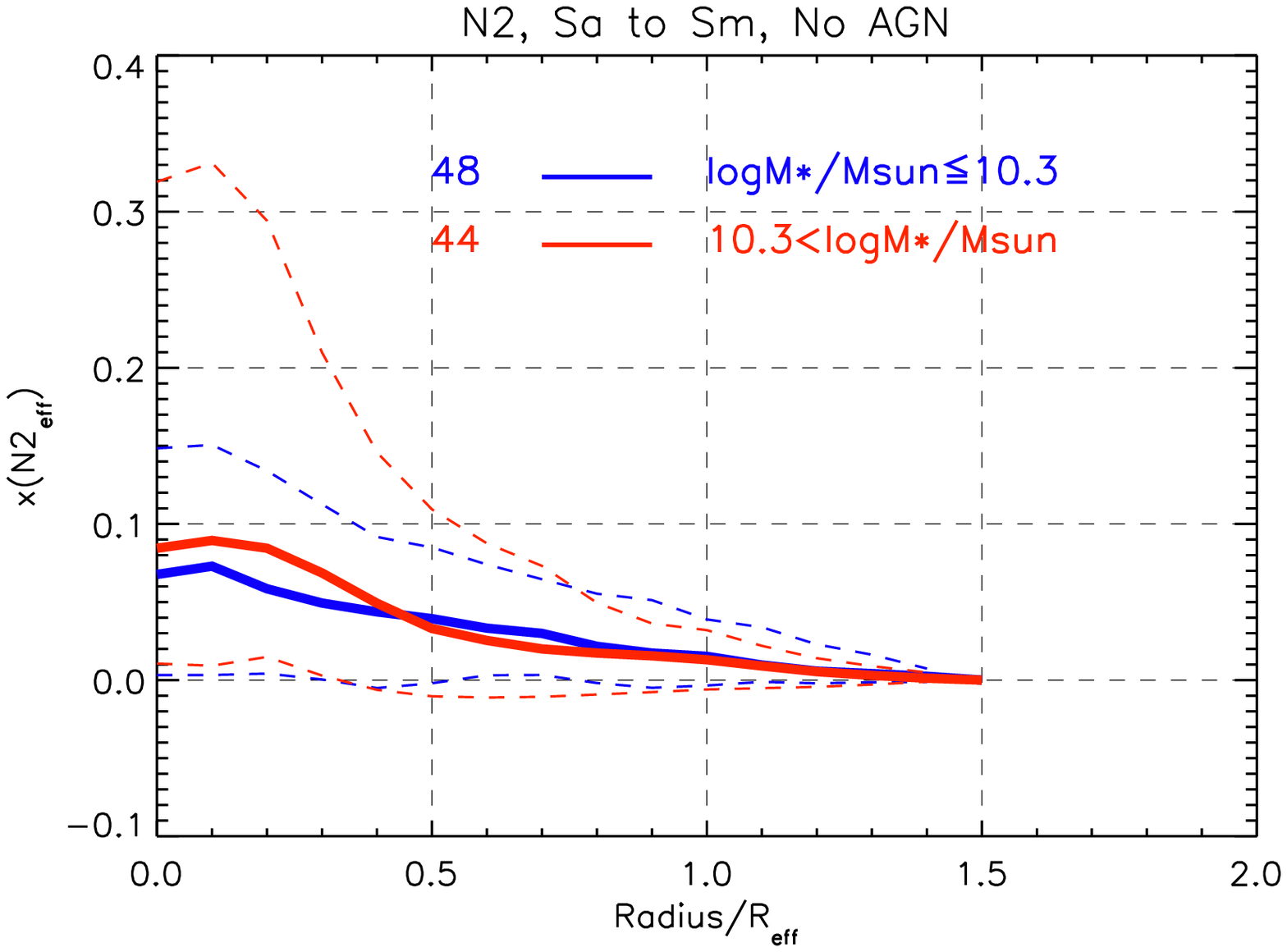}
\caption{\xnhap~(left) and \xnhae~(right) for galaxies with different stellar masses.
Solid lines correspond to median values. Dashed lines contain 68.2\% of the distribution.
Galaxies with $\log M^{*}/M_{\odot} \leq 10.3$ and $\log M^{*}/M_{\odot} > 10.3$ are represented in blue and red respectively.
The numbers within the plot box indicate the sizes of the samples.\label{n2ha_mass}}
\end{figure}

Figure~\ref{n2ha_incli} shows \xnhap~ and \xnhae~ for different inclinations.
\xnhap~and \xnhae~behave in a similar way as \xhahbp~and \xhahbe, that is, they show a mild decline from the inner parts of the galaxies and do not show clear differences related to the inclination of galaxies.
\xnhap~and \xnhae~show central values reach maxima of $\approx 1.2-1.3$ and larger dispersions around these central values in the inner regions of the galaxies compared to the outskirts.
The meaning of this trend is that observing galaxies through small apertures results in values of N2 larger than the ones obtained with the integrated values.
Figure~\ref{n2ha_type} shows \xnhap~and \xnhae~ for different morphological types.
The behavior of \xnhap~and \xnhae~is similar to the one previously described.
Figure~\ref{n2ha_mass} shows \xnhap~and \xnhae~ for different stellar masses which shows a marginal difference for small apertures with more massive galaxies showing larger values of \xnhap~ and \xnhae~ than less massive ones. However, again this difference is much smalles than the dispersions around the median values of \xnhap~ and \xnhae.
As we will show in the next section, the radial change of the N2 growth curves reflects in a radial change in the oxygen abundances.

\subsubsection{O3N2 growth curves}

\begin{figure}
\plottwo{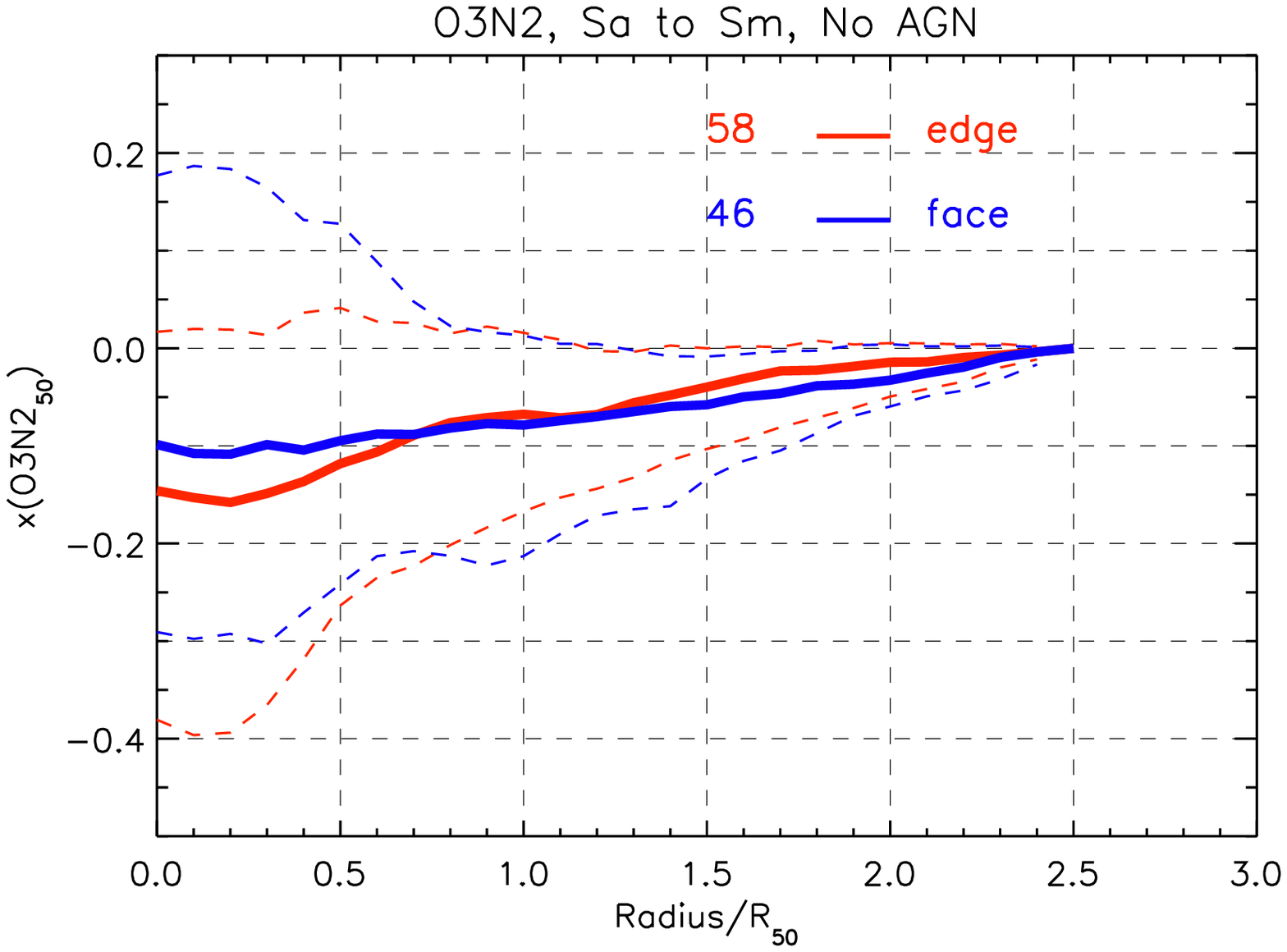}{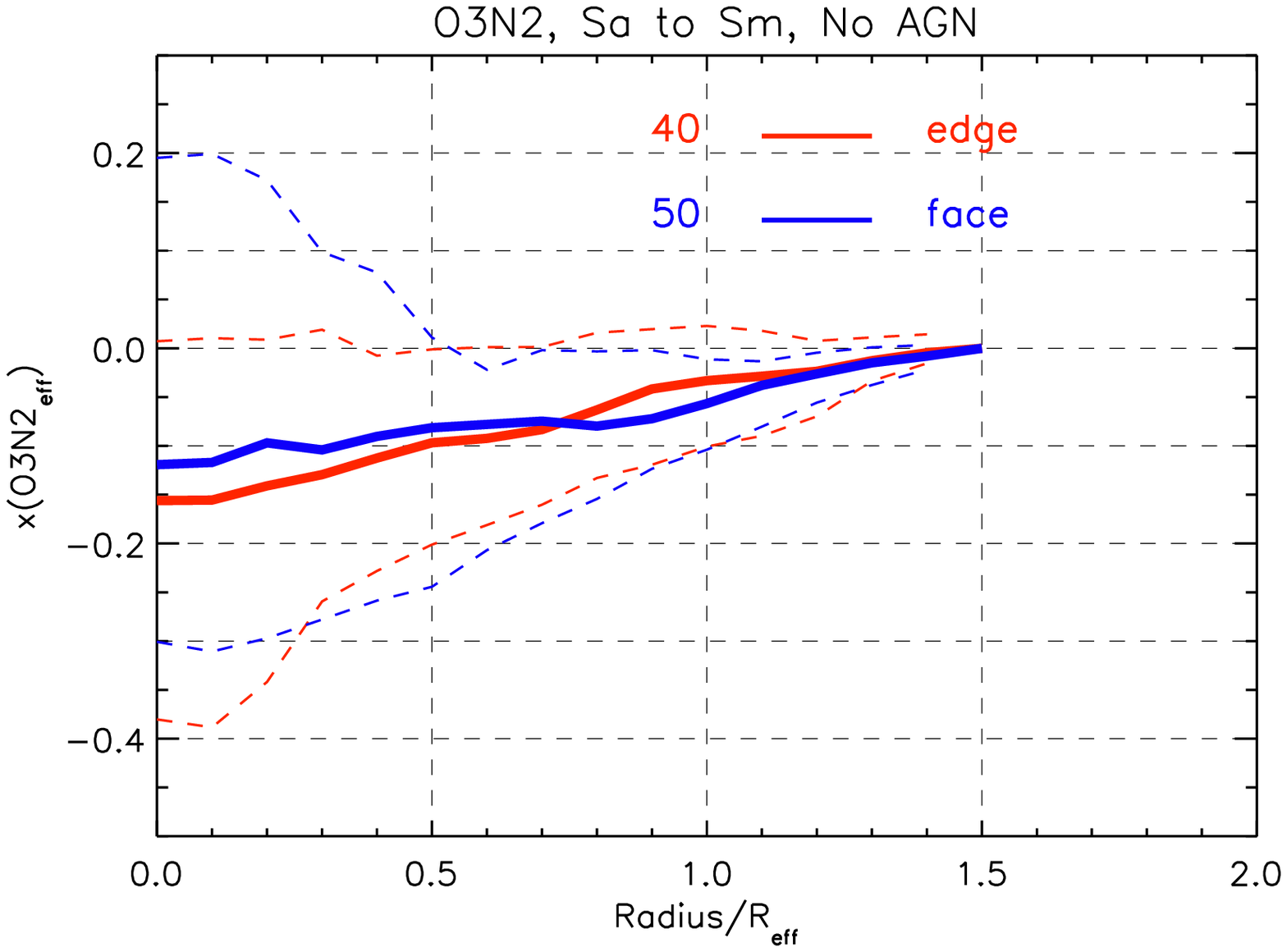}
\caption{\xonp~(left) and \xone~(right) for galaxies with different inclination.
Solid lines correspond to median values. Dashed lines contain 68.2\% of the distribution.
Face (blue) and edge (red) galaxies have $b/a > 0.4$ and $b/a \leq 0.4$ respectively. 
The numbers within the plot box indicate the sizes of the samples.\label{o3n2_incli}}
\end{figure}

\begin{figure}
\plottwo{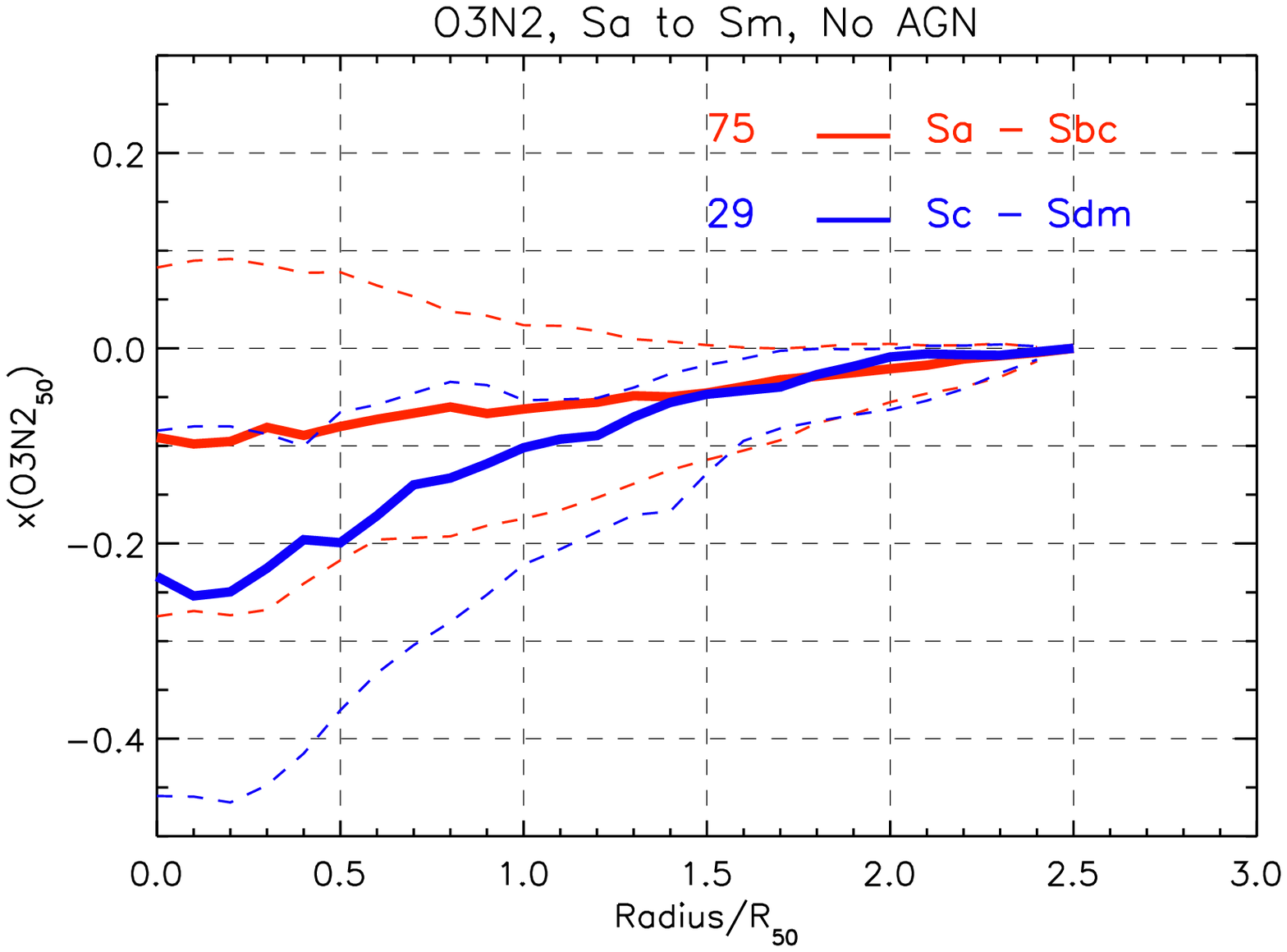}{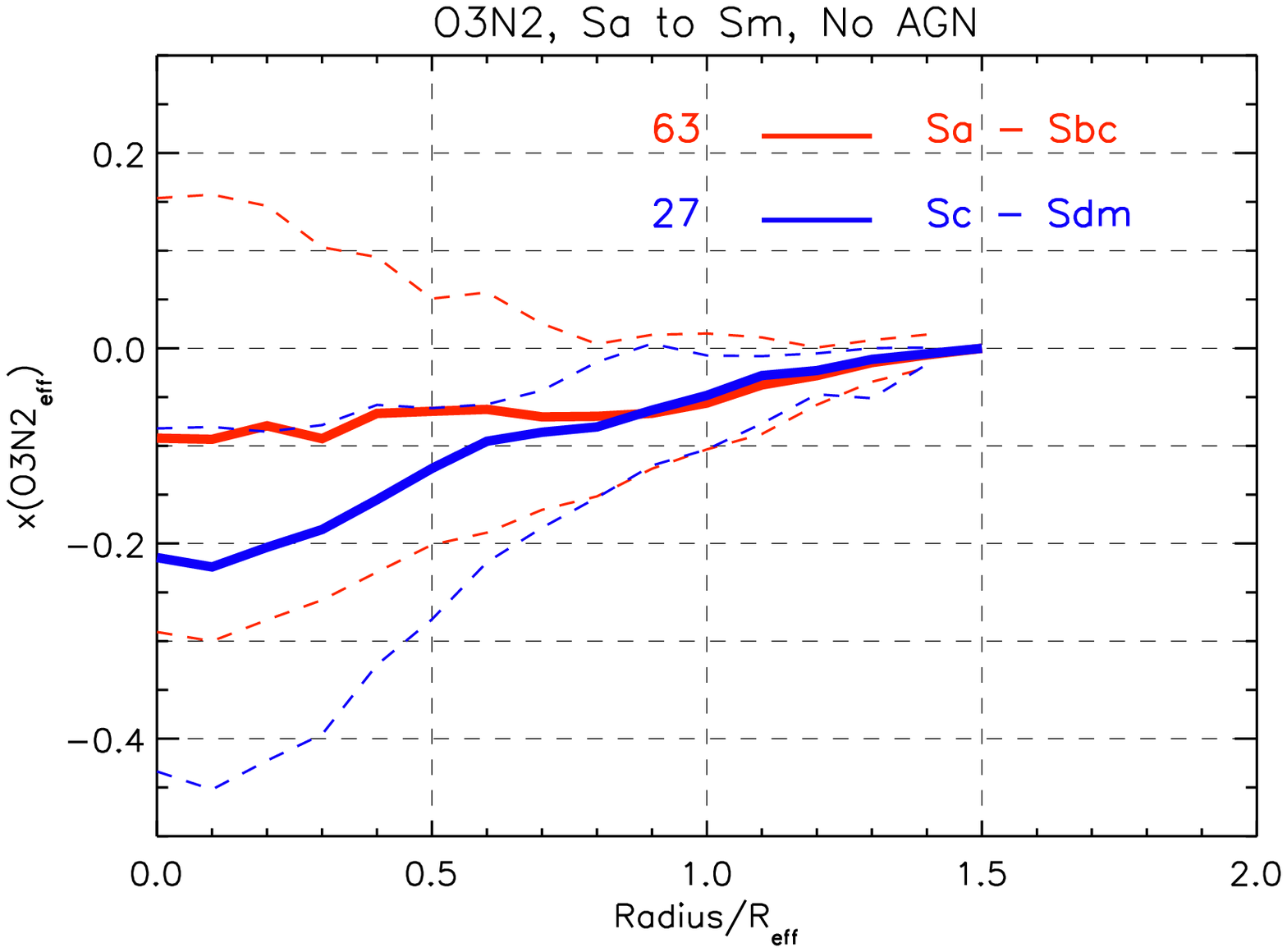}
\caption{\xonp~(left) and \xone~(right) for galaxies with different morphological types.
Solid lines correspond to median values. Dashed lines contain 68.2\% of the distribution.
Sa-Sbc and Sc-Sm galaxies are represented in blue and red respectively.
The numbers within the plot box indicate the sizes of the samples.\label{o3n2_type}}
\end{figure}

\begin{figure}
\plottwo{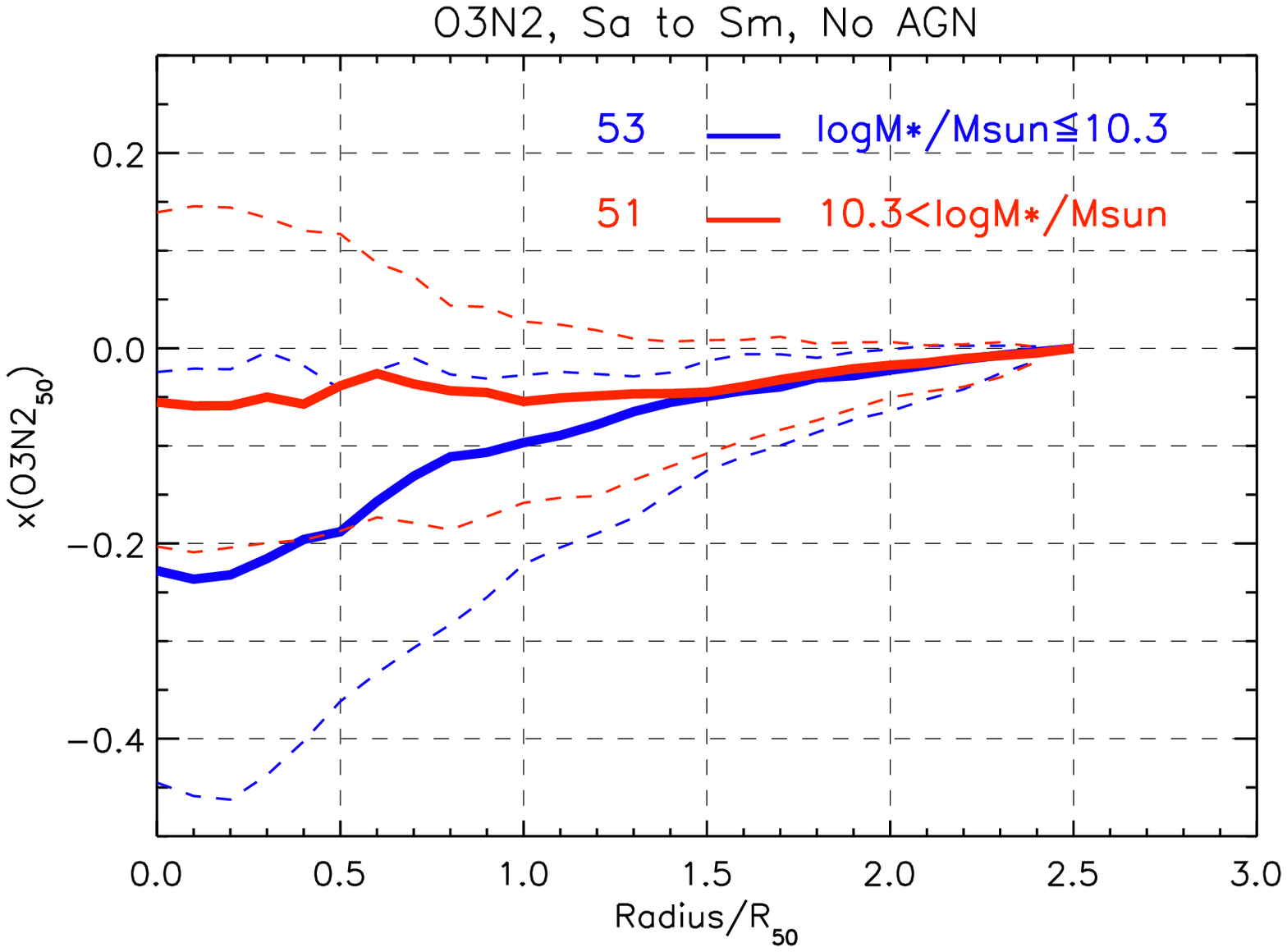}{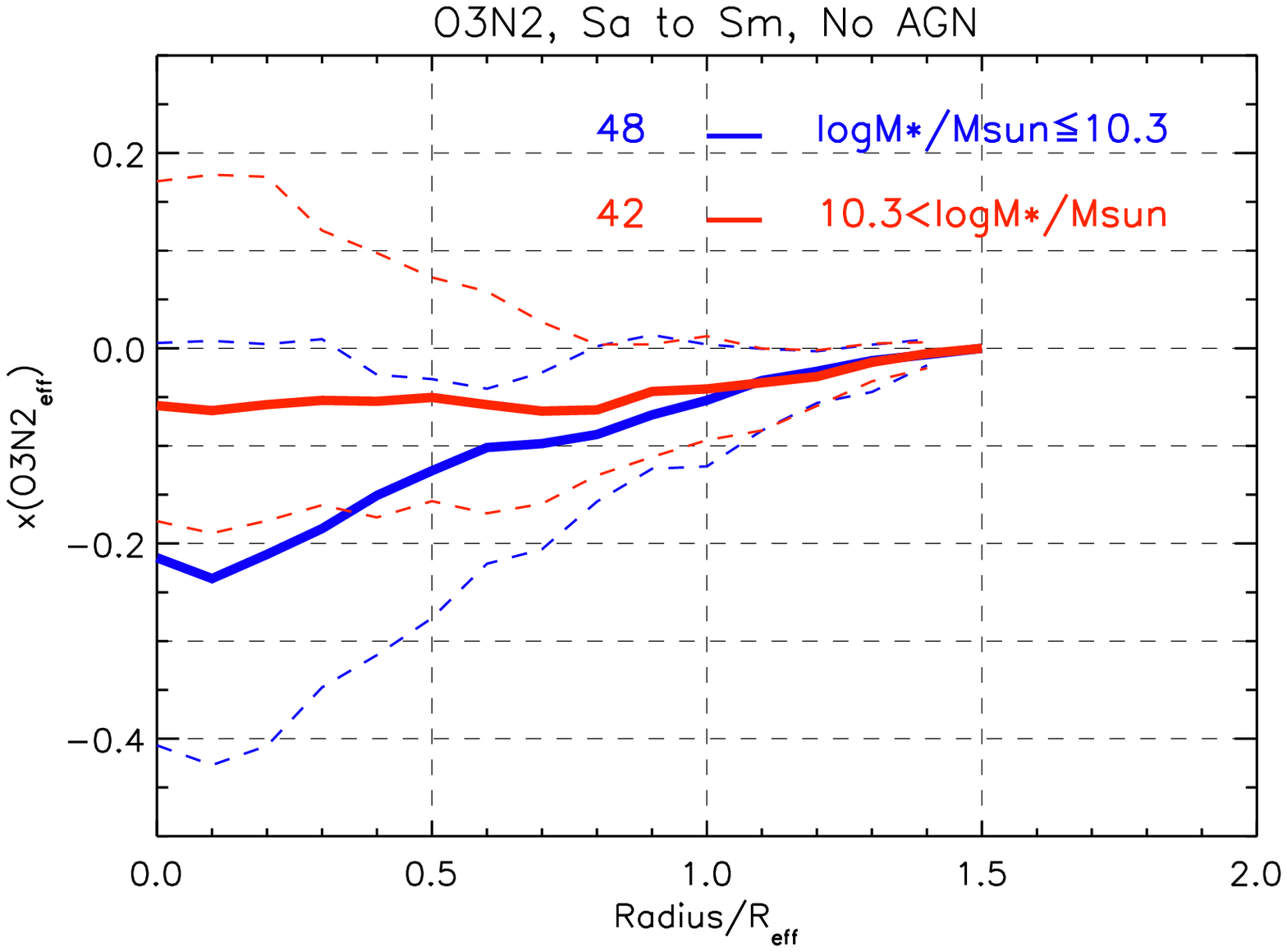}
\caption{\xonp~(left) and \xone~(right) for galaxies with different stellar masses.
Solid lines correspond to median values. Dashed lines contain 68.2\% of the distribution.
Galaxies with $\log M^{*}/M_{\odot} \leq 10.3$ and $\log M^{*}/M_{\odot} > 10.3$ are represented in blue and red respectively.
The numbers within the plot box indicate the sizes of the samples.\label{o3n2_mass}}
\end{figure}

Figure~\ref{o3n2_incli} shows \xonp~ and \xone~ for different inclinations.
Following this figure, \xonp~and \xone~show an almost linear increasing trend from the inner regions to the outskirts of the galaxies, and no difference with the inclination of the galaxy is apparent.
This means that observing galaxies through small apertures results in values of O3N2 smaller than those obtained with the integrated values.
Figure~\ref{o3n2_type} shows \xonp~and \xone~ for different morphological types.
In this case, Sa-Sbc galaxies show larger values of \xonp~and \xone~ for apertures with $R/$\rp~$\leq 1.2$~and $R/$\re~$\leq 0.7$~respectively than Sc-Sm galaxies, although the difference is lower than the dispersions around the median values.
Figure~\ref{o3n2_mass} shows \xonp~and \xone~ for different stellar masses.
Similarly to the previous case, galaxies with $\log M^{*}/M_{\odot} > 10.3$ present larger values of \xonp~and \xone~ for apertures with $R/$\rp~$\leq 1.2$~and $R/$\re~$\leq 0.7$~respectively than galaxies with $\log M^{*}/M_{\odot} \leq 10.3$, but as commented above, this difference is smalles than the dispersions around the median values of \xonp~ and \xone.
In this case also the radial change of the O3N2 growth curves can be interpreted in terms of a radial change of the oxygen abundanes.

Tables~\ref{ha_type_tab} to \ref{o3n2_mass_re_tab} show the numerical values of the median growth curves and the $1\sigma$ confidence intervals corresponding to the growth curves previously shown.
These tables also contain the fits to all spiral galaxies considered in this work.

As a summary of the present section we show in Table~\ref{poly_fits} the coefficients of the fits to 5th order polynomials of the aperture corrections previously discussed. 

\subsection{Corrections for fixed physical apertures}

\begin{figure}
\plottwo{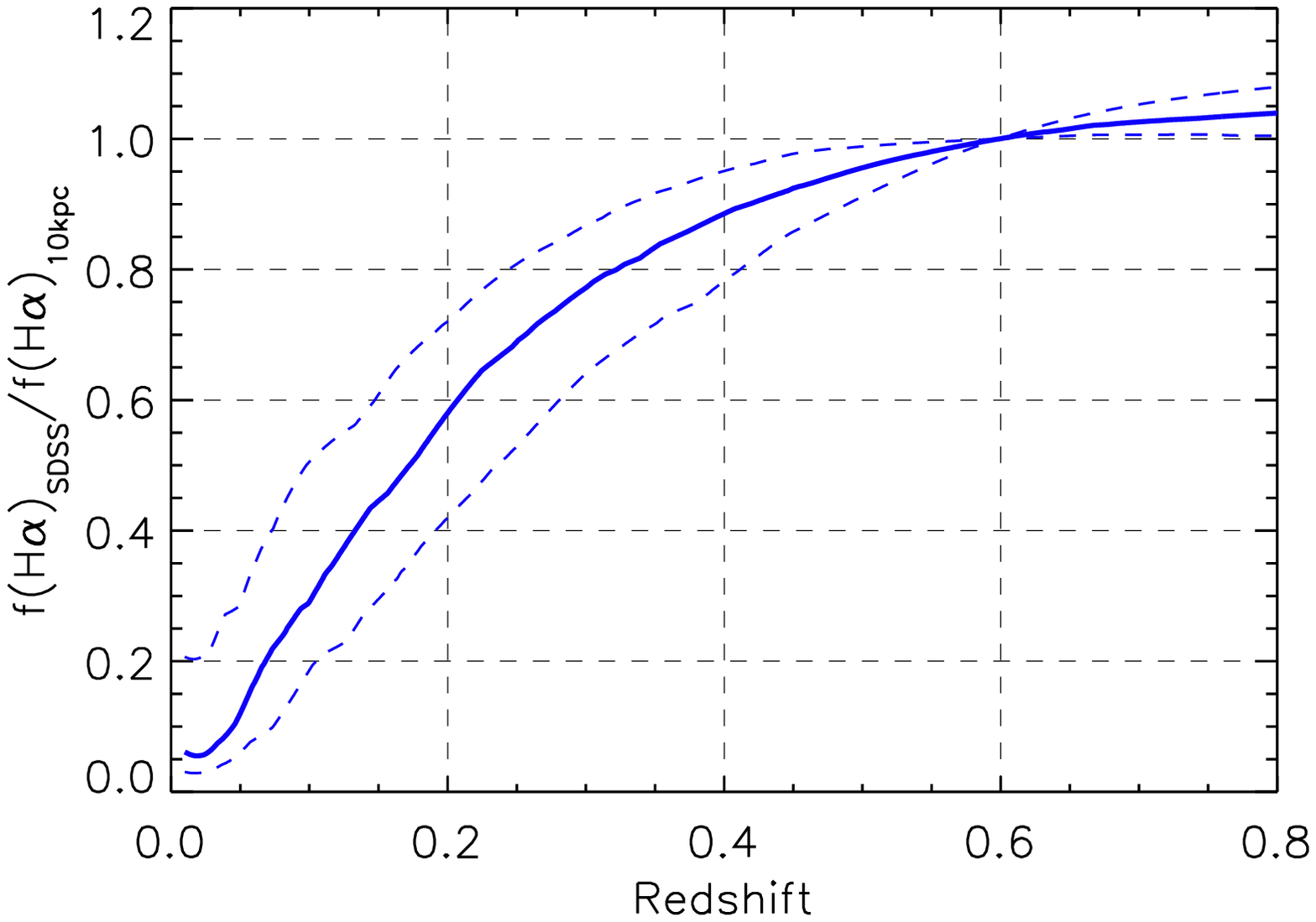}{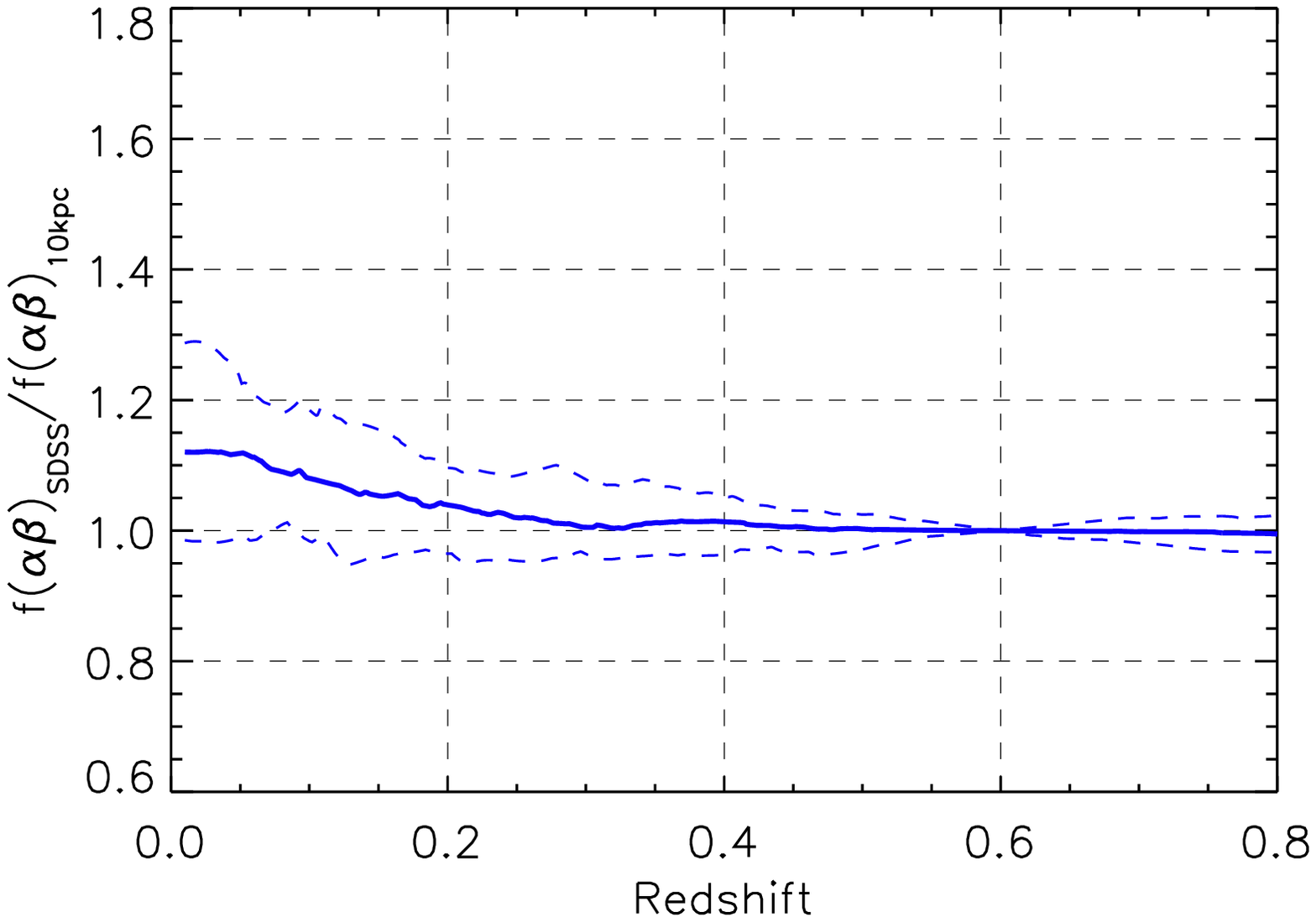}
\plottwo{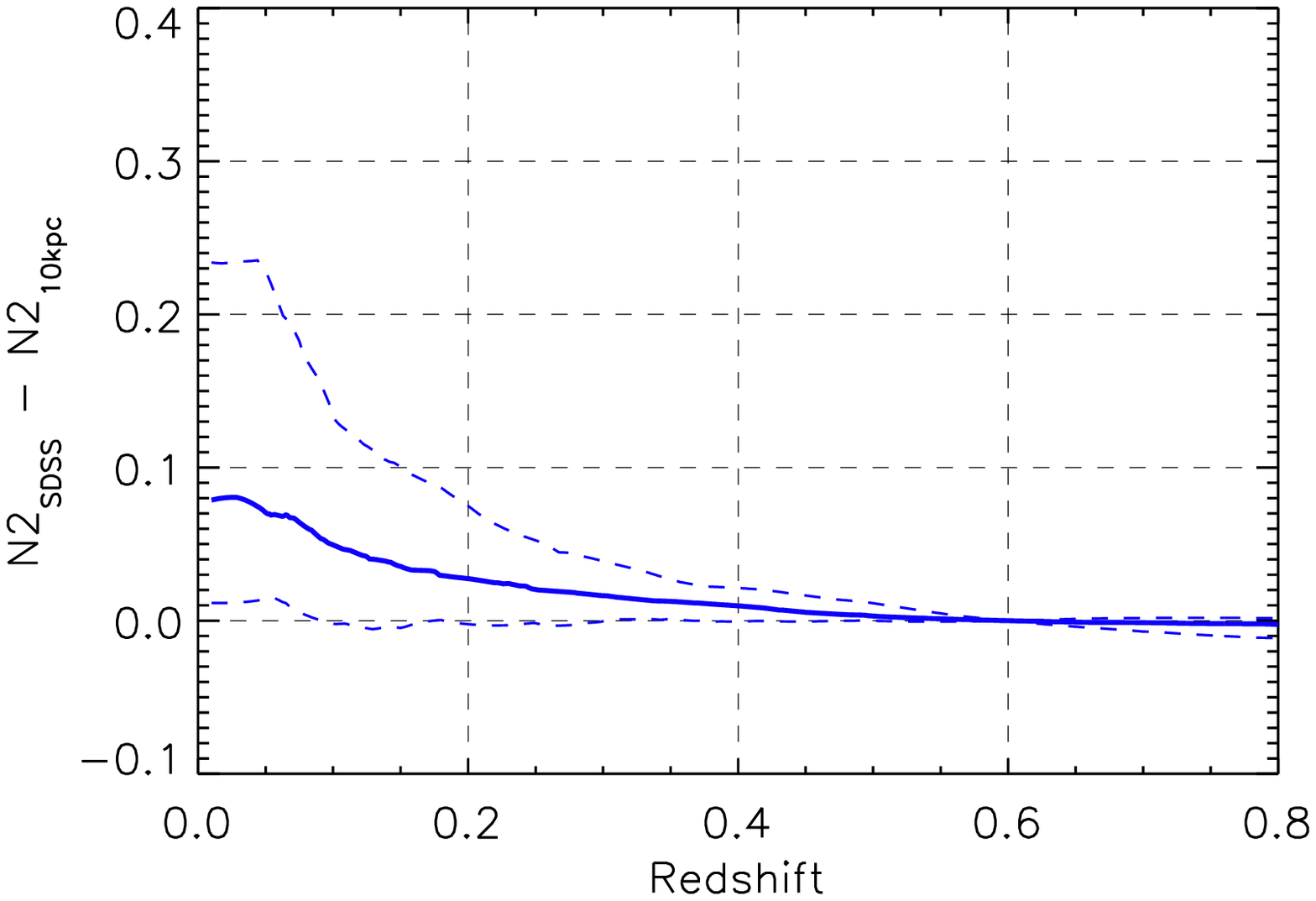}{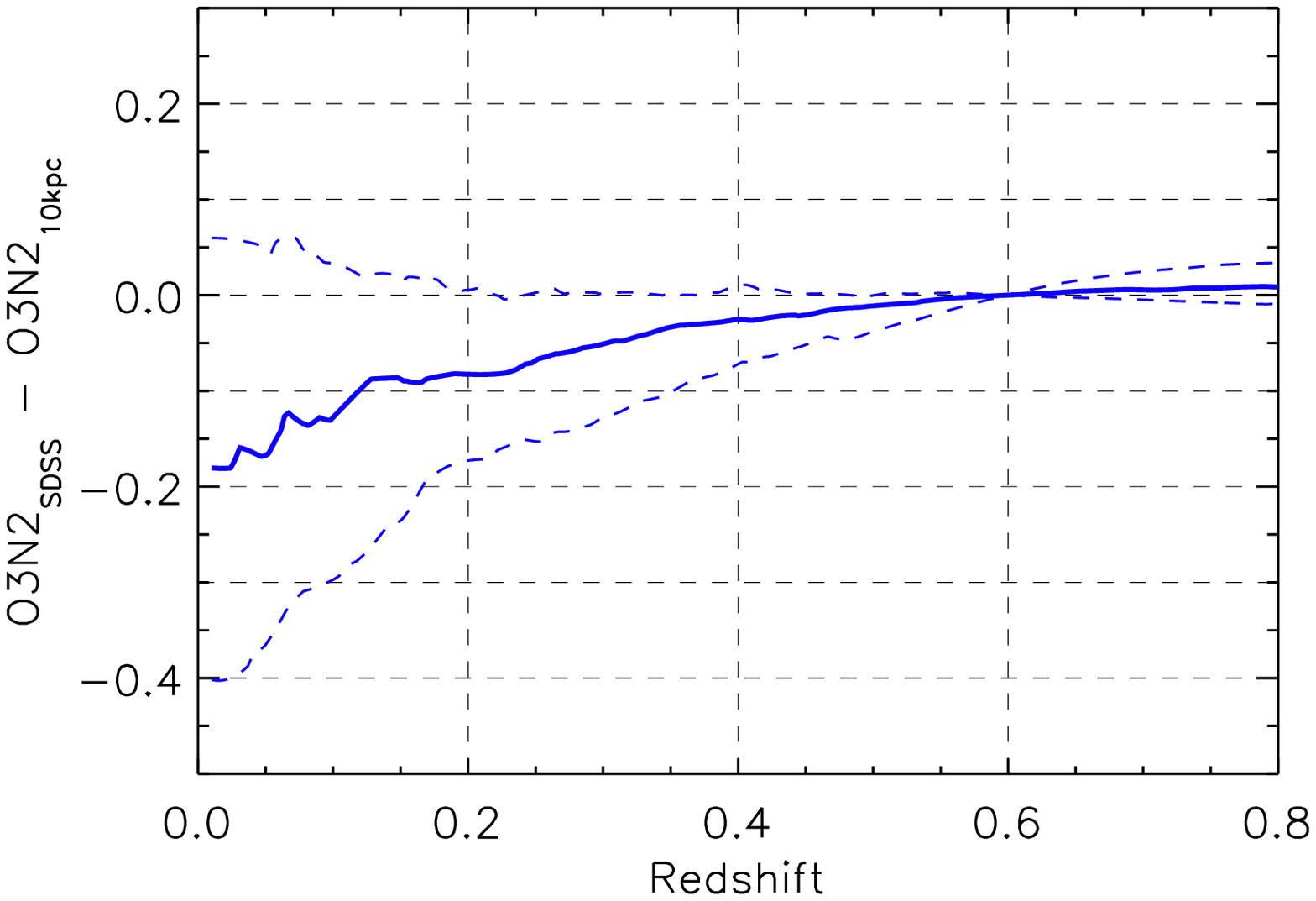}
\caption{Ratio of the H$\alpha$ flux (top left), H$\alpha$/H$\beta$ (top right), N2 (bottom left) and O3N2 (bottom right) contained in the SDSS aperture at different redshifts to the corresponding values within a circular aperture of 10~kpc diameter as a function of redshift for the (52) CALIFA spirals whose 10~kpc diameter aperture is completely covered by PMAS/PPAK.
Lower dashed, solid and upper dashed lines correspond to 15.86\%, 50\% and 84.14\% of the distributions.\label{aper12_sdss}}
\end{figure}

\begin{figure}
\plottwo{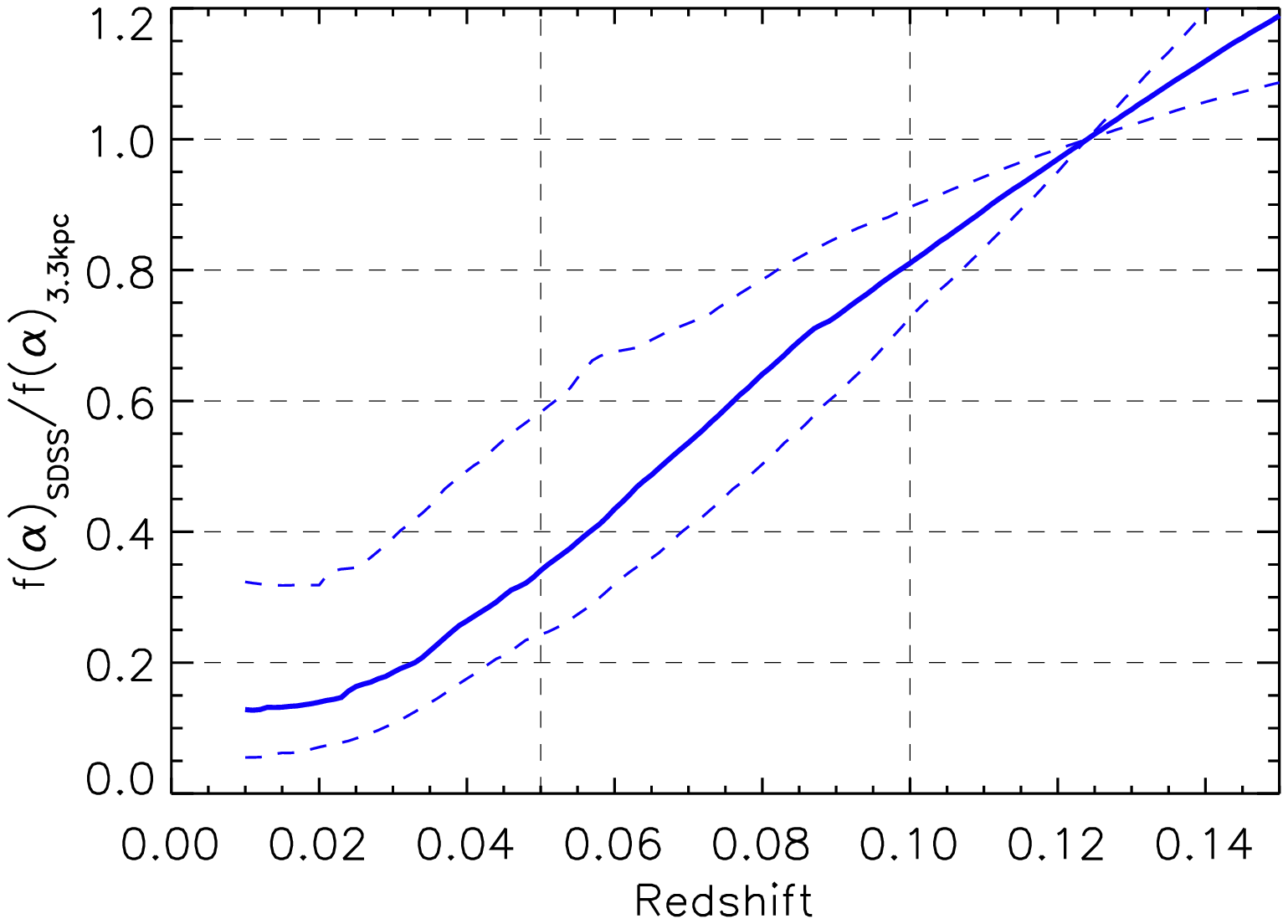}{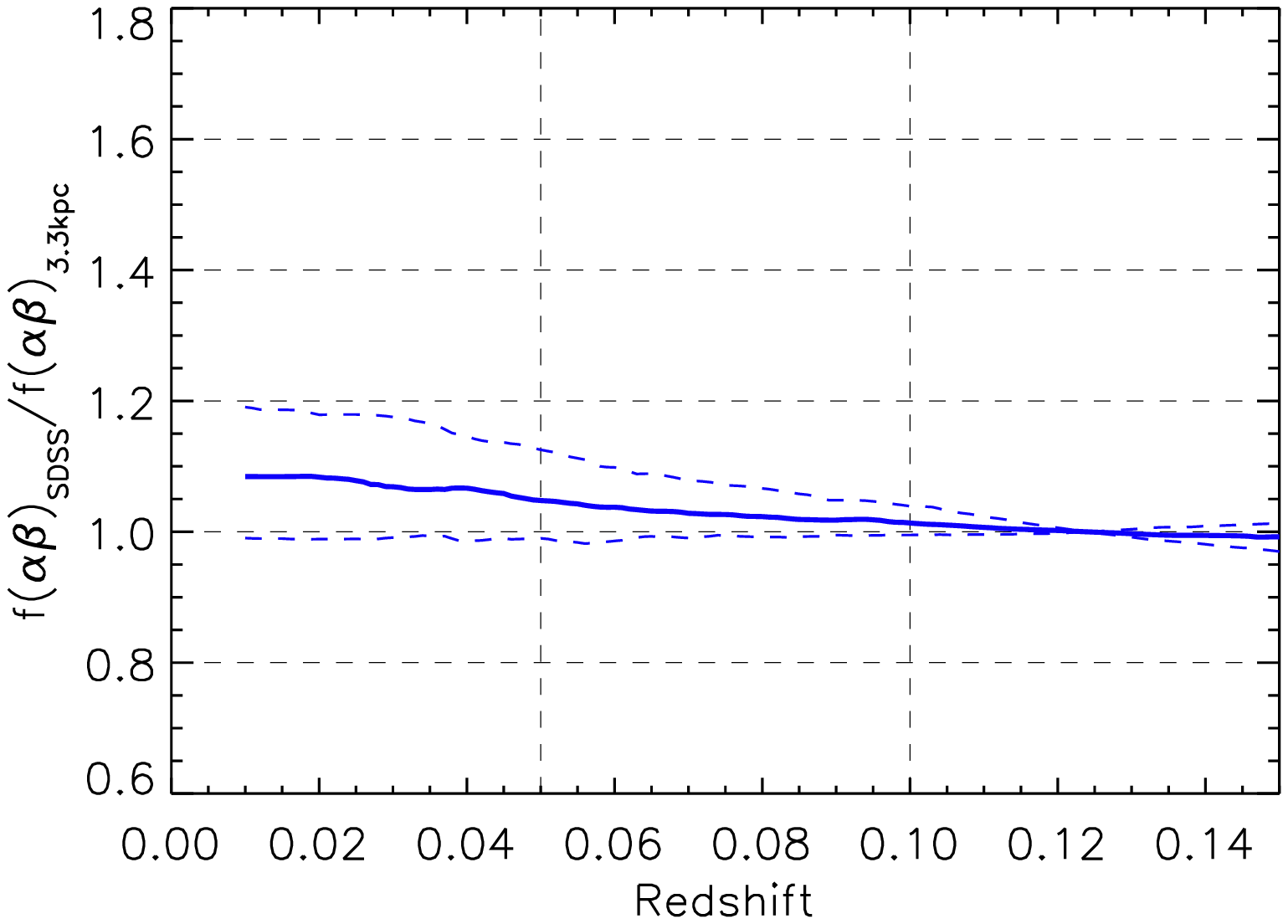}
\plottwo{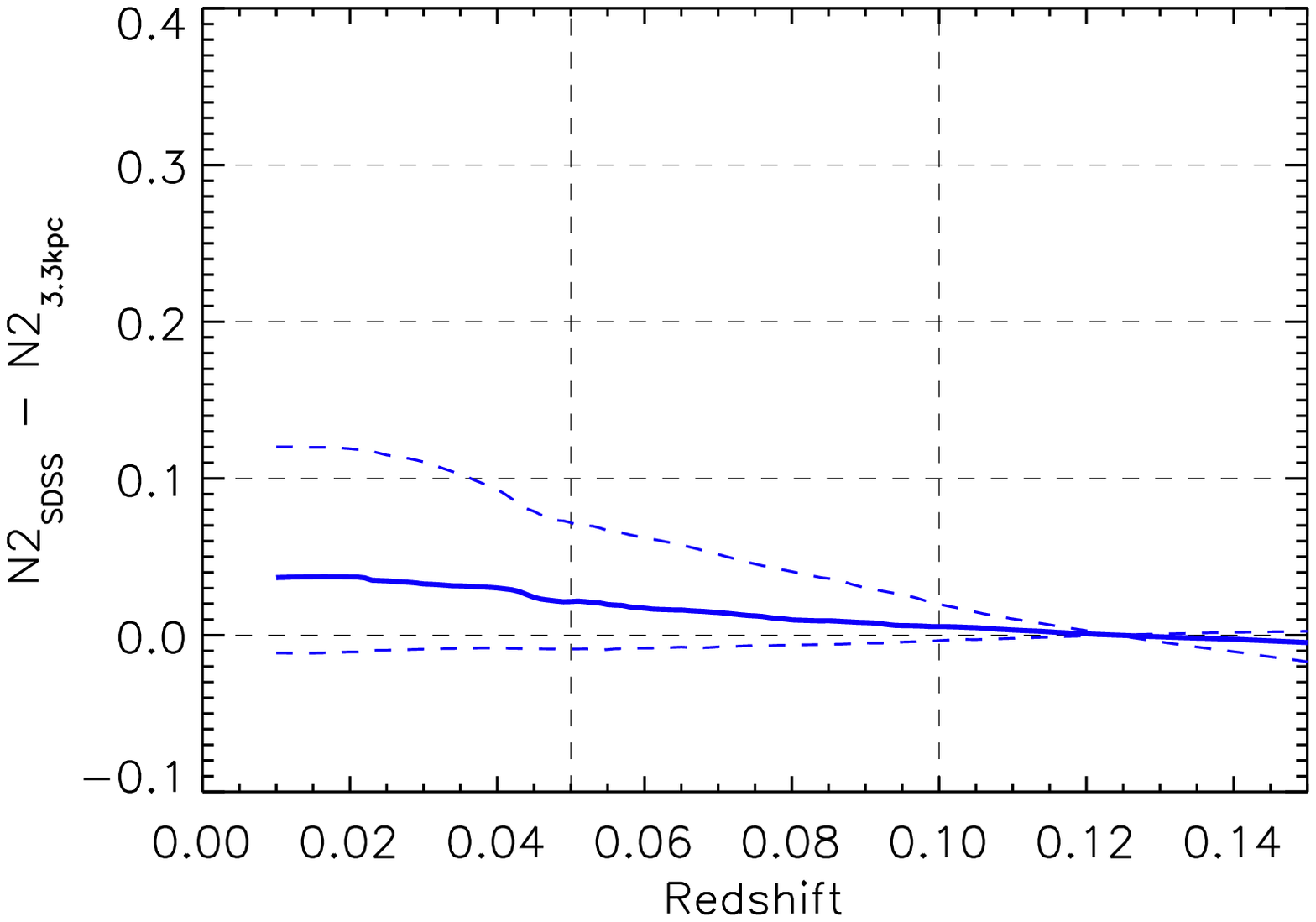}{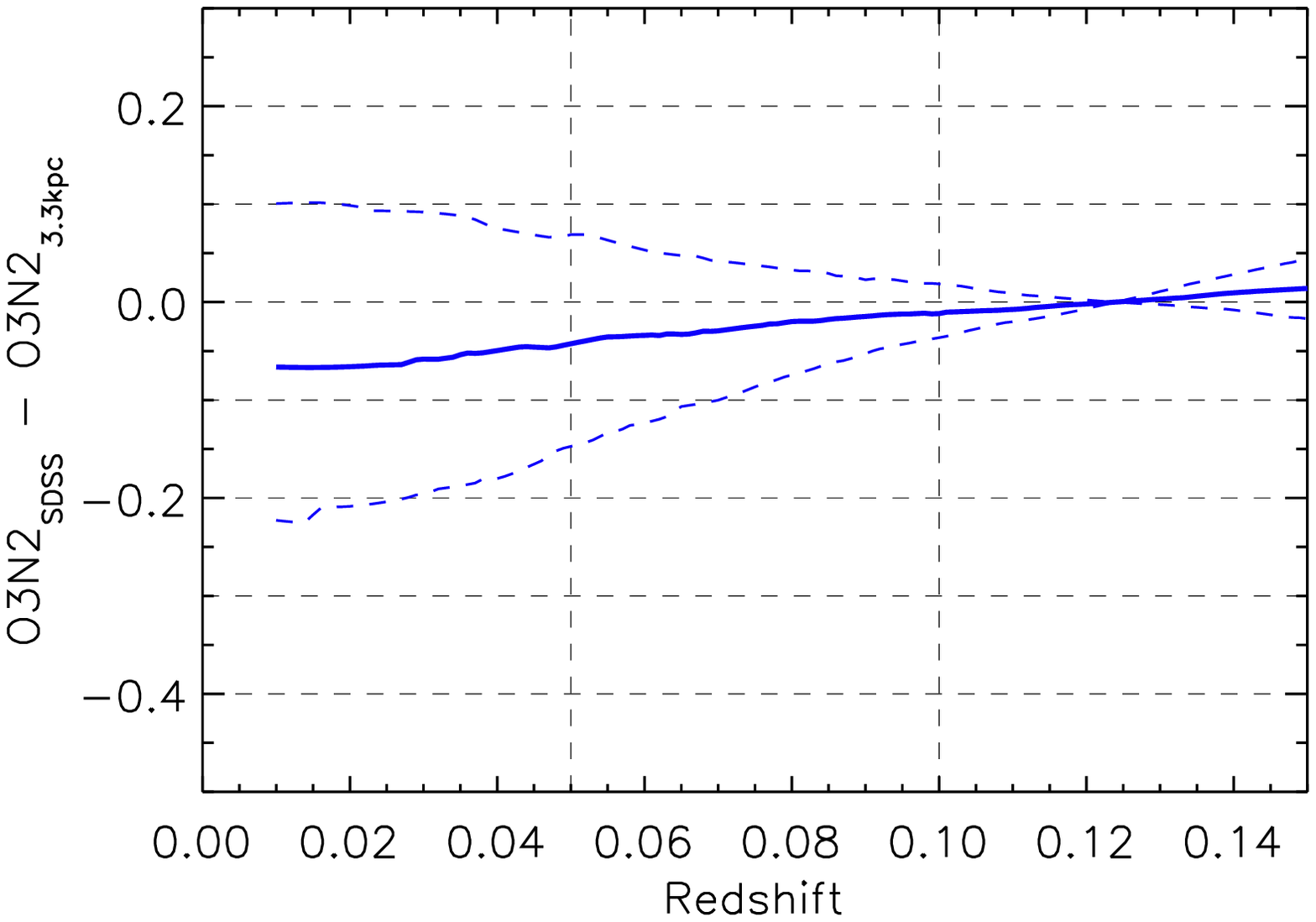}
\caption{Ratio of the H$\alpha$ flux (top left), H$\alpha$/H$\beta$ (top right), N2 (bottom left) and O3N2 (bottom right) contained in the SDSS aperture at different redshifts to the corresponding values within a circular aperture of 3.3~kpc diameter as a function of redshift for the (96) CALIFA spirals whose 3.3~kpc diameter aperture is completely covered by PMAS/PPAK.
Lower dashed, solid and upper dashed lines correspond to 15.86\%, 50\% and 84.14\% of the distributions.\label{aper6_sdss}}
\end{figure}

The large field-of-view covered by the CALIFA data allows us to predict the fraction of flux enclosed within a given angular aperture with respect to the flux enclosed within a fixed physical aperture at different redshifts.
In our case, we have focused on the SDSS and SAMI fields-of-view. As the median value of \re~ of our sample (spirals from Sa to Sdm excluding AGNs) is $\approx 6.68$~kpc, and based on \xhae~ shown in figure~\ref{ha_40}, we will estimate the quantities of interest within two circular apertures, one containing most of the flux of the galaxy (10~kpc radius, containing on average 90\% of the total H$\alpha$ flux) and another one enclosing the flux in the central region (3.3~kpc radius, containing on average 30\% of the total H$\alpha$ flux).
Then for each of the CALIFA spirals we measure each of the relevant quantities within the circular aperture subtended by the SDSS fiber if the galaxy was placed at different redshifts.
Figure~\ref{aper12_sdss} gives the median values of the aperture corrections with respect to an aperture of 10~kpc radius for the $f$(H$\alpha$), $f$(H$\alpha$)/$f$(H$\beta$), N2 and O3N2.
The SDSS fiber covers an aperture of $\approx 10$~kpc at a redshift of $z \approx 0.6$.
The trends with redshift shown by the aperture corrections are similar to the ones obtained in previous sections as a function of \rp~ and \re. 
Figure~\ref{aper6_sdss} also provides corrections for a physical aperture of 3.3~kpc radius.
In this case, the SDSS fiber covers this aperture at a redshift of $z \approx 0.12$.

\begin{figure}
\plottwo{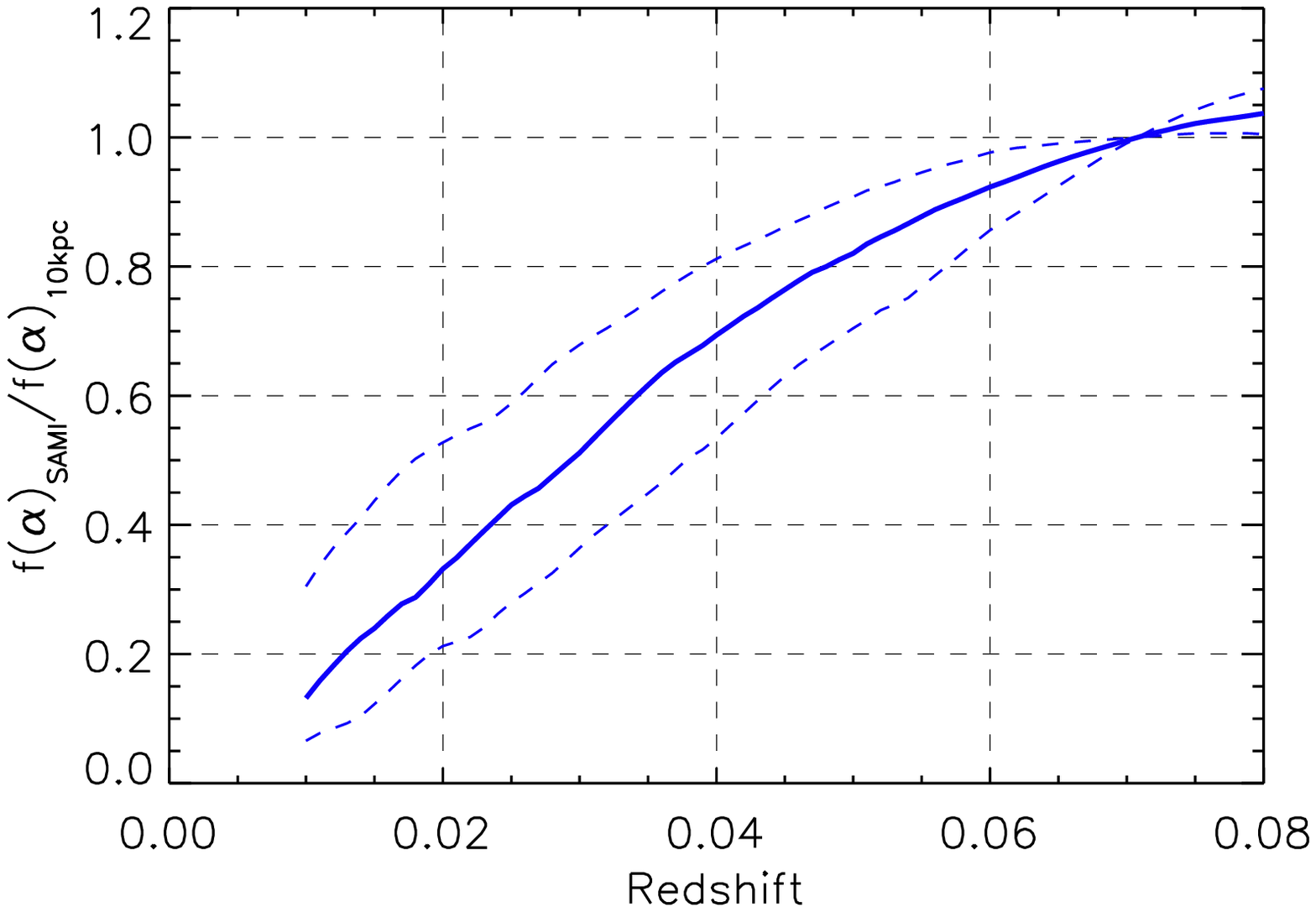}{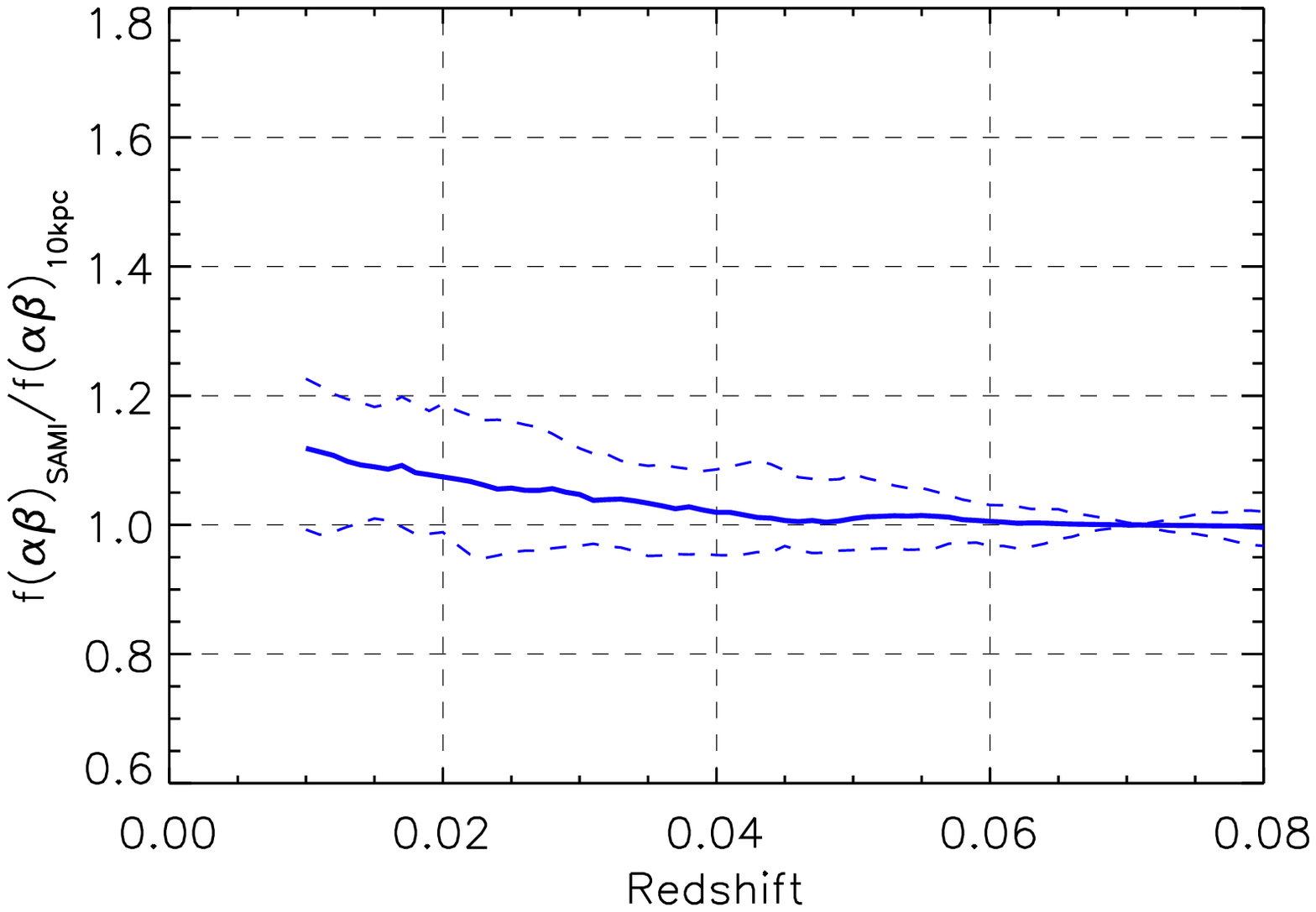}
\plottwo{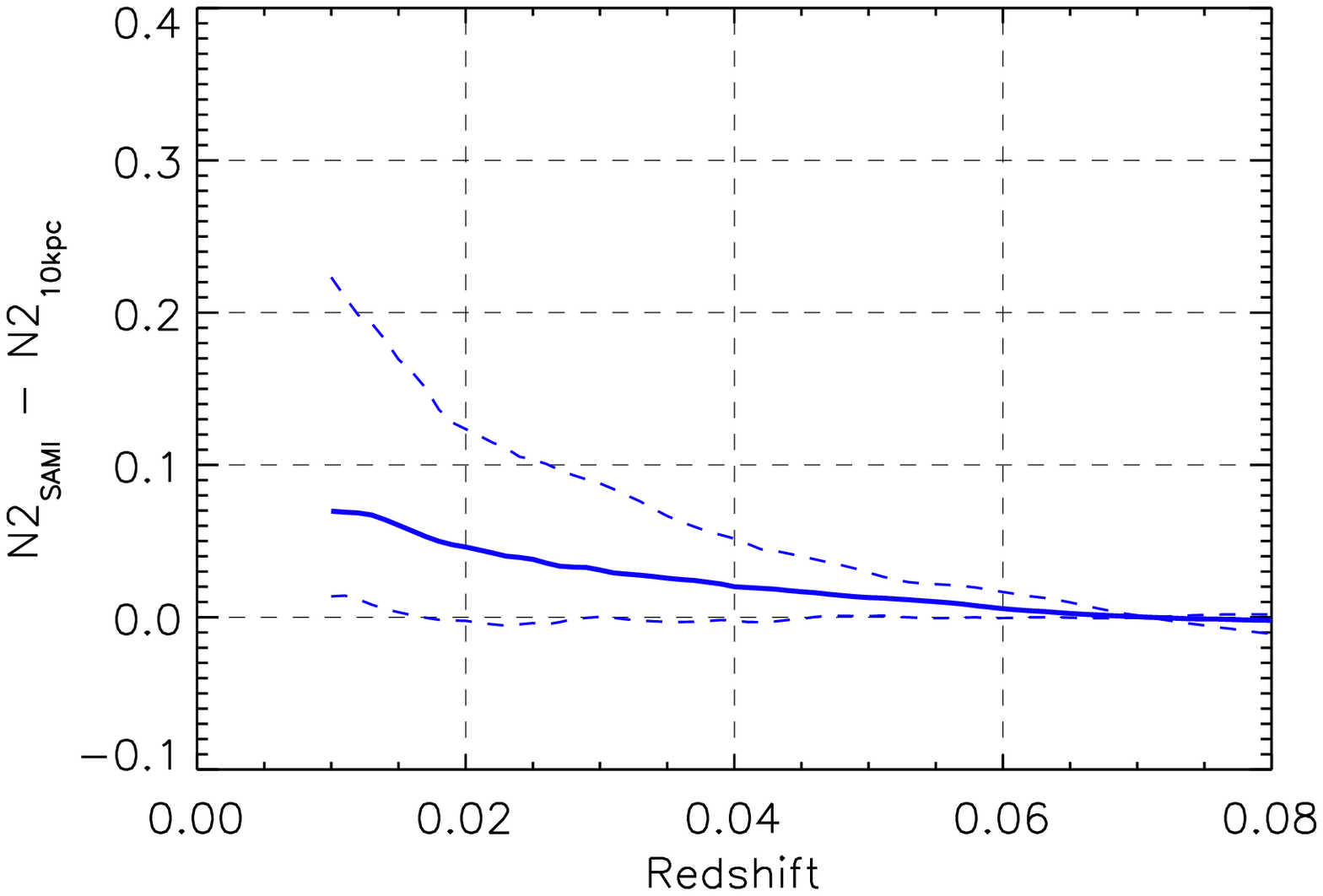}{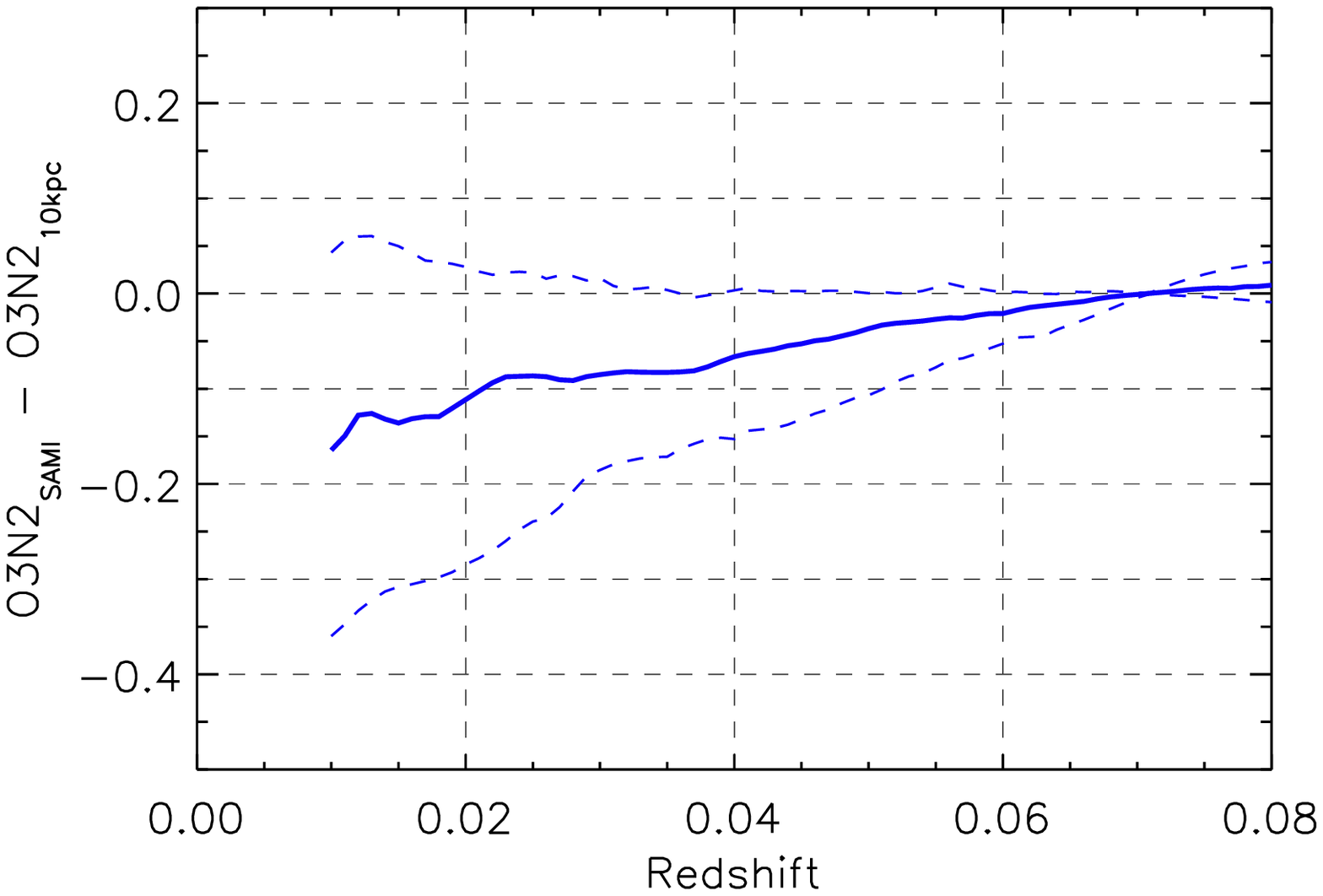}
\caption{Ratio of the H$\alpha$ flux (top left), H$\alpha$/H$\beta$ (top right), N2 (bottom left) and O3N2 (bottom right) contained in the SAMI aperture at different redshifts to the corresponding values within a circular aperture of 10~kpc diameter as a function of redshift for the (52) CALIFA spirals whose 10~kpc diameter aperture is completely covered by PMAS/PPAK.
Lower dashed, solid and upper dashed lines correspond to 15.86\%, 50\% and 84.14\% of the distributions.\label{aper12_sami}}
\end{figure}

\begin{figure}
\plottwo{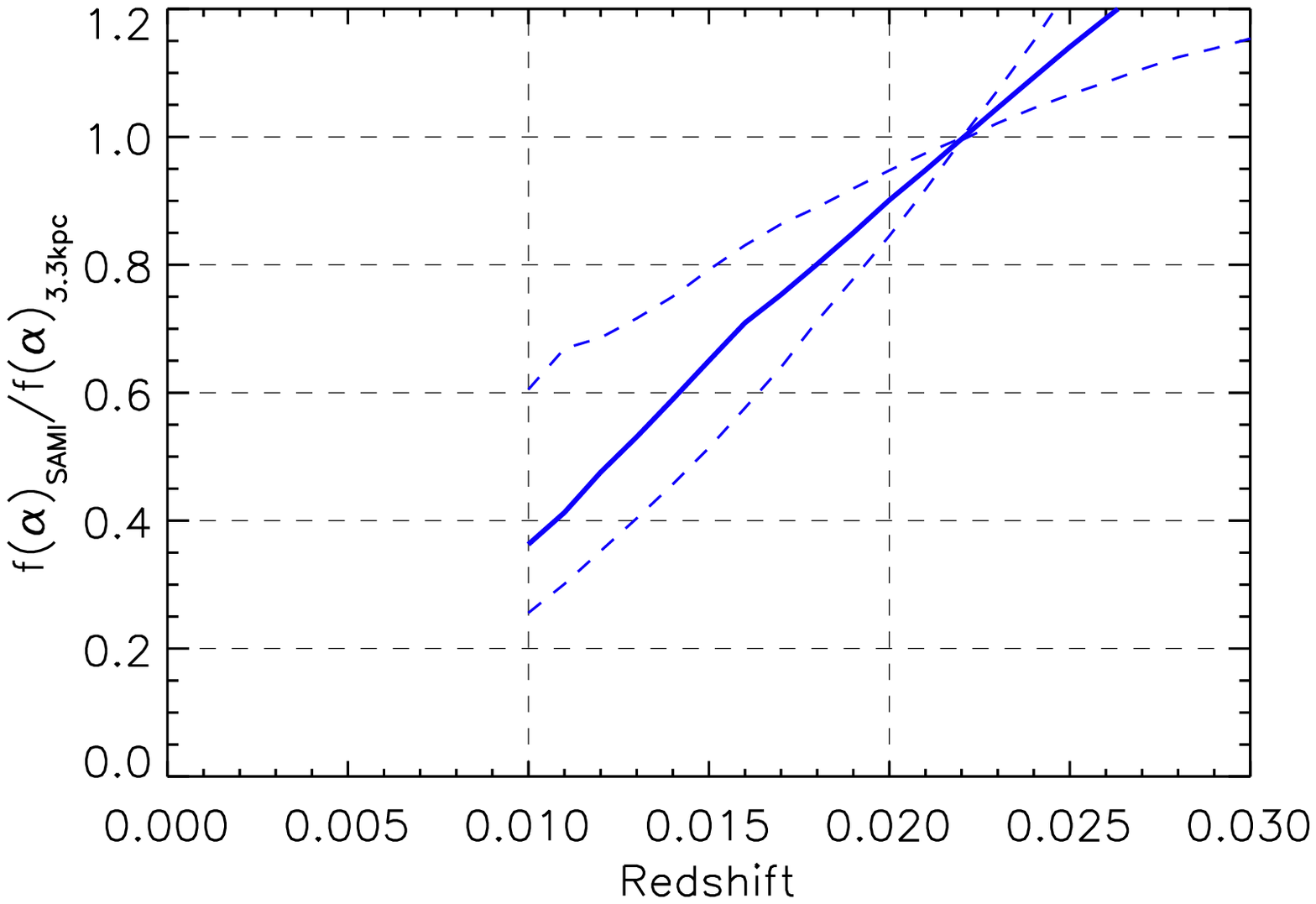}{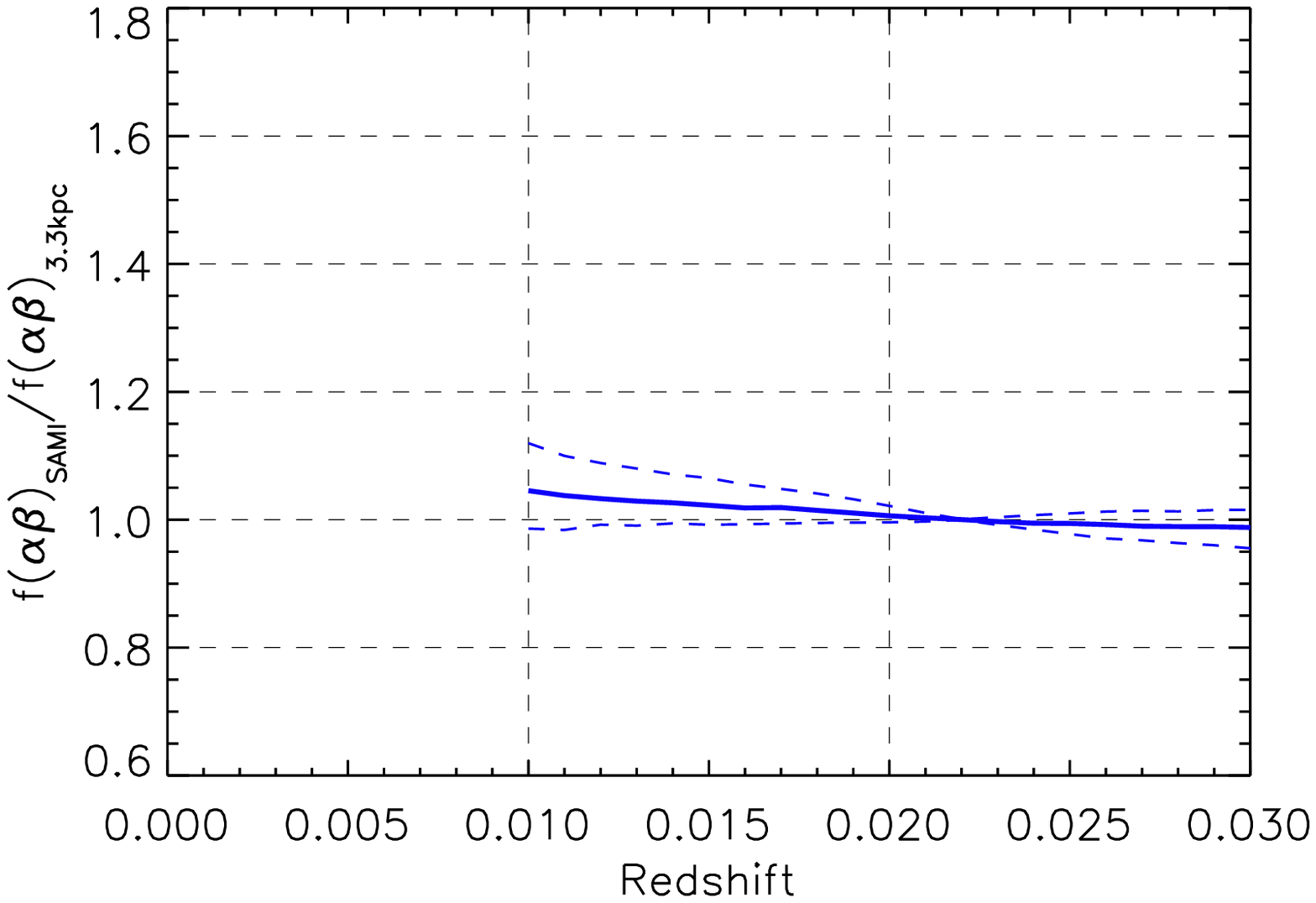}
\plottwo{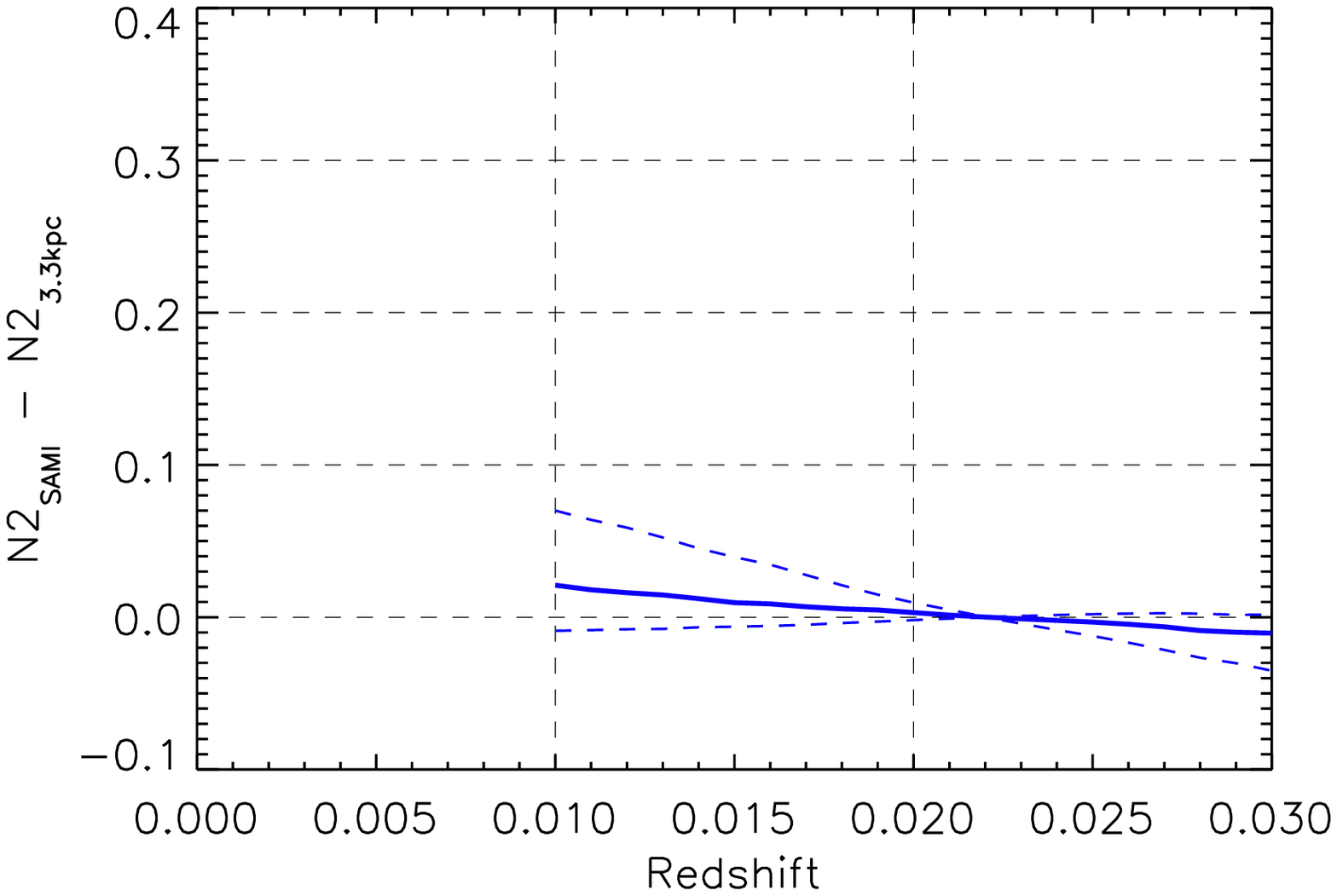}{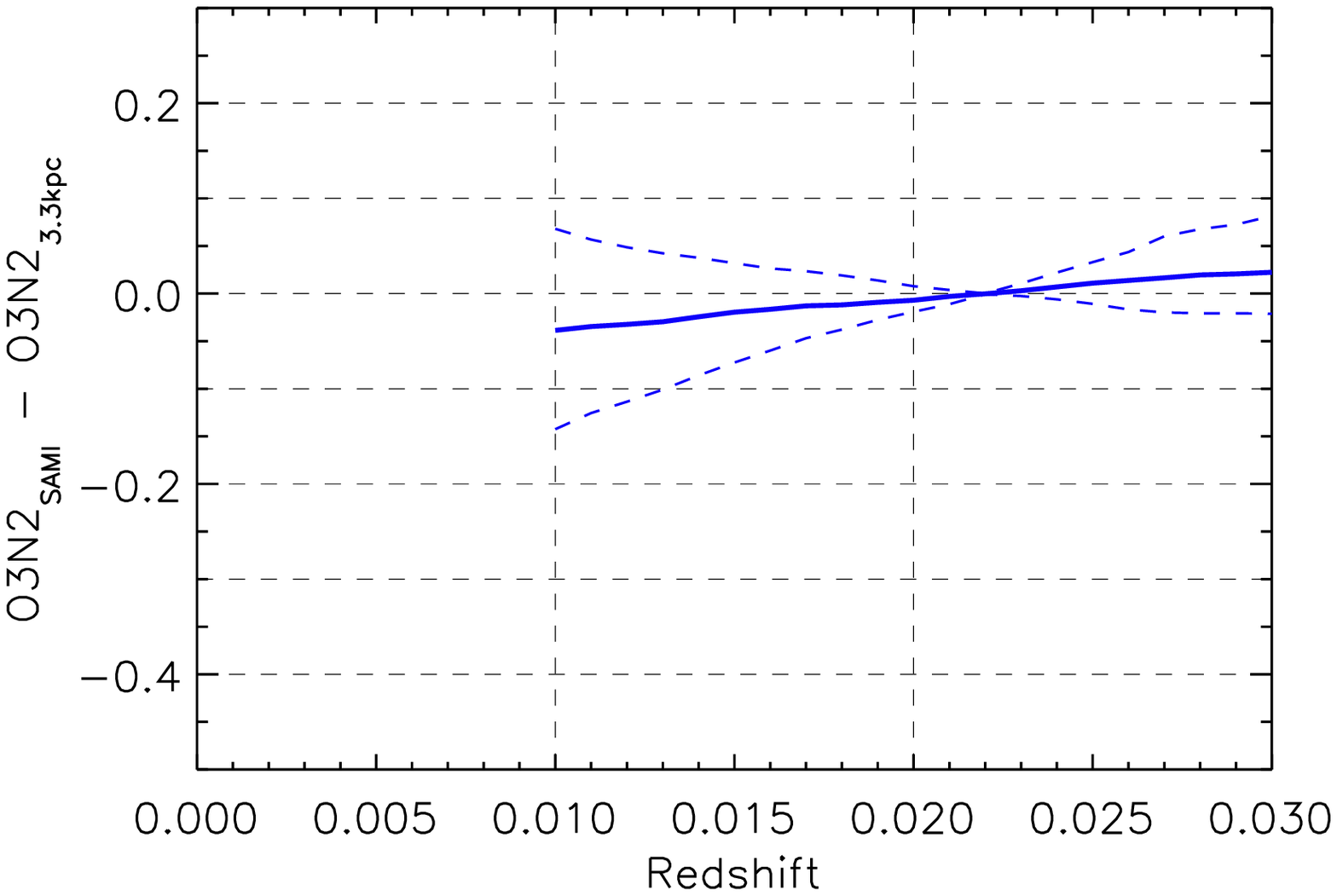}
\caption{Ratio of the H$\alpha$ flux (top left), H$\alpha$/H$\beta$ (top right), N2 (bottom left) and O3N2 (bottom right) contained in the SAMI aperture at different redshifts to the corresponding values within a circular aperture of 3.3~kpc diameter as a function of redshift for the (96) CALIFA spirals whose 3.3~kpc diameter aperture is completely covered by PMAS/PPAK.
Lower dashed, solid and upper dashed lines correspond to 15.86\%, 50\% and 84.14\% of the distributions.\label{aper6_sami}}
\end{figure}

We also provide corrections with respect to these two fixed apertures for the SAMI case (a circular bundle of fibers of 15~arcsec diameter).
The results are shown in Figs.~\ref{aper12_sami} and~\ref{aper6_sami}.
The larger size of the SAMI aperture compared to the SDSS one reduces the range of applicability of our aperture corrections since it results in complete coverage of the 10 and 3.3~kpc apertures at redshifts of $z \approx 0.07$ and 0.02 respectively.

Tables~\ref{aper12_sdss_ha_tab} to~\ref{aper12_sdss_o3n2_tab} list the median values and 1-$\sigma$ dispersions of the aperture corrections for fixed apertures of 10~kpc and 3.3~kpc diameter, focused on the cases of the SDSS and SAMI surveys.

\section{Discussion}

In previous sections we have shown the growth curves of some emission line fluxes and line ratios, relevant for the estimation of Star Formation Rates (SFRs), extinction and characteristic oxygen abundances of spiral galaxies.
These growth curves provide with aperture corrections for these quantities that should be considered in a statistical sense, i.e. applicable to large statistical samples of spiral galaxies.
The effect of the aperture on H$\alpha$ luminosity, H$\alpha$/H$\beta$ ratio, N2 and O3N2 is found to be statistically uncorrelated to the inclination, morphological type and stellar mass of galaxies, except for some marginal differences always below the typical dispersions of the median growth curves for small apertures.

\begin{figure}
\plottwo{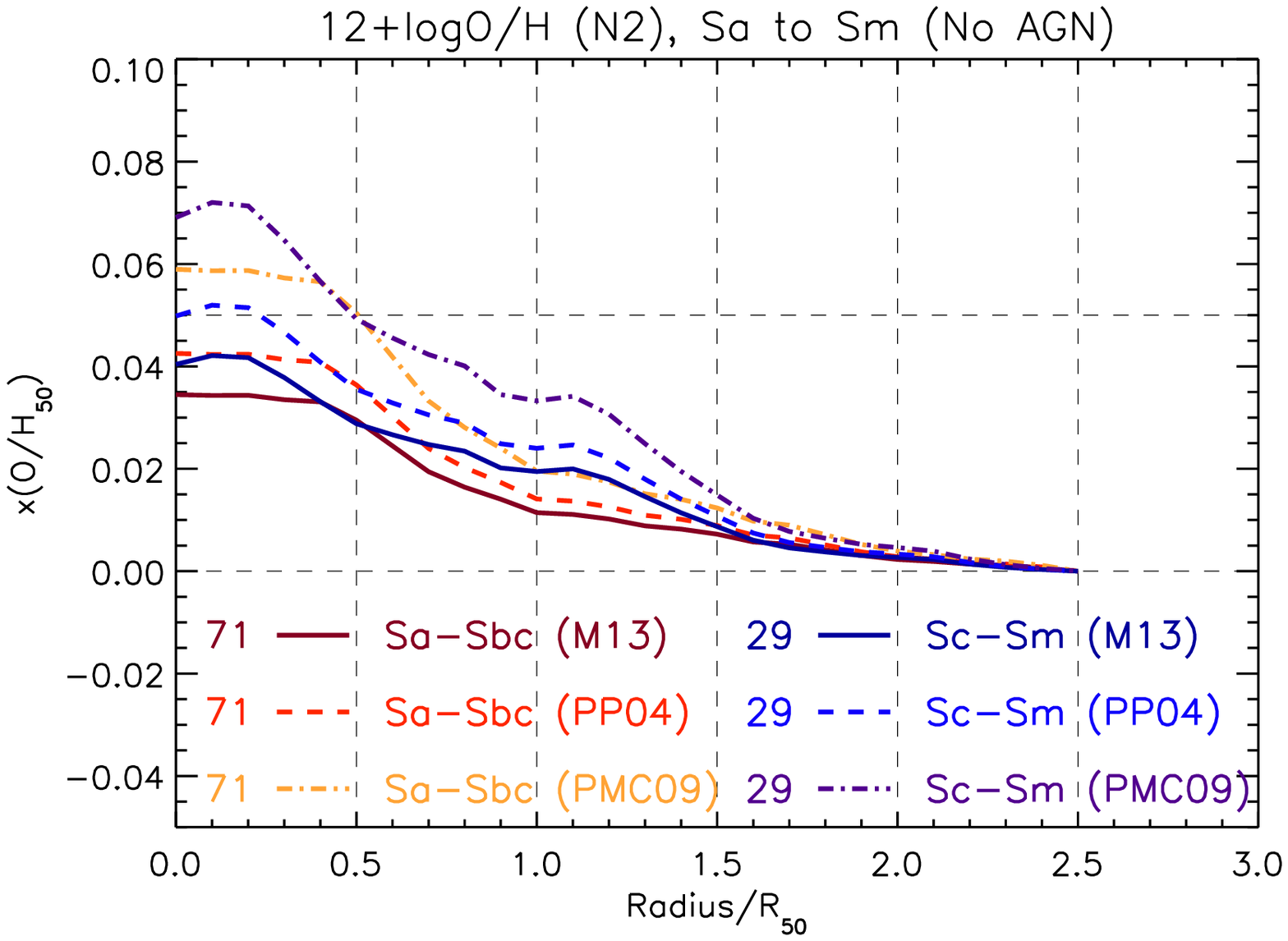}{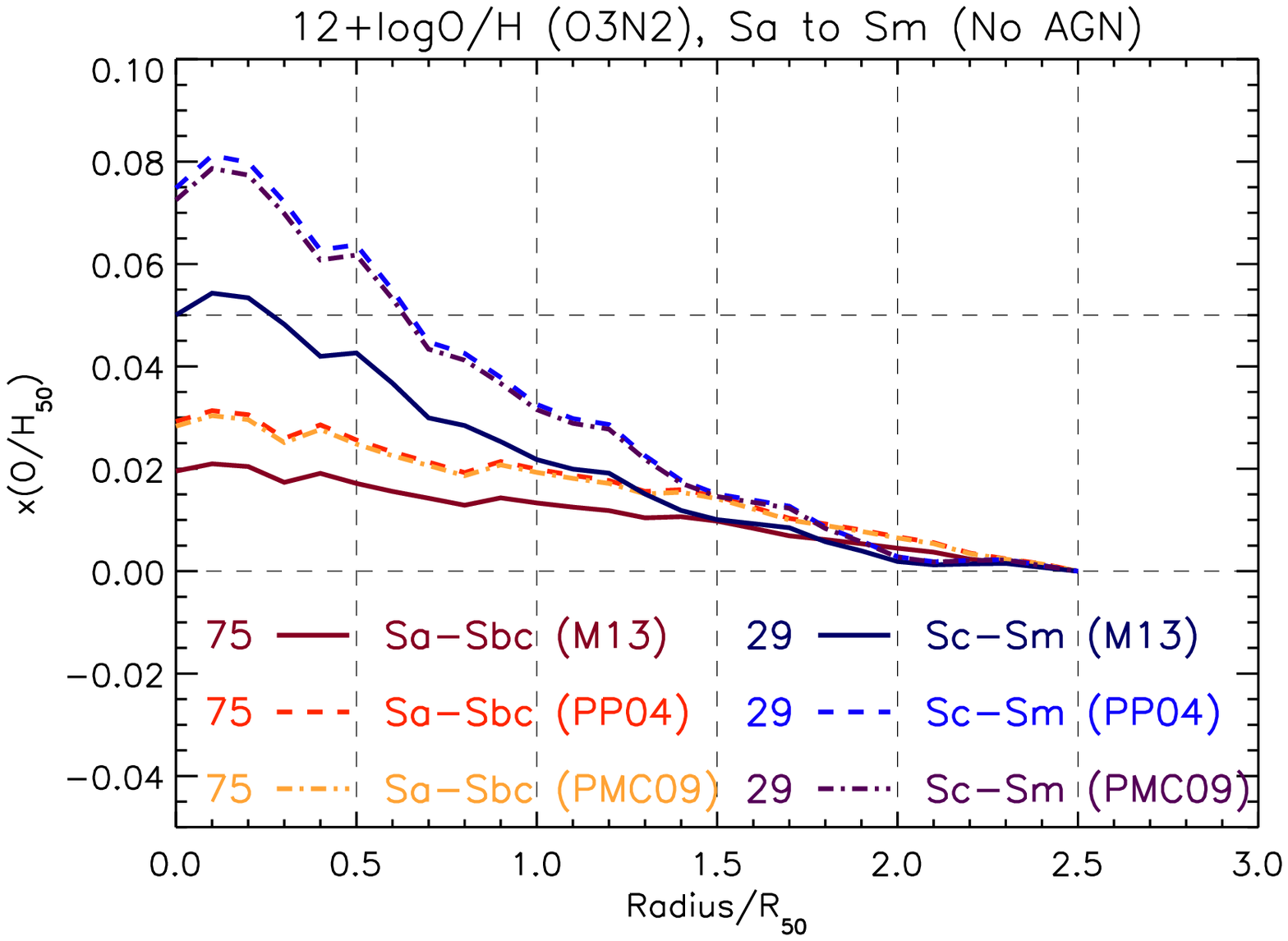}
\caption{{\bf Left:} \xohp~for two different morphological type bins, and the three empirical calibrations of N2: PP04, PMC09 and M13.
{\bf Right:} \xohp~for two different morphological type bins, and the three empirical calibrations of O3N2: PP04, PMC09 and M13.\label{oh_type_compameta}}
\end{figure}

It is interesting noting that the aperture effect found for N2 and O3N2 can be interpreted as an aperture effect in the oxygen abundances.
As a first step to illustrate the aperture effect on the oxygen abundances we show in Fig.~\ref{oh_type_compameta} the median growth curves of the logarithmic oxygen abundance as a function of \rp~, \xohp\footnote{Given that the oxygen abundance is usually given as a logarithmic quantity, the definition of \xohp~ and \xohe~ at a given radius $r$ is, as in the case of N2 and O3N2, $\log$(O/H)$_{r} - \log$(O/H)$_{int}$, where $\log$(O/H)$_{int}$ corresponds to the value of the logarithmic oxygen abundance measured within the largest aperture considered, 36'', assumed to be the integrated value.}, derived using 
the calibrations by Marino et al. (2013, M13), Pettini \& Pagel (2004, PP04) and P\'{e}rez-Montero \& Contini (2009, PMC09), for N2 and O3N2 respectively.
As it can be seen, the effect of the aperture marginally depends on the calibration used both for N2 and O3N2 showing a maximum difference of $\approx 0.02-0.03$~dex, which corresponds to 5-7.5\%. Although the dispersions around the median values are not shown in the figures (for clarity), they are much lower than this difference.

\begin{figure}
\plottwo{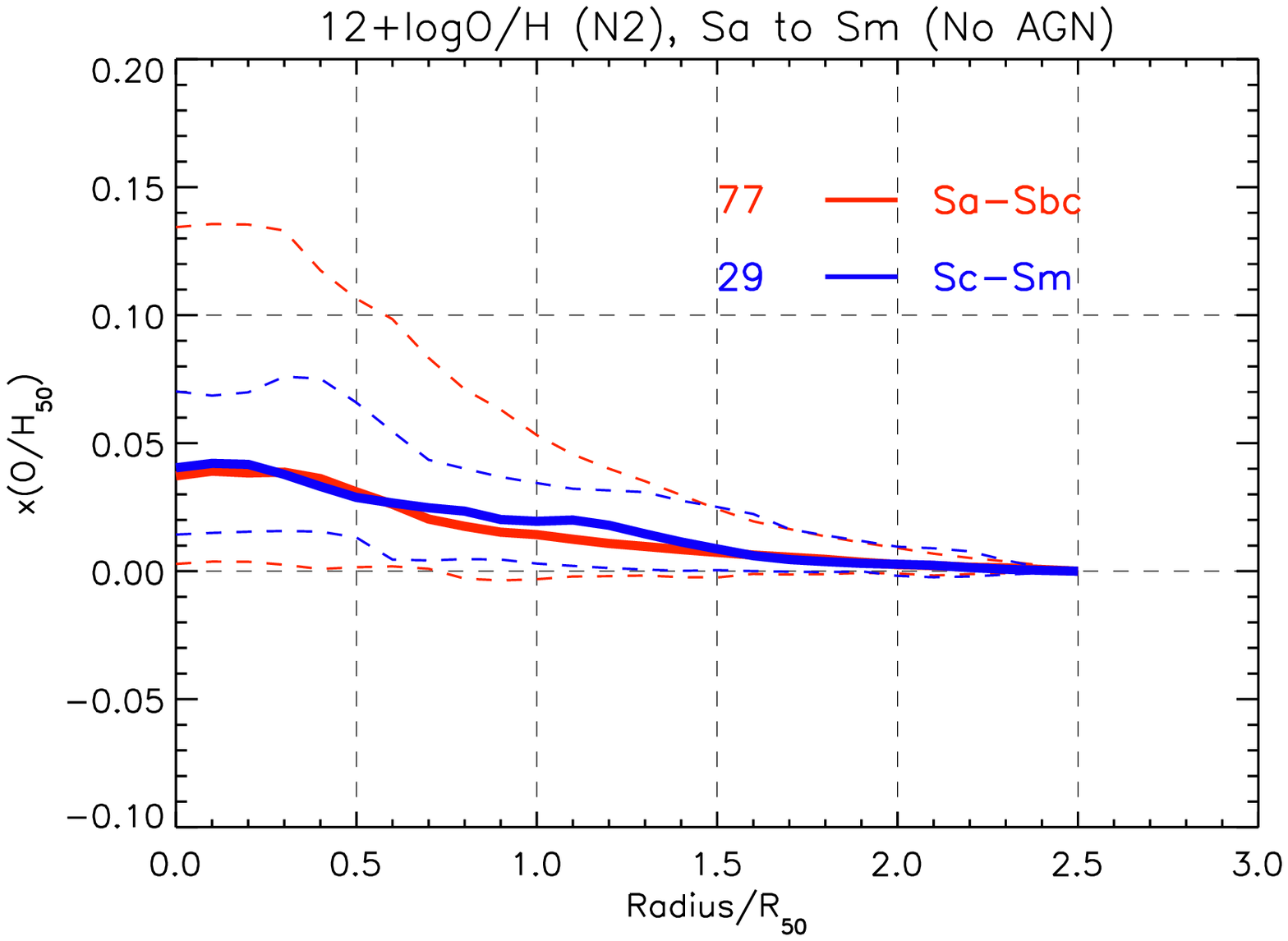}{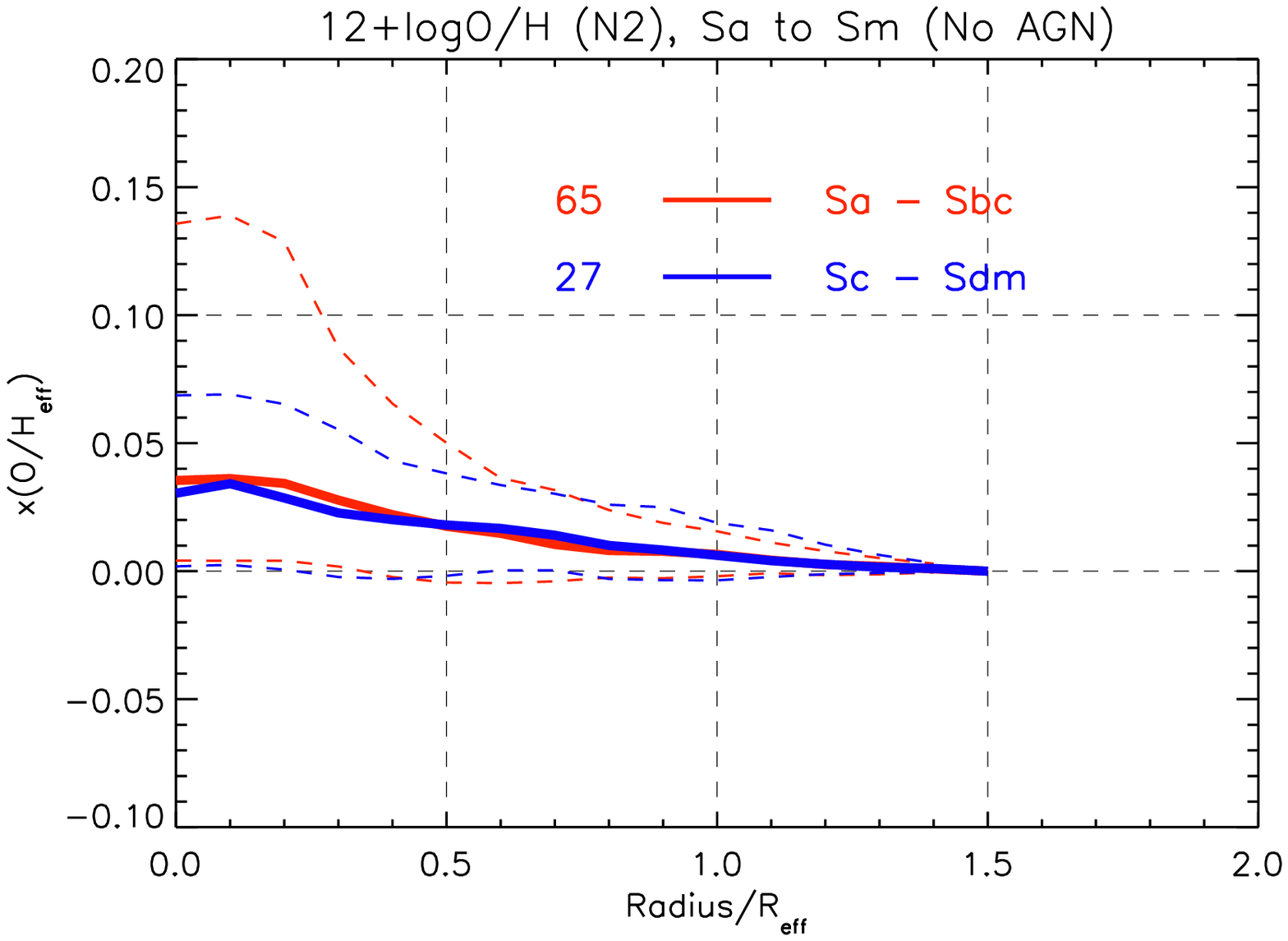}
\caption{{\bf Left:} \xohp~for two different morphological type bins, estimated with N2 (M13). Solid lines correspond to median values. Dashed lines contain 68.2\% of the distribution.
Sa-Sbc and Sc-Sm galaxies are represented in blue and red respectively.
{\bf Right:} Same as left for \xohe.\label{oh_n2ha_type}}
\end{figure}

\begin{figure}
\plottwo{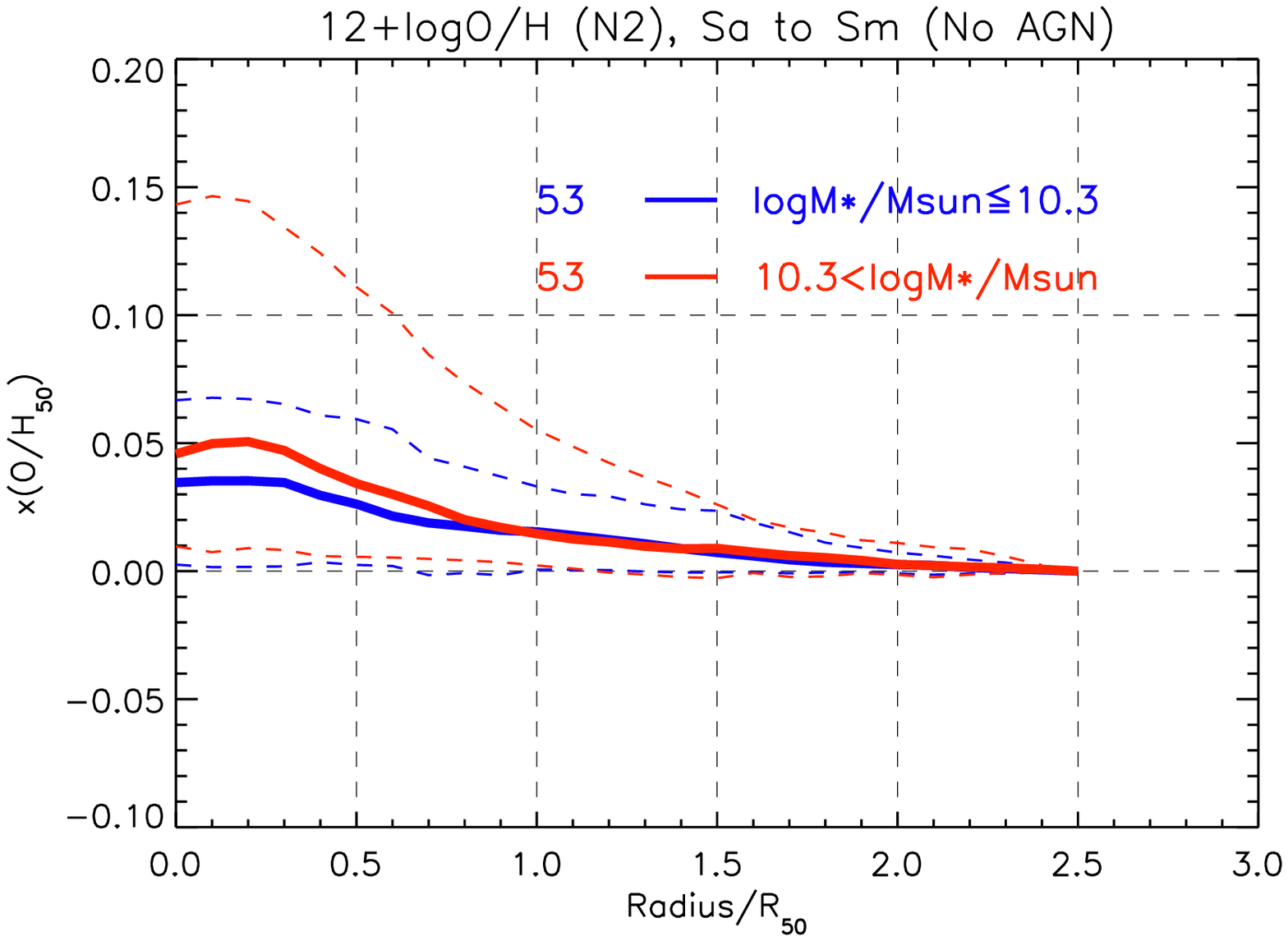}{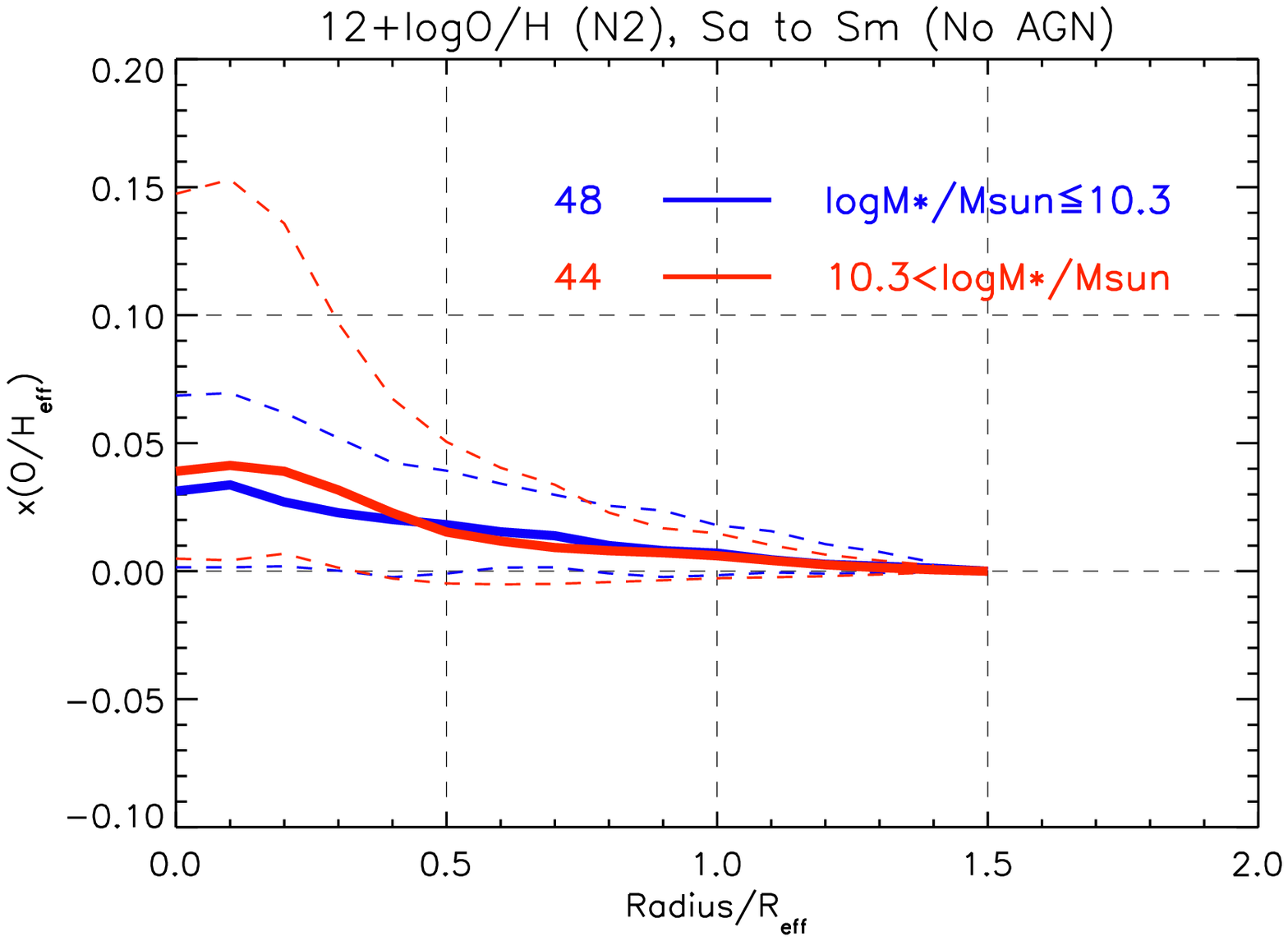}
\caption{{\bf Left:} \xohp~for two different stellar mass bins, estimated with N2 (M13). Solid lines correspond to median values. Dashed lines contain 68.2\% of the distribution.
Galaxies with $\log M^{*}/M_{\odot} \leq 10.3$ and $\log M^{*}/M_{\odot} > 10.3$ are represented in blue and red respectively.
{\bf Right:} Same as left for \xohe.\label{oh_n2ha_mass}}
\end{figure}

\begin{figure}
\plottwo{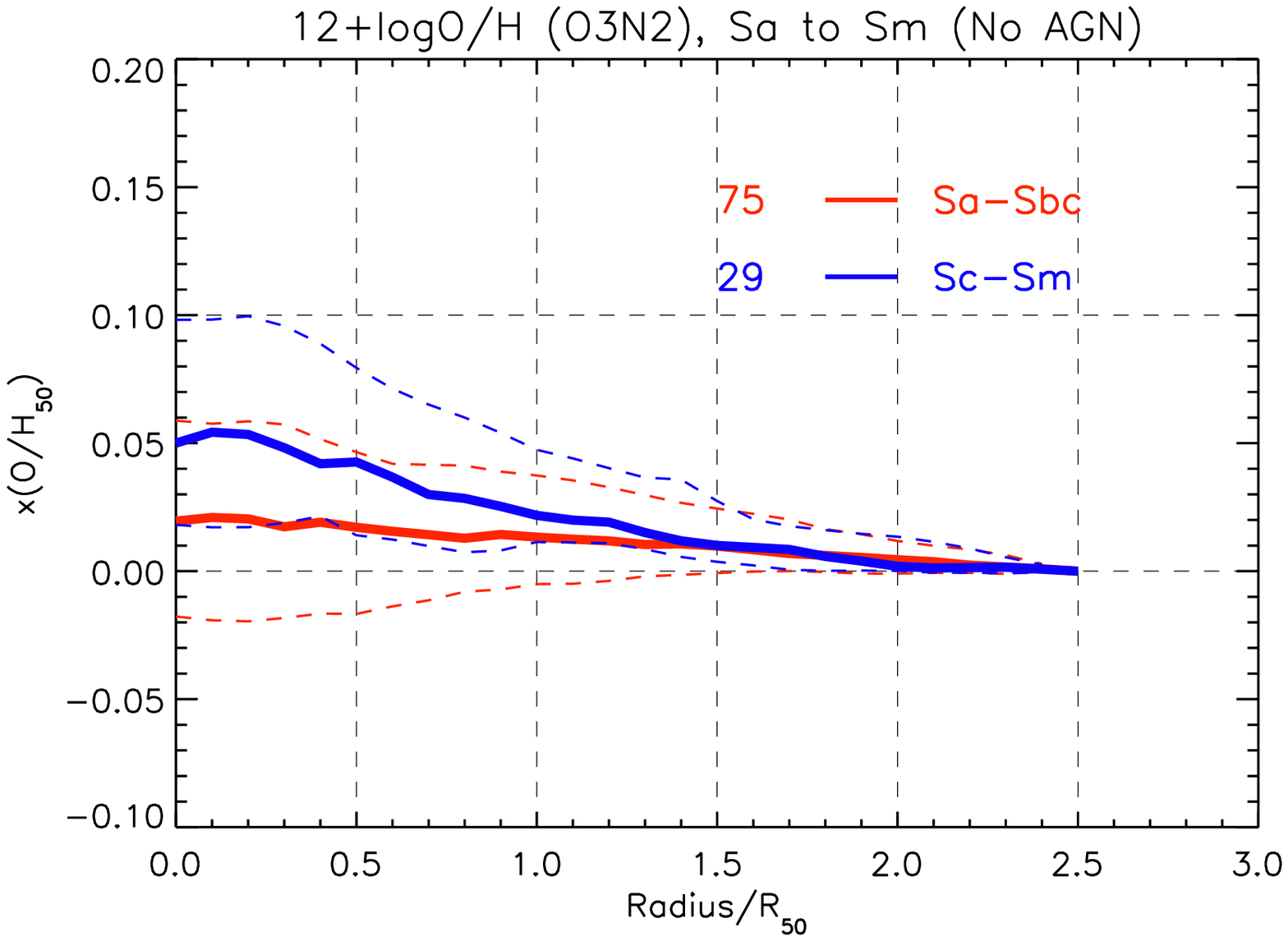}{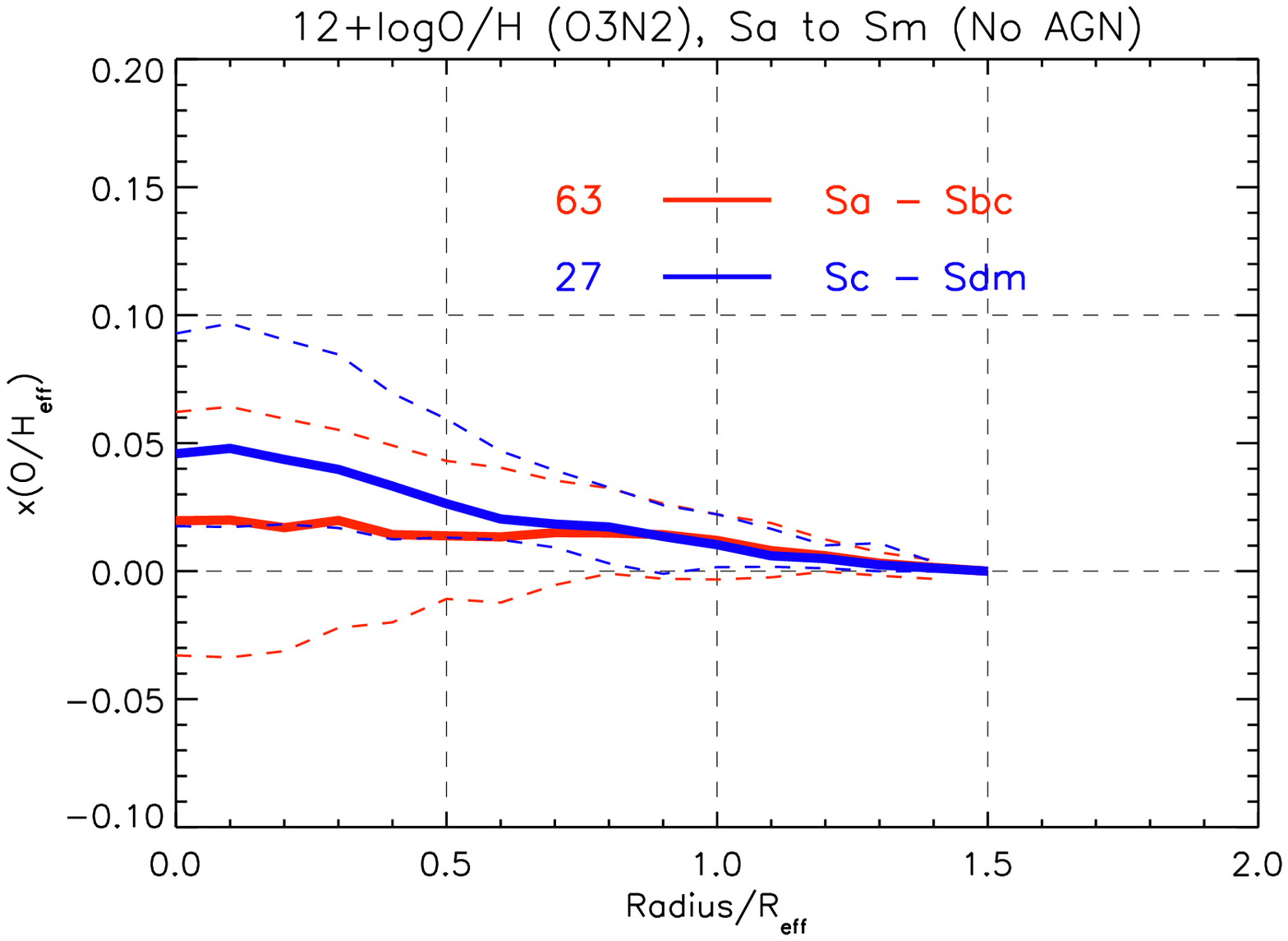}
\caption{{\bf Left:} \xohp~for two different morphological type bins, estimated with O3N2 (M13). Solid lines correspond to median values. Dashed lines contain 68.2\% of the distribution.
Sa-Sbc and Sc-Sm galaxies are represented in blue and red respectively.
{\bf Right:} Same as left for \xohe.\label{oh_o3n2_type}}
\end{figure}

\begin{figure}
\plottwo{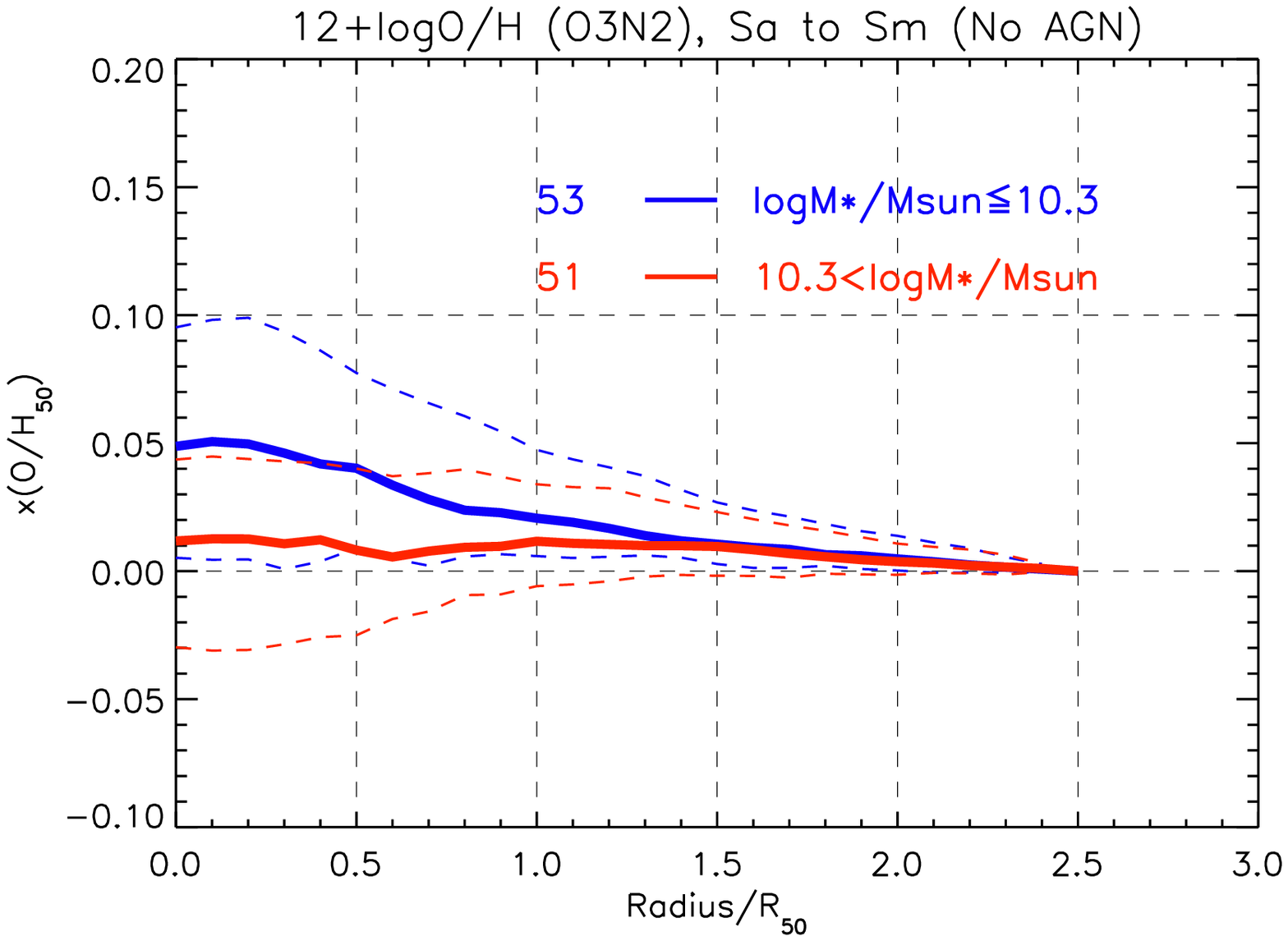}{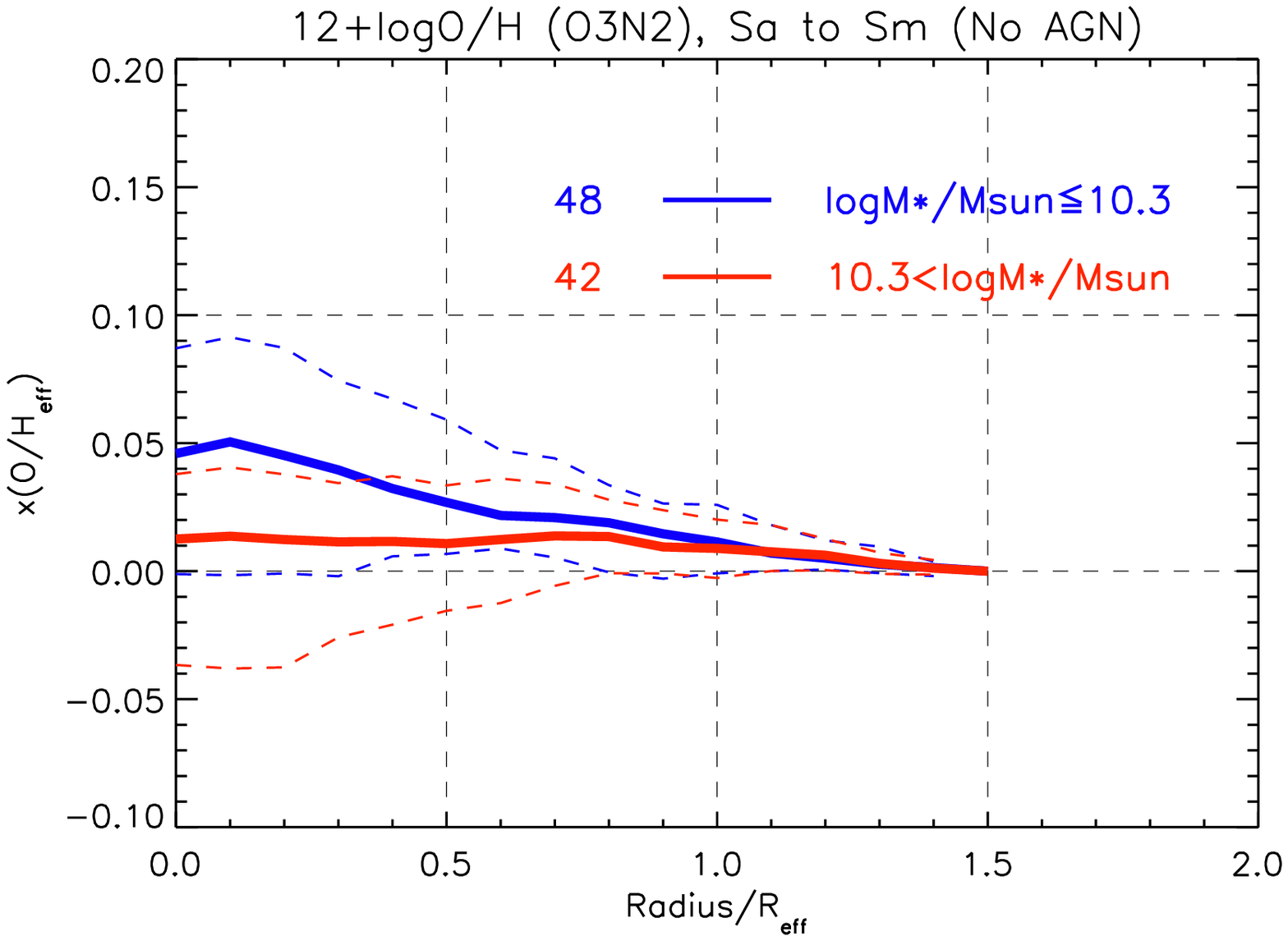}
\caption{{\bf Left:} \xohp~for two different stellar mass bins, estimated with O3N2 (M13). Solid lines correspond to median values. Dashed lines contain 68.2\% of the distribution.
Galaxies with $\log M^{*}/M_{\odot} \leq 10.3$ and $\log M^{*}/M_{\odot} > 10.3$ are represented in blue and red respectively.
{\bf Right:} Same as left for \xohe.\label{oh_o3n2_mass}}
\end{figure}

Figures~\ref{oh_n2ha_type} to ~\ref{oh_o3n2_mass} show \xohp~and \xohe~ estimated from N2 and O3N2 using the M13 calibrations.
As shown in the figures, the effect of the aperture for both indicators is maximal for small apertures and only marginal differences (maximum values of $\approx 0.04$~dex) with
morphological type and stellar mass are seeing, much lower than the typical dispersions around the median values. 
However, what is more relevant is the dispersions around the median values, than can reach maximal values of up to $+25$\% 
for small apertures when deriving oxygen abundances of early-type and/or high mass spirals using N2.
We keep in mind that using other calibrations for the oxygen abundances could result in even larger dispersions around the 
median values. In particular, as shown in figure~\ref{oh_type_compameta}, using for example the PMC09 calibrations would result in dispersions $\approx 5$\% larger than using the M13 calibrations.
This means that when the fluxes are obtained through very small apertures, the aperture corrected oxygen abundances for individual galaxies could be very uncertain and far from the real values. 
We stress again that the median aperture corrections for oxygen abundances (and other relevant quantities reported in this work) must be applied to large samples of galaxies and interpreted in a statistical sense.

\subsection{A sample case: SDSS galaxies at $0.02 \leq z \leq 0.3$}

In this subsection we apply the results of our work on a sample of galaxies from the SDSS survey and quantify 
the average effect of the aperture resulting when estimating the oxygen abundance from the flux measured through the SDSS fiber.
For this we will make use of the corrections derived using \rp~ and split in two bins of stellar masses, 
since these two parameters are available for all the SDSS galaxies.

\begin{figure}
\plotone{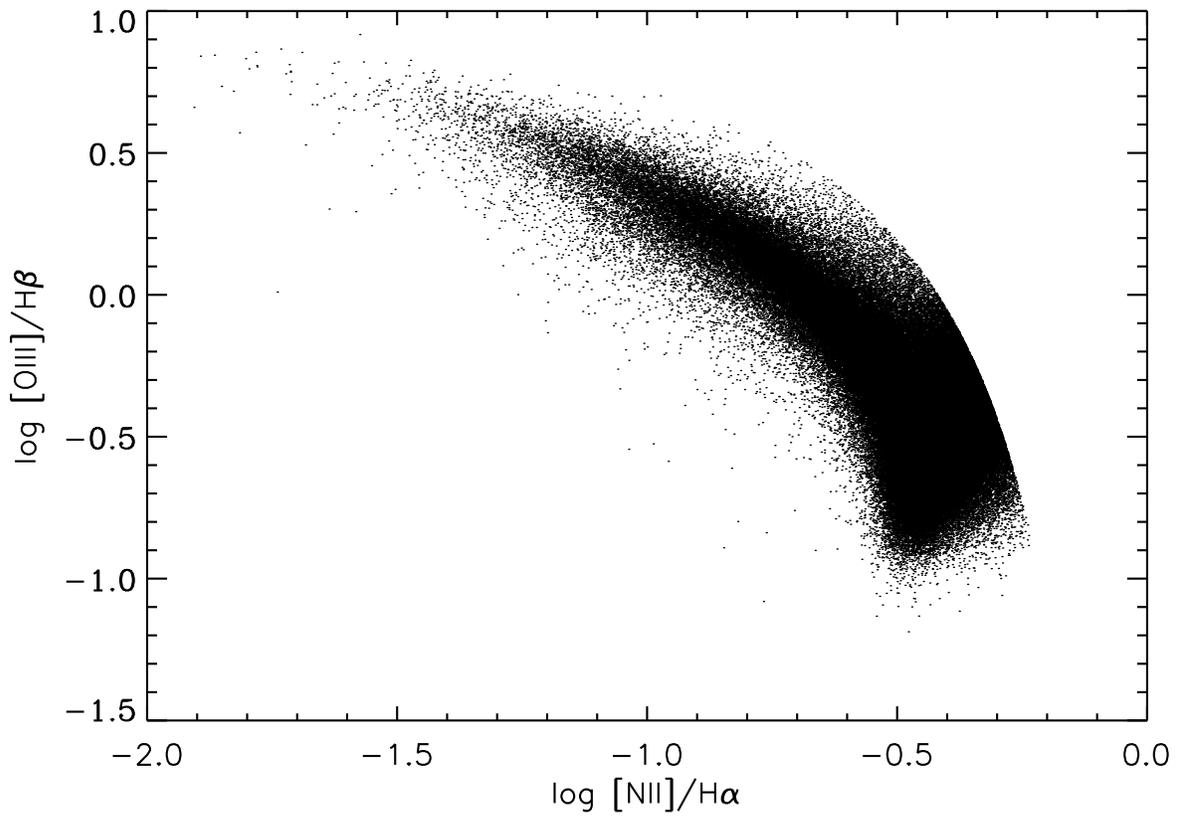}
\caption{BPT diagram corresponding to the SDSS sample.\label{sdss_bpt}}
\end{figure}

The sample of galaxies is the MPA-JHU sample (Kauffmann et al. 2003, Brinchmann et al. 2004, Tremonti et al. 2004, and Salim et al. 2007), 
that provides stellar masses and uses spectroscopy from SDSS-DR7 (Abazajian et al. 2009), 
and photometry complemented from SDSS-DR12 (Alam et al. 2015), satisfying the following criteria:
\begin{itemize}
\item Redshift in the range $0.02 \leq z \leq 0.3$.
\item Signal-to-noise ratio of the emission lines H$\alpha$, H$\beta$, [\ion{N}{2}]~$\lambda$6583 and [\ion{O}{3}]~$\lambda$5007 larger than 3.
\item Stellar mass in the range $8.5 \leq \log M^{*}/M_{\odot} \leq 11.5$.
\item Signal-to-noise ratio of the half-light Petrosian radius in the $r'$ band ($R_{50}$) larger than 3.
\end{itemize}
Next we remove those galaxies classified as QSO by SDSS.
We also remove those galaxies not classified as star-forming according to the BPT diagram (Baldwin et al. 1981) of Kauffmann et al. (2003) 
and the condition EW(H$\alpha$) $\ge 3$~\AA\ as indicated in the WHAN diagram of Cid Fernandes et al. (2011).
Finally, we keep only the galaxies for which $0.3 \leq 1.5''/R_{50} \leq 2.5$, as it is the range of 
applicability of the corrections as indicated in IP13.
Figure~\ref{sdss_bpt} shows the BPT diagram of the SDSS sample.

\begin{figure}
\plottwo{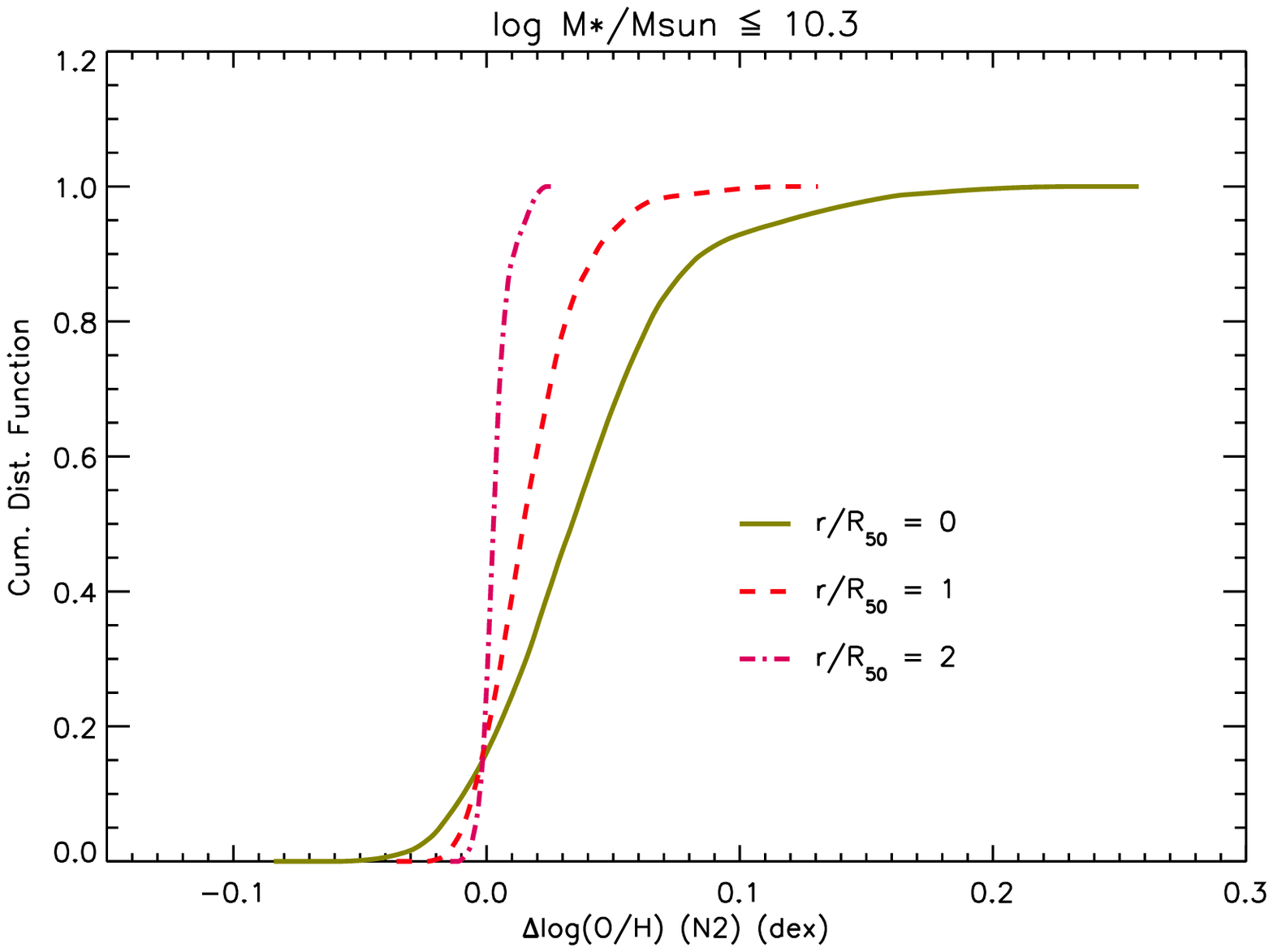}{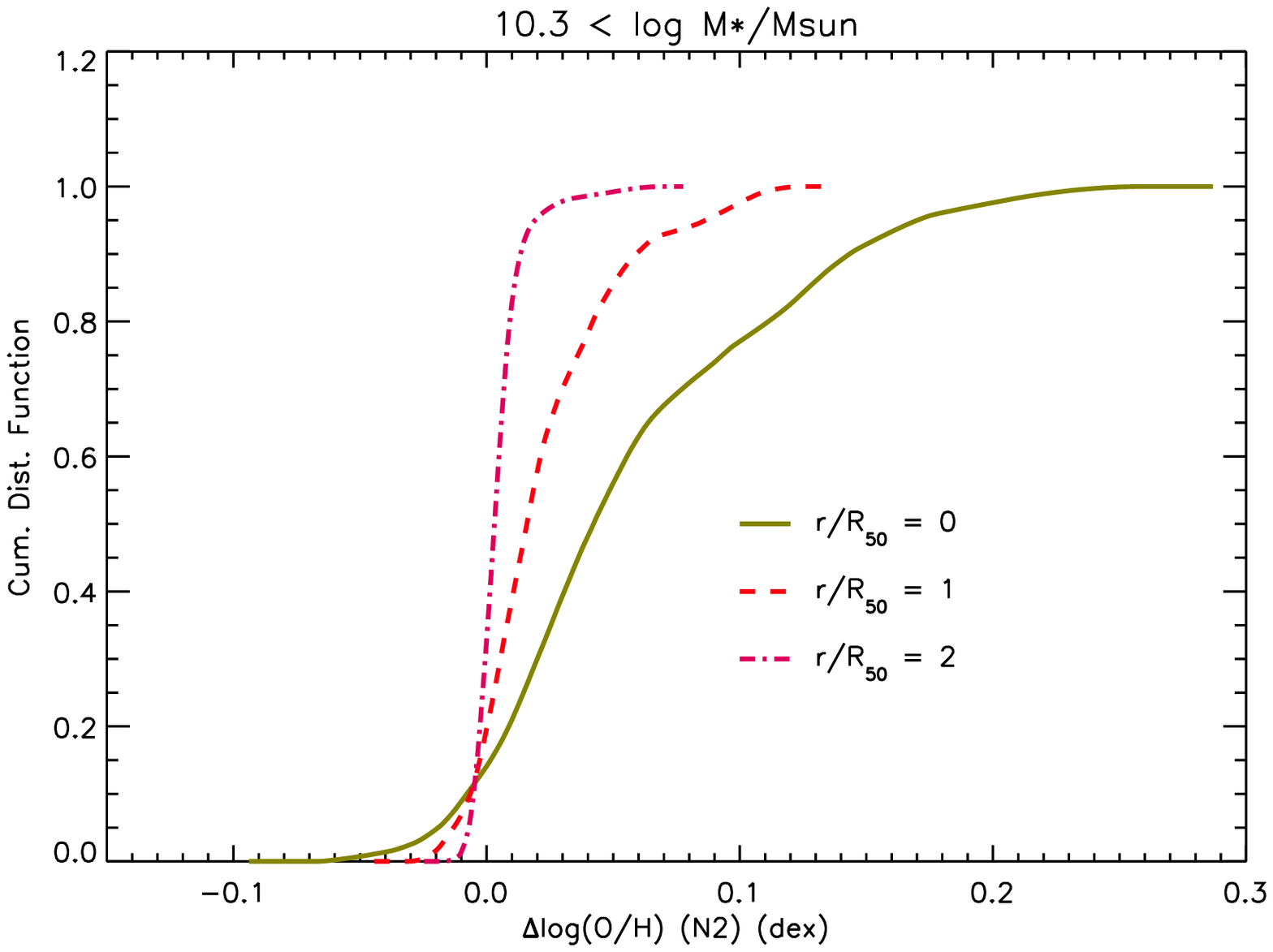}
\caption{{\bf Left:} Cumulative Distribution Function (CDF) of \xohp~estimated with N2, for galaxies with $\log M^{*}/M_{\odot} \leq 10.3$ for apertures of three different radii. 
{\bf Right:} Cumulative Distribution Function (CDF) of \xohp~estimated with N2, for galaxies with $\log M^{*}/M_{\odot} > 10.3$ for apertures of three different radii.\label{cdf_oh_n2ha}}
\end{figure}

\begin{figure}
\plottwo{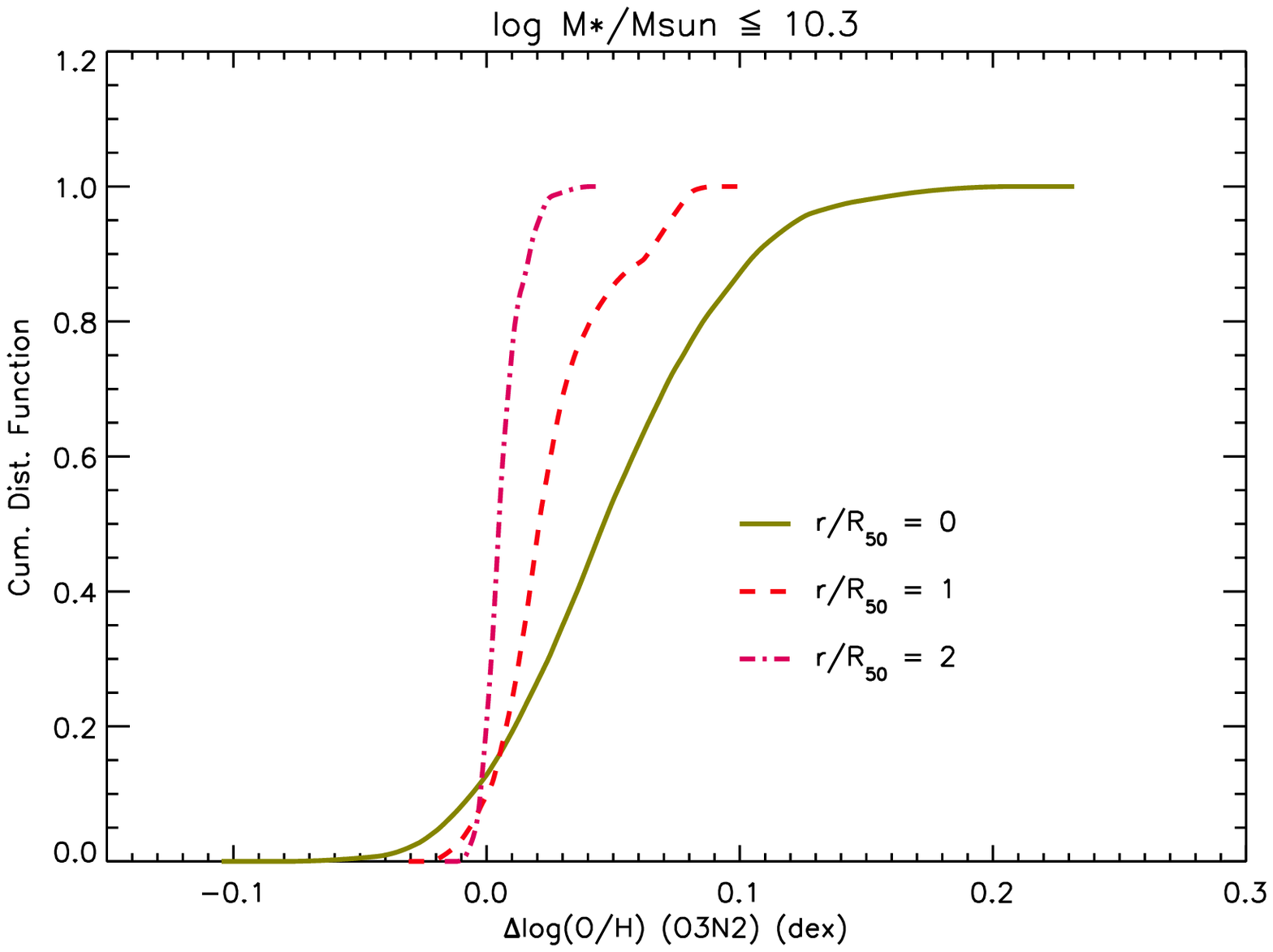}{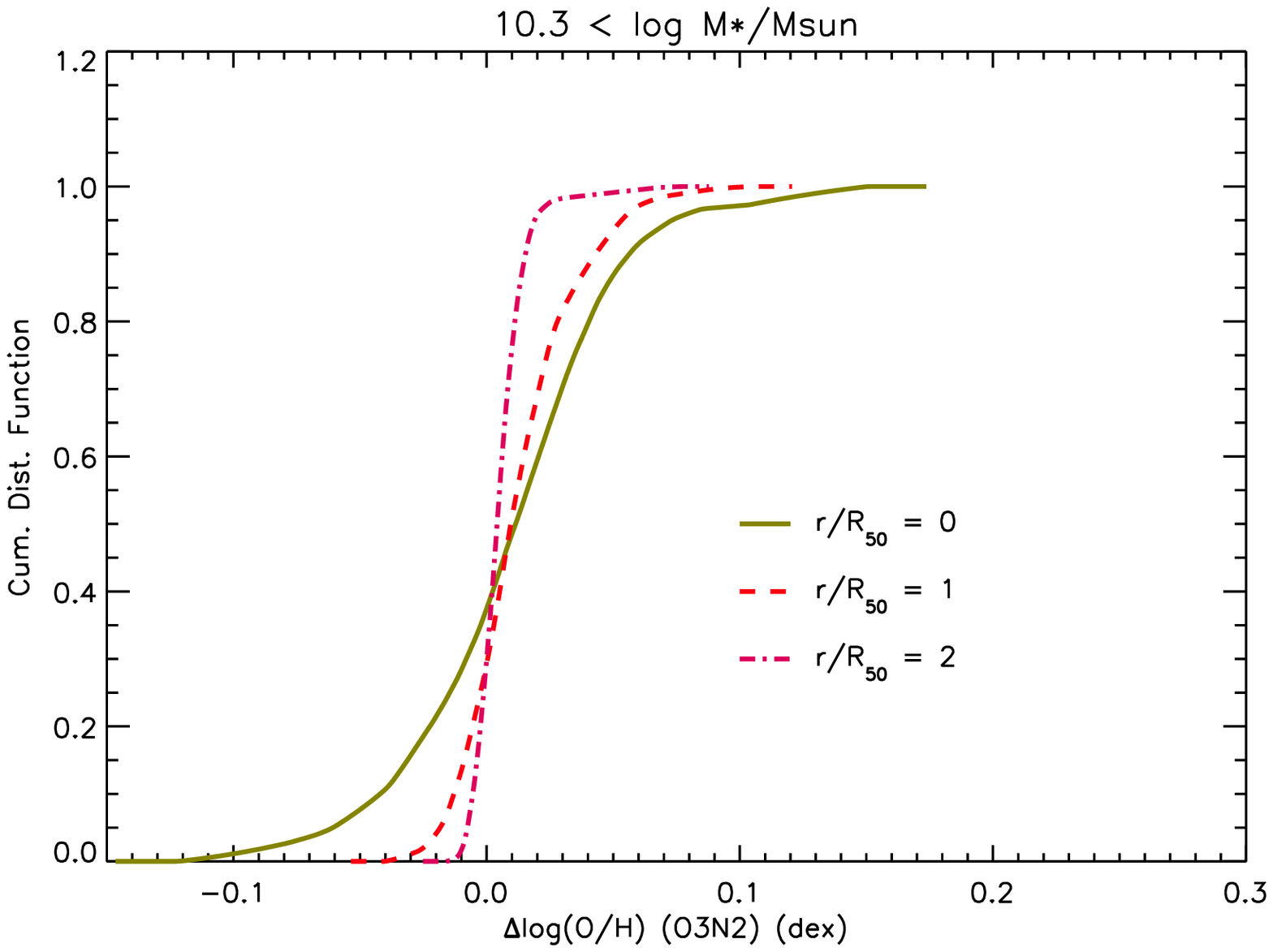}
\caption{{\bf Left:} Cumulative Distribution Function (CDF) of \xohp~estimated with O3N2, for galaxies with $\log M^{*}/M_{\odot} \leq 10.3$ for apertures of three different radii. 
{\bf Right:} Cumulative Distribution Function (CDF) of \xohp~estimated with O3N2, for galaxies with $\log M^{*}/M_{\odot} > 10.3$ for apertures of three different radii.\label{cdf_oh_o3n2}}
\end{figure}

In order to be as much realistic as possible, we want to take into account the large dispersion observed around the median values of the growth curves for small apertures.
For this, we first construct approximate cumulative distribution functions (CDFs) of the distribution of the individual O/H growth curves at each radius 
for the two stellar mass bins previously considered in Figs.~\ref{oh_n2ha_mass} and~\ref{oh_o3n2_mass}.
Figures~\ref{cdf_oh_n2ha} and~\ref{cdf_oh_o3n2} show the CDFs at three different radii normalized to~\rp.
As it can be seen, the larger the radius normalized with respect to~\rp, the closer the distribution remains to zero, as it was observed in the previous section.
Several Monte Carlo tests were produced to verify that the random distributions following these CDFs are consistent with being drawn from the same parental distribution as the original distributions from the CALIFA galaxies.

Then we are ready to quantify the aperture effect with a sample of galaxies from SDSS for which the stellar mass, \rp, and the fluxes of the required emission lines are well known. 
To each of the galaxies we will apply a value of the correction corresponding to its stellar mass and coverage of the SDSS fiber (normalized to \rp) taken at random from the corresponding CDF.
The difference between the oxygen abundances estimated from the flux of the SDSS fiber and that obtained using the aperture correction as described above was computed for each galaxy using the two indicators N2 and O3N2.
Then, the median value of these differences was computed for five stellar mass bins: $\log M^{*}/M_{\odot} \in$~[8.5,9.1], [9.1,9.7], [9.7,10.3], [10.3,10.9] and [10.9,11.5].
We repeated this process a total of 25 times, and computed the average value of the median of the differences for each of these four mass bins.
Table~\ref{abun_monte} shows the results of this simulation split in six redshift bins: $z \in$~[0.02,0.05], [0.05,0.10], [0.10,0.15], [0.15,0.20], [0.20,0.25] and [0.25,0.30].

In this study the aperture effect depends on two different aspects: on the one side, the fraction of galaxy covered by the SDSS fiber is partly driven by the redshift of the galaxies. On the other hand, the aperture effect depends on the stellar mass of the galaxy. Thus, table~\ref{abun_monte} must show a combination of these effects.
The first thing we note is that the maximum average aperture effect measured with these two calibrations at the redshifts and stellar mass ranges probed is of the order of $\approx 0.047$~dex ($\approx 11$\%), which as previously said, is much smaller than the typical uncertainties of any of the two empirical calibrations.
The average corrections when using the N2 calibration tend to increase with stellar mass at all redshift ranges probed, and for a given stellar mass, the average correction decreases whn redshift increases.
Regarding the O3N2 calibration, it reaches a maximum for stellar masses $9.7 < \log M^{*}/M_{\odot} < 10.3$ at all redshift ranges, and for a given mass range it tends to decrease as redshift increases.

We also remark that in this work we have studied the effect of the aperture on the indicators used to estimate the oxygen abundance (N2 and O3N2). 
However, it is out of the scope of this paper enter in the discussion about the goodness of the integrated values of these two indicators as proxies for the oxygen abundance, 
since they contain emission not only from star-forming regions but also from the diffuse ionized gas of the disks of spiral galaxies.
It has been reported in the literature that the oxygen abundance estimated from integrated fluxes of emission lines 
approximately equals the corresponding abundance of star-forming regions at the galactocentric distance $0.4 \times R_{opt}$ (Pilyugin et al. 2004; Moustakas et al. 2006).
Although this relation links these two quantities, still further information is required in order to get information about the chemical evolution of spiral galaxies.

\section{Conclusions}

This paper presents the growth curves taken through circular apertures of several emission line fluxes for a sample of spiral galaxies from the CALIFA project.
These curves allow to study the effect of estimating galactic properties from the light of a circular aperture covering only partially the spatial extent of a galaxy instead of using the integrated light.
The main conclusions arising from this work are the following:
\begin{itemize}
\item Whereas the median H$\alpha$ growth curves are insensitive to inclination, morphology
and total stellar mass of the studied galaxies, our analysis documents
a strong dependence of the registered H$\alpha$ luminosity on the aperture
size (consequently, also on redshift).
\item The median H$\alpha$/H$\beta$ growth curves are insensitive to the inclination, morphological types and stellar masses of the galaxies, and shows a mild trend with a maximum for small apertures smoothly decreasing towards large apertures, which means that the extinction is overestimated on average when observing galaxies through small apertures.
\item The median N2 growth curves are also insensitive to the inclinations, morphological types and stellar masses of the galaxies, showing a decreasing trend towards large apertures.
This means that the oxygen abundance is overestimated when observing through small apertures.
\item The median O3N2 growth curves are insensitive to the inclinations, and show higher values for galaxies of earlier morphological types and $\log M^{*}/M_{\odot} > 10.3$ at small radii apertures.
Again, this means that estimating oxygen abundances through small apertures tends to overestimate the abundances.
\item When applying our aperture corrections to a sample of SDSS galaxies with $0.02 \leq z \leq 0.3$, 
the average corrected oxygen abundances derived from the SDSS fluxes may result enhanced by a maximum of $\approx 11$\% with respect to the fiber-based ones.
\end{itemize}

This work has shown that although the median aperture corrections for oxygen abundance are always small, the dispersions around these median values become very large for small apertures, thus preventing the use of aperture corrections for studies of individual galaxies unless a very large uncertainty is assumed.
A more detailed study on the aperture effects on some individual emission line ratios will be presented in a forthcoming paper.

\acknowledgments

This study made use of the data provided by the Calar
Alto Legacy Integral Field Area (CALIFA) survey ({\tt {http://califa.caha.es/}}). 
The CALIFA collaboration would like to thank the IAA-CSIC and
MPIA-MPG as major partners of the observatory, and CAHA itself, for the
unique access to telescope time and support in manpower and infrastructures.
The CALIFA collaboration also thanks the CAHA staff for the dedication to
this project. Based on observations collected at the Centro Astron\'{o}mico Hispano
Alem\'{a}n (CAHA) at Calar Alto, operated jointly by the Max-Planck-Institut f\"{u}r
Astronomie and the Instituto de Astrof\'{\i}sica de Andaluc\'{\i}a (CSIC). We thank the
Viabilidad, Dise\~{n}o, Acceso y Mejora funding program ICTS-2009-10, for supporting
the initial developement of this project.
JIP, JVM, CK, EPM and SDP acknowledge financial support from the Spanish MINECO under
grant AYA2010-21887-C04-01, and from Junta de Andaluc\'{\i}a Excellence Project
PEX2011-FQM7058.
CCT and ACM also thank the support from the Plan Nacional de 
Investigaci\'{o}n y Desarrollo funding program AYA2013-46724-P.
JMG acknowledges support by Funda\c{c}\~{a}o para a Ci\^{e}ncia e a Tecnologia (FCT) through the
Fellowship SFRH/BPD/66958/2009 and POPH/FSE (EC) by FEDER funding through
the program Programa Operacional de Factores de Competitividade
(COMPETE). PP is supported by FCT through the Investigador FCT Contract
No. IF/01220/2013 and POPH/FSE (EC) by FEDER funding through the program
COMPETE. JMG\&PP also acknowledge support by FCT under project
FCOMP-01-0124-FEDER-029170 (Reference FCT PTDC/FIS-AST/3214/2012), funded
by FCT-MEC (PIDDAC) and FEDER (COMPETE).
They also acknowledge support by the exchange programme
‘Study of Emission-Line Galaxies with Integral-Field Spectroscopy’
(SELGIFS, FP7-PEOPLE-2013-IRSES-612701), funded by the EU through the
IRSES scheme.
FFRO acknowledges the exchange programme ‘Study of Emission-Line Galaxies with Integral-Field Spectroscopy’ (SELGIFS, FP7-PEOPLE-2013-IRSES-612701), funded by the EU through the IRSES scheme.
Support for LG is provided by the Ministry of Economy, Development, and Tourism's 
Millennium Science Initiative through grant IC120009, awarded to The Millennium Institute of Astrophysics, MAS. 
LG acknowledges support by CONICYT through FONDECYT grant 3140566.
This research has made use of
the NASA/IPAC Extragalactic Database (NED) which is operated by the Jet
Propulsion Laboratory, California Institute of Technology, under contract with
the National Aeronautics and Space Administration.
We acknowledge the usage of the HyperLeda database ({\tt {http://leda.univ-lyon1.fr}}).
Funding for SDSS-III has been provided by the Alfred P. Sloan Foundation,
the Participating Institutions, the National Science Foundation, and the
U.S. Department of Energy Office of Science. SDSS-III is managed by the
Astrophysical Research Consortium for the Participating Institutions of
the SDSS-III Collaboration including the University of Arizona, the
Brazilian Participation Group, Brookhaven National Laboratory, Carnegie
Mellon University, University of Florida, the French Participation Group,
the German Participation Group, Harvard University, the Instituto de
Astrof\'{\i}sica de Canarias, the Michigan State/Notre Dame/JINA Participation
Group, Johns Hopkins University, Lawrence Berkeley National Laboratory,
Max Planck Institute for Astrophysics, Max Planck Institute for
Extraterrestrial Physics, New Mexico State University, New York
University, Ohio State University, Pennsylvania State University,
University of Portsmouth, Princeton University, the Spanish Participation
Group, University of Tokyo, University of Utah, Vanderbilt University,
University of Virginia, University of Washington, and Yale University. The
SDSS-III web site is {\tt {http://www.sdss3.org}}.

\clearpage

\clearpage

\clearpage

\clearpage

\clearpage

\clearpage

\clearpage

\clearpage

\clearpage

\clearpage

\clearpage

\clearpage

\clearpage

\clearpage

\begin{table}
\begin{center}
\caption{Comparison of the H$\alpha$ growth curve and average dispersion of IP13 with the ones obtained in this work for all the spiral galaxies disregarding the stellar mass, morphological types and inclination.\label{compa}}
\begin{tabular}{ccccc}
\tableline\tableline
$r$/\rp & \xhap & $\sigma$(\xhap) & \xhap & $\sigma$(\xhap) \\    
        & \multicolumn{2}{c}{(IP13)} & \multicolumn{2}{c}{(This work)} \\
\tableline
0.3 & 0.091 & 0.074 & 0.085 & 0.060 \\
0.5 & 0.194 & 0.119 & 0.182 & 0.102 \\
0.7 & 0.301 & 0.156 & 0.283 & 0.126 \\
0.9 & 0.419 & 0.158 & 0.397 & 0.132 \\
1.1 & 0.530 & 0.146 & 0.506 & 0.134 \\
1.3 & 0.629 & 0.130 & 0.620 & 0.121 \\
1.5 & 0.718 & 0.117 & 0.715 & 0.121 \\
1.7 & 0.798 & 0.101 & 0.801 & 0.104 \\
1.9 & 0.854 & 0.079 & 0.865 & 0.078 \\
2.1 & 0.917 & 0.070 & 0.920 & 0.050 \\
2.3 & 0.961 & 0.039 & 0.965 & 0.027 \\
\tableline
\end{tabular}
\end{center}
\end{table}

\clearpage

\begin{table}
\begin{center}
\caption{Description of the sizes of the two subsamples after each of the filters imposed.\label{samples}}
\begin{tabular}{llll}
\tableline\tableline
\multicolumn{2}{c}{$S_{50}$} & \multicolumn{2}{c}{$S_{\rm eff}$} \\
\tableline
Initial sample & 402 & Initial sample & 394$\dagger$ \\
Excluding Es and S0s & 301 & Excluding Es and S0s & 296 \\
Excluding AGNs & 212 & Excluding AGNs & 208 \\
\rp~$\leq 14.4$'' & 171 & \re~$\leq 24$'' & 138 \\
$f$(H$\alpha$)~$\ge 0$, $\sigma$(H$\alpha$)/$f$(H$\alpha$)~$\le 0.333$ & 165 & $f$(H$\alpha$)~$\ge 0$, $\sigma$(H$\alpha$)/$f$(H$\alpha$)~$\le 0.333$ & 133 \\
\tableline
\end{tabular}

$\dagger$ The initial number of galaxies in $S_{\rm eff}$ is different than that in $S_{50}$ because \re~could not be determined for 8 galaxies, and thus they are excluded from the analysis.
\end{center}
\end{table}

\clearpage

\begin{table}
\begin{center}
\caption{Basic properties of the samples $S_{50}$ and $S_{\rm eff}$: Median values of the stellar masses, \rp~and \re; Number of galaxies split in two bins of morphological types.\label{samples_prop}}
\begin{tabular}{ccc}
\tableline\tableline
 & $S_{50}$ & $S_{\rm eff}$ \\
\tableline
$\langle \log M^{*}/M_{\odot} \rangle$ & 10.25 & 10.25 \\
$\langle$\rp $\rangle$ ('')          & 11.04 & \\
$\langle$\re $\rangle$ ('')          &       & 20.31 \\
Sa$+$Sab$+$Sb$+$Sbc & 107 & 85 \\
Sc$+$Scd$+$Sd$+$Sdm$+$Sm & 85 & 48 \\
\tableline
\end{tabular}
\end{center}
\end{table}

\clearpage

\begin{table}
\begin{center}
\caption{\xhap~ for two bins of morphological types assuming the same weight for all the galaxies.
(1) Aperture radius in units of \rp.
(2)-(4) Values of $x$(H$\alpha$) corresponding to 15.86\%, 50\% and 84.14\% of the distribution for Sa-Sbc galaxies.
(5)-(7) Values of $x$(H$\alpha$) corresponding to 15.86\%, 50\% and 84.14\% of the distribution for Sc-Sdm galaxies.
(8)-(10) Values of $x$(H$\alpha$) corresponding to 15.86\%, 50\% and 84.14\% of the distribution for all galaxies.\label{ha_type_tab}}
\begin{tabular}{cccccccccc}
\tableline\tableline
$r$/\rp & \multicolumn{3}{c}{Sa-Sbc} & \multicolumn{3}{c}{Sc-Sdm}  & \multicolumn{3}{c}{All}\\    
        & $-\sigma$ & \xhap & $+\sigma$ & $-\sigma$ & \xhap & $+\sigma$ & $-\sigma$ & \xhap & $+\sigma$ \\
\tableline
    0.00 &    0.000 &    0.000 &    0.000 &    0.000 &    0.000 &    0.000  &    0.000 &    0.000 &    0.000 \\
    0.10 &    0.007 &    0.021 &    0.066 &    0.009 &    0.018 &    0.032  &    0.008 &    0.020 &    0.057 \\
    0.20 &    0.022 &    0.046 &    0.131 &    0.027 &    0.045 &    0.073  &    0.024 &    0.046 &    0.114 \\
    0.30 &    0.040 &    0.082 &    0.200 &    0.053 &    0.086 &    0.135  &    0.049 &    0.085 &    0.169 \\
    0.40 &    0.072 &    0.130 &    0.262 &    0.084 &    0.132 &    0.213  &    0.077 &    0.130 &    0.246 \\
    0.50 &    0.105 &    0.182 &    0.333 &    0.128 &    0.187 &    0.284  &    0.111 &    0.182 &    0.316 \\
\end{tabular}
\end{center}
\end{table}

\clearpage

\begin{table}
\begin{center}
\caption{\xhae~ for two bins of morphological types assuming the same weight for all the galaxies.
(1) Aperture radius in units of \re.
(2)-(4) Values of $x$(H$\alpha$) corresponding to 15.86\%, 50\% and 84.14\% of the distribution for Sa-Sbc galaxies.
(5)-(7) Values of $x$(H$\alpha$) corresponding to 15.86\%, 50\% and 84.14\% of the distribution for Sc-Sdm galaxies.
(8)-(10) Values of $x$(H$\alpha$) corresponding to 15.86\%, 50\% and 84.14\% of the distribution for all galaxies.
\label{ha_type_re_tab}}
\begin{tabular}{cccccccccc}
\tableline\tableline
$r$/\re & \multicolumn{3}{c}{Sa-Sbc} & \multicolumn{3}{c}{Sc-Sdm}  & \multicolumn{3}{c}{All} \\    
        & $-\sigma$ & \xhae & $+\sigma$ & $-\sigma$ & \xhae & $+\sigma$  $-\sigma$ & \xhae & $+\sigma$ \\
\tableline
    0.00 &    0.000 &    0.000 &    0.000 &    0.000 &    0.000 &    0.000  &    0.000 &    0.000 &    0.000 \\
    0.10 &    0.016 &    0.038 &    0.142 &    0.018 &    0.040 &    0.080  &    0.017 &    0.039 &    0.107 \\
    0.20 &    0.049 &    0.109 &    0.237 &    0.057 &    0.100 &    0.189  &    0.054 &    0.107 &    0.227 \\
    0.30 &    0.103 &    0.210 &    0.364 &    0.113 &    0.184 &    0.310  &    0.103 &    0.195 &    0.335 \\
    0.40 &    0.173 &    0.284 &    0.471 &    0.189 &    0.271 &    0.425  &    0.177 &    0.280 &    0.457 \\
    0.50 &    0.254 &    0.388 &    0.589 &    0.264 &    0.365 &    0.531  &    0.256 &    0.385 &    0.561 \\
\end{tabular}
\end{center}
\end{table}

\clearpage

\begin{table}
\begin{center}
\caption{\xhap~ for two bins of stellar masses assuming the same weight for all the galaxies.
(1) Aperture radius in units of \rp.
(2)-(4) Values of $x$(H$\alpha$) corresponding to 15.86\%, 50\% and 84.14\% of the distribution for $\log M^{*} \leq 10.3$ galaxies.
(5)-(7) Values of $x$(H$\alpha$) corresponding to 15.86\%, 50\% and 84.14\% of the distribution for $\log M^{*} > 10.3$ galaxies.\label{ha_mass_tab}}
\begin{tabular}{ccccccc}
\tableline\tableline
$r$/\rp & \multicolumn{3}{c}{$\log M^{*} \leq 10.3$} & \multicolumn{3}{c}{$\log M^{*} > 10.3$} \\    
        & $-\sigma$ & \xhap & $+\sigma$ & $-\sigma$ & \xhap & $+\sigma$ \\
\tableline
    0.00 &    0.000 &    0.000 &    0.000 &    0.000 &    0.000 &    0.000 \\
    0.10 &    0.009 &    0.020 &    0.039 &    0.007 &    0.020 &    0.066 \\
    0.20 &    0.027 &    0.046 &    0.083 &    0.021 &    0.048 &    0.132 \\
    0.30 &    0.052 &    0.085 &    0.146 &    0.039 &    0.084 &    0.202 \\
    0.40 &    0.083 &    0.133 &    0.231 &    0.070 &    0.130 &    0.263 \\
    0.50 &    0.120 &    0.186 &    0.298 &    0.104 &    0.182 &    0.331 \\
\end{tabular}
\end{center}
\end{table}

\clearpage

\begin{table}
\begin{center}
\caption{\xhae~ for two bins of stellar masses assuming the same weight for all the galaxies.
(1) Aperture radius in units of \re.
(2)-(4) Values of $x$(H$\alpha$) corresponding to 15.86\%, 50\% and 84.14\% of the distribution for $\log M^{*} \leq 10.3$ galaxies.
(5)-(7) Values of $x$(H$\alpha$) corresponding to 15.86\%, 50\% and 84.14\% of the distribution for $\log M^{*} > 10.3$ galaxies.\label{ha_mass_re_tab}}
\begin{tabular}{ccccccc}
\tableline\tableline
$r$/\re & \multicolumn{3}{c}{$\log M^{*} \leq 10.3$} & \multicolumn{3}{c}{$\log M^{*} > 10.3$} \\    
        & $-\sigma$ & \xhae & $+\sigma$ & $-\sigma$ & \xhae & $+\sigma$ \\
\tableline
    0.00 &    0.000 &    0.000 &    0.000 &    0.000 &    0.000 &    0.000 \\
    0.10 &    0.019 &    0.040 &    0.091 &    0.016 &    0.035 &    0.114 \\
    0.20 &    0.057 &    0.110 &    0.229 &    0.046 &    0.096 &    0.226 \\
    0.30 &    0.116 &    0.203 &    0.375 &    0.103 &    0.180 &    0.332 \\
    0.40 &    0.191 &    0.293 &    0.464 &    0.165 &    0.263 &    0.452 \\
    0.50 &    0.273 &    0.404 &    0.587 &    0.235 &    0.366 &    0.557 \\
    0.60 &    0.357 &    0.498 &    0.667 &    0.322 &    0.465 &    0.638 \\
    0.70 &    0.441 &    0.590 &    0.732 &    0.415 &    0.561 &    0.718 \\
    0.80 &    0.525 &    0.657 &    0.815 &    0.536 &    0.643 &    0.775 \\
    0.90 &    0.612 &    0.744 &    0.871 &    0.612 &    0.713 &    0.831 \\
    1.00 &    0.683 &    0.803 &    0.905 &    0.685 &    0.784 &    0.867 \\
    1.10 &    0.768 &    0.851 &    0.933 &    0.771 &    0.850 &    0.922 \\
    1.20 &    0.841 &    0.906 &    0.966 &    0.843 &    0.898 &    0.938 \\
    1.30 &    0.903 &    0.946 &    0.982 &    0.902 &    0.938 &    0.970 \\
    1.40 &    0.956 &    0.977 &    0.994 &    0.957 &    0.973 &    0.990 \\
    1.50 &    1.000 &    1.000 &    1.000 &    1.000 &    1.000 &    1.000 \\
\tableline
\end{tabular}
\end{center}
\end{table}

\clearpage

\begin{table}
\begin{center}
\caption{\xhahbp~ for two bins of morphological types assuming the same weight for all the galaxies.
(1) Aperture radius in units of \rp.
(2)-(4) Values of \xhahbp~ corresponding to 15.86\%, 50\% and 84.14\% of the distribution for Sa-Sbc galaxies.
(5)-(7) Values of \xhahbp~ corresponding to 15.86\%, 50\% and 84.14\% of the distribution for Sc-Sdm galaxies.
(8)-(10) Values of \xhahbp~ corresponding to 15.86\%, 50\% and 84.14\% of the distribution for all galaxies.
\label{hahb_type_tab}}
\begin{tabular}{cccccccccc}
\tableline\tableline
$r$/\rp & \multicolumn{3}{c}{Sa-Sbc} & \multicolumn{3}{c}{Sc-Sdm}  & \multicolumn{3}{c}{All} \\    
        & $-\sigma$ & \xhahbp & $+\sigma$ & $-\sigma$ & \xhahbp & $+\sigma$  & $-\sigma$ & \xhahbp & $+\sigma$ \\
\tableline
    0.00 &    1.006 &    1.144 &    1.391 &    0.965 &    1.087 &    1.234  &    0.992 &    1.123 &    1.333 \\
    0.10 &    0.996 &    1.153 &    1.404 &    0.962 &    1.092 &    1.244  &    0.985 &    1.123 &    1.351 \\
    0.20 &    0.998 &    1.149 &    1.399 &    0.964 &    1.094 &    1.241  &    0.984 &    1.126 &    1.352 \\
    0.30 &    1.010 &    1.153 &    1.369 &    0.987 &    1.092 &    1.233  &    0.993 &    1.118 &    1.328 \\
    0.40 &    1.007 &    1.132 &    1.304 &    0.995 &    1.079 &    1.220  &    0.996 &    1.104 &    1.284 \\
    0.50 &    0.992 &    1.113 &    1.260 &    0.994 &    1.059 &    1.200  &    0.993 &    1.083 &    1.234 \\
\end{tabular}
\end{center}
\end{table}

\clearpage

\begin{table}
\begin{center}
\caption{\xhahbe~ for two bins of morphological types assuming the same weight for all the galaxies.
(1) Aperture radius in units of \re.
(2)-(4) Values of \xhahbe~ corresponding to 15.86\%, 50\% and 84.14\% of the distribution for Sa-Sbc galaxies.
(5)-(7) Values of \xhahbe~ corresponding to 15.86\%, 50\% and 84.14\% of the distribution for Sc-Sdm galaxies.
(8)-(10) Values of \xhahbe~ corresponding to 15.86\%, 50\% and 84.14\% of the distribution for all galaxies.
\label{hahb_type_re_tab}}
\begin{tabular}{cccccccccc}
\tableline\tableline
$r$/\re & \multicolumn{3}{c}{Sa-Sbc} & \multicolumn{3}{c}{Sc-Sdm}  & \multicolumn{3}{c}{All} \\    
        & $-\sigma$ & \xhahbe & $+\sigma$ & $-\sigma$ & \xhahbe & $+\sigma$  & $-\sigma$ & \xhahbe & $+\sigma$ \\
\tableline
    0.00 &    0.968 &    1.114 &    1.369 &    0.940 &    1.101 &    1.196  &    0.965 &    1.109 &    1.320 \\
    0.10 &    0.974 &    1.116 &    1.387 &    0.933 &    1.102 &    1.192  &    0.958 &    1.110 &    1.330 \\
    0.20 &    0.966 &    1.110 &    1.311 &    0.941 &    1.093 &    1.208  &    0.954 &    1.104 &    1.295 \\
    0.30 &    0.968 &    1.085 &    1.216 &    0.974 &    1.067 &    1.210  &    0.969 &    1.076 &    1.213 \\
    0.40 &    0.963 &    1.080 &    1.201 &    0.965 &    1.047 &    1.123  &    0.965 &    1.064 &    1.180 \\
    0.50 &    0.981 &    1.064 &    1.173 &    0.970 &    1.044 &    1.107  &    0.975 &    1.055 &    1.140 \\
\end{tabular}
\end{center}
\end{table}

\clearpage

\begin{table}
\begin{center}
\caption{\xhahbp~ for two bins of stellar masses assuming the same weight for all the galaxies.
(1) Aperture radius in units of \rp.
(2)-(4) Values of \xhahbp~ corresponding to 15.86\%, 50\% and 84.14\% of the distribution for $\log M^{*} \leq 10.3$ galaxies.
(5)-(7) Values of \xhahbp~ corresponding to 15.86\%, 50\% and 84.14\% of the distribution for $\log M^{*} > 10.3$ galaxies.\label{hahb_mass_tab}}
\begin{tabular}{ccccccc}
\tableline\tableline
$r$/\rp & \multicolumn{3}{c}{$\log M^{*} \leq 10.3$} & \multicolumn{3}{c}{$\log M^{*} > 10.3$} \\    
        & $-\sigma$ & \xhahbp & $+\sigma$ & $-\sigma$ & \xhahbp & $+\sigma$ \\
\tableline
    0.00 &    0.989 &    1.102 &    1.257 &    1.005 &    1.178 &    1.441 \\
    0.10 &    0.981 &    1.100 &    1.267 &    0.998 &    1.181 &    1.461 \\
    0.20 &    0.982 &    1.101 &    1.263 &    1.000 &    1.181 &    1.461 \\
    0.30 &    0.993 &    1.101 &    1.255 &    1.003 &    1.167 &    1.430 \\
    0.40 &    0.996 &    1.081 &    1.216 &    1.005 &    1.149 &    1.340 \\
    0.50 &    0.995 &    1.063 &    1.199 &    0.992 &    1.124 &    1.287 \\
\end{tabular}
\end{center}
\end{table}

\clearpage

\begin{table}
\begin{center}
\caption{\xhahbe~ for two bins of stellar masses assuming the same weight for all the galaxies.
(1) Aperture radius in units of \re.
(2)-(4) Values of \xhahbe~ corresponding to 15.86\%, 50\% and 84.14\% of the distribution for $\log M^{*} \leq 10.3$ galaxies.
(5)-(7) Values of \xhahbe~ corresponding to 15.86\%, 50\% and 84.14\% of the distribution for $\log M^{*} > 10.3$ galaxies.\label{hahb_mass_re_tab}}
\begin{tabular}{ccccccc}
\tableline\tableline
$r$/\re & \multicolumn{3}{c}{$\log M^{*} \leq 10.3$} & \multicolumn{3}{c}{$\log M^{*} > 10.3$} \\    
        & $-\sigma$ & \xhahbe & $+\sigma$ & $-\sigma$ & \xhahbe & $+\sigma$ \\
\tableline
    0.00 &    0.941 &    1.094 &    1.226 &    0.969 &    1.145 &    1.370 \\
    0.10 &    0.941 &    1.099 &    1.239 &    0.975 &    1.146 &    1.393 \\
    0.20 &    0.941 &    1.084 &    1.219 &    0.984 &    1.126 &    1.320 \\
    0.30 &    0.964 &    1.057 &    1.203 &    0.995 &    1.096 &    1.232 \\
    0.40 &    0.956 &    1.047 &    1.144 &    0.979 &    1.090 &    1.210 \\
    0.50 &    0.973 &    1.044 &    1.115 &    0.981 &    1.073 &    1.179 \\
    0.60 &    0.963 &    1.036 &    1.093 &    0.968 &    1.064 &    1.166 \\
    0.70 &    0.956 &    1.030 &    1.093 &    0.959 &    1.039 &    1.145 \\
    0.80 &    0.958 &    1.020 &    1.102 &    0.959 &    1.023 &    1.122 \\
    0.90 &    0.962 &    1.018 &    1.102 &    0.952 &    1.023 &    1.117 \\
    1.00 &    0.971 &    1.016 &    1.080 &    0.962 &    1.020 &    1.097 \\
    1.10 &    0.967 &    1.010 &    1.071 &    0.952 &    1.018 &    1.093 \\
    1.20 &    0.959 &    1.009 &    1.065 &    0.960 &    1.009 &    1.069 \\
    1.30 &    0.972 &    1.004 &    1.051 &    0.958 &    1.006 &    1.054 \\
    1.40 &    0.986 &    1.003 &    1.025 &    0.977 &    1.002 &    1.028 \\
    1.50 &    1.000 &    1.000 &    1.000 &    1.000 &    1.000 &    1.000 \\
\tableline
\end{tabular}
\end{center}
\end{table}

\clearpage

\begin{table}
\begin{center}
\caption{\xnhap~ for two bins of morphological types assuming the same weight for all the galaxies.
(1) Aperture radius in units of \rp.
(2)-(4) Values of \xnhap~ corresponding to 15.86\%, 50\% and 84.14\% of the distribution for Sa-Sbc galaxies.
(5)-(7) Values of \xnhap~ corresponding to 15.86\%, 50\% and 84.14\% of the distribution for Sc-Sdm galaxies.
(8)-(10) Values of \xnhap~ corresponding to 15.86\%, 50\% and 84.14\% of the distribution for all galaxies.
\label{n2ha_type_tab}}
\begin{tabular}{cccccccccc}
\tableline\tableline
$r$/\rp & \multicolumn{3}{c}{Sa-Sbc} & \multicolumn{3}{c}{Sc-Sdm}  & \multicolumn{3}{c}{All} \\    
        & $-\sigma$ & \xnhap & $+\sigma$ & $-\sigma$ & \xnhap & $+\sigma$  & $-\sigma$ & \xnhap & $+\sigma$ \\
\tableline
    0.00 &    0.005 &    0.075 &    0.249 &    0.031 &    0.087 &    0.152  &    0.006 &    0.081 &    0.221 \\
    0.10 &    0.006 &    0.074 &    0.254 &    0.032 &    0.091 &    0.148  &    0.008 &    0.082 &    0.223 \\
    0.20 &    0.006 &    0.074 &    0.255 &    0.033 &    0.090 &    0.151  &    0.008 &    0.082 &    0.223 \\
    0.30 &    0.004 &    0.072 &    0.248 &    0.034 &    0.082 &    0.165  &    0.007 &    0.080 &    0.200 \\
    0.40 &    0.001 &    0.072 &    0.226 &    0.034 &    0.072 &    0.163  &    0.004 &    0.072 &    0.189 \\
    0.50 &    0.000 &    0.064 &    0.196 &    0.028 &    0.062 &    0.143  &    0.005 &    0.063 &    0.160 \\
\end{tabular}
\end{center}
\end{table}

\clearpage

\begin{table}
\begin{center}
\caption{\xnhae~ for two bins of morphological types assuming the same weight for all the galaxies.
(1) Aperture radius in units of \re.
(2)-(4) Values of \xnhae~ corresponding to 15.86\%, 50\% and 84.14\% of the distribution for Sa-Sbc galaxies.
(5)-(7) Values of \xnhae~ corresponding to 15.86\%, 50\% and 84.14\% of the distribution for Sc-Sdm galaxies.
(8)-(10) Values of \xnhae~ corresponding to 15.86\%, 50\% and 84.14\% of the distribution for all galaxies.
\label{n2ha_type_re_tab}}
\begin{tabular}{cccccccccc}
\tableline\tableline
$r$/\re & \multicolumn{3}{c}{Sa-Sbc} & \multicolumn{3}{c}{Sc-Sdm}  & \multicolumn{3}{c}{All} \\    
        & $-\sigma$ & \xnhae & $+\sigma$ & $-\sigma$ & \xnhae & $+\sigma$  & $-\sigma$ & \xnhae & $+\sigma$ \\
\tableline
    0.00 &    0.008 &    0.073 &    0.252 &    0.004 &    0.066 &    0.149  &    0.006 &    0.072 &    0.170 \\
    0.10 &    0.007 &    0.076 &    0.259 &    0.005 &    0.074 &    0.150  &    0.006 &    0.075 &    0.178 \\
    0.20 &    0.008 &    0.068 &    0.236 &    0.001 &    0.062 &    0.141  &    0.005 &    0.064 &    0.171 \\
    0.30 &    0.002 &    0.052 &    0.169 &   -0.005 &    0.049 &    0.120  &    0.001 &    0.051 &    0.150 \\
    0.40 &   -0.006 &    0.045 &    0.127 &   -0.007 &    0.044 &    0.093  &   -0.006 &    0.044 &    0.104 \\
    0.50 &   -0.010 &    0.032 &    0.102 &   -0.004 &    0.039 &    0.083  &   -0.008 &    0.033 &    0.092 \\
\end{tabular}
\end{center}
\end{table}

\clearpage

\begin{table}
\begin{center}
\caption{\xnhap~ for two bins of stellar masses assuming the same weight for all the galaxies.
(1) Aperture radius in units of \rp.
(2)-(4) Values of \xnhap~ corresponding to 15.86\%, 50\% and 84.14\% of the distribution for $\log M^{*} \leq 10.3$ galaxies.
(5)-(7) Values of \xnhap~ corresponding to 15.86\%, 50\% and 84.14\% of the distribution for $\log M^{*} > 10.3$ galaxies.\label{n2ha_mass_tab}}
\begin{tabular}{ccccccc}
\tableline\tableline
$r$/\rp & \multicolumn{3}{c}{$\log M^{*} \leq 10.3$} & \multicolumn{3}{c}{$\log M^{*} > 10.3$} \\    
        & $-\sigma$ & \xnhap & $+\sigma$ & $-\sigma$ & \xnhap & $+\sigma$ \\
\tableline
    0.00 &    0.006 &    0.075 &    0.145 &    0.013 &    0.081 &    0.258 \\
    0.10 &    0.003 &    0.076 &    0.147 &    0.014 &    0.085 &    0.266 \\
    0.20 &    0.004 &    0.076 &    0.146 &    0.014 &    0.083 &    0.269 \\
    0.30 &    0.004 &    0.075 &    0.141 &    0.014 &    0.084 &    0.264 \\
    0.40 &    0.008 &    0.064 &    0.132 &    0.002 &    0.083 &    0.249 \\
    0.50 &    0.005 &    0.057 &    0.129 &    0.004 &    0.071 &    0.218 \\
\end{tabular}
\end{center}
\end{table}

\clearpage

\begin{table}
\begin{center}
\caption{\xnhae~ for two bins of stellar masses assuming the same weight for all the galaxies.
(1) Aperture radius in units of \re.
(2)-(4) Values of \xnhae~ corresponding to 15.86\%, 50\% and 84.14\% of the distribution for edge-on $\log M^{*} \leq 10.3$ galaxies.
(5)-(7) Values of \xnhae~ corresponding to 15.86\%, 50\% and 84.14\% of the distribution for face-on $\log M^{*} > 10.3$ galaxies.\label{n2ha_mass_re_tab}}
\begin{tabular}{ccccccc}
\tableline\tableline
$r$/\re & \multicolumn{3}{c}{$\log M^{*} \leq 10.3$} & \multicolumn{3}{c}{$\log M^{*} > 10.3$ } \\    
        & $-\sigma$ & \xnhae & $+\sigma$ & $-\sigma$ & \xnhae & $+\sigma$ \\
\tableline
    0.00 &    0.003 &    0.068 &    0.148 &    0.009 &    0.077 &    0.293 \\
    0.10 &    0.003 &    0.073 &    0.151 &    0.007 &    0.078 &    0.300 \\
    0.20 &    0.004 &    0.058 &    0.134 &    0.008 &    0.077 &    0.249 \\
    0.30 &    0.000 &    0.049 &    0.113 &   -0.000 &    0.056 &    0.188 \\
    0.40 &   -0.005 &    0.044 &    0.092 &   -0.008 &    0.042 &    0.131 \\
    0.50 &   -0.002 &    0.039 &    0.085 &   -0.012 &    0.029 &    0.101 \\
    0.60 &    0.003 &    0.033 &    0.074 &   -0.012 &    0.020 &    0.074 \\
    0.70 &    0.003 &    0.030 &    0.065 &   -0.012 &    0.019 &    0.064 \\
    0.80 &   -0.002 &    0.022 &    0.055 &   -0.010 &    0.015 &    0.048 \\
    0.90 &   -0.005 &    0.017 &    0.051 &   -0.009 &    0.013 &    0.031 \\
    1.00 &   -0.003 &    0.015 &    0.039 &   -0.007 &    0.012 &    0.026 \\
    1.10 &   -0.001 &    0.010 &    0.034 &   -0.006 &    0.008 &    0.019 \\
    1.20 &   -0.002 &    0.006 &    0.023 &   -0.005 &    0.005 &    0.014 \\
    1.30 &   -0.001 &    0.004 &    0.016 &   -0.003 &    0.002 &    0.008 \\
    1.40 &   -0.001 &    0.002 &    0.007 &   -0.001 &    0.001 &    0.005 \\
    1.50 &    0.000 &    0.000 &    0.000 &    0.000 &    0.000 &    0.000 \\
\tableline
\end{tabular}
\end{center}
\end{table}

\clearpage

\begin{table}
\begin{center}
\caption{\xonp~ for two bins of morphological types assuming the same weight for all the galaxies.
(1) Aperture radius in units of \rp.
(2)-(4) Values of \xonp~ corresponding to 15.86\%, 50\% and 84.14\% of the distribution for Sa-Sbc galaxies.
(5)-(7) Values of \xonp~ corresponding to 15.86\%, 50\% and 84.14\% of the distribution for Sc-Sdm galaxies.
(8)-(10) Values of \xonp~ corresponding to 15.86\%, 50\% and 84.14\% of the distribution for all galaxies.
\label{o3n2_type_tab}}
\begin{tabular}{cccccccccc}
\tableline\tableline
$r$/\rp & \multicolumn{3}{c}{Sa-Sbc} & \multicolumn{3}{c}{Sc-Sdm}  & \multicolumn{3}{c}{All} \\    
        & $-\sigma$ & \xonp & $+\sigma$ & $-\sigma$ & \xonp & $+\sigma$  & $-\sigma$ & \xonp & $+\sigma$ \\
\tableline
    0.00 &   -0.275 &   -0.091 &    0.083 &   -0.459 &   -0.234 &   -0.084  &   -0.368 &   -0.127 &    0.063 \\
    0.10 &   -0.269 &   -0.098 &    0.090 &   -0.459 &   -0.254 &   -0.080  &   -0.385 &   -0.133 &    0.061 \\
    0.20 &   -0.274 &   -0.095 &    0.092 &   -0.465 &   -0.250 &   -0.080  &   -0.379 &   -0.128 &    0.060 \\
    0.30 &   -0.268 &   -0.081 &    0.085 &   -0.448 &   -0.225 &   -0.088  &   -0.348 &   -0.123 &    0.063 \\
    0.40 &   -0.241 &   -0.089 &    0.077 &   -0.415 &   -0.196 &   -0.100  &   -0.299 &   -0.125 &    0.061 \\
    0.50 &   -0.217 &   -0.080 &    0.078 &   -0.371 &   -0.199 &   -0.066  &   -0.251 &   -0.111 &    0.052 \\
\end{tabular}
\end{center}
\end{table}

\clearpage

\begin{table}
\begin{center}
\caption{\xone~ for two bins of morphological types assuming the same weight for all the galaxies.
(1) Aperture radius in units of \re.
(2)-(4) Values of \xone~ corresponding to 15.86\%, 50\% and 84.14\% of the distribution for Sa-Sbc galaxies.
(5)-(7) Values of \xone~ corresponding to 15.86\%, 50\% and 84.14\% of the distribution for Sc-Sdm galaxies.
(8)-(10) Values of \xone~ corresponding to 15.86\%, 50\% and 84.14\% of the distribution for all galaxies.
\label{o3n2_type_re_tab}}
\begin{tabular}{cccccccccc}
\tableline\tableline
$r$/\re & \multicolumn{3}{c}{Sa-Sbc} & \multicolumn{3}{c}{Sc-Sdm}  & \multicolumn{3}{c}{All} \\    
        & $-\sigma$ & \xone & $+\sigma$ & $-\sigma$ & \xone & $+\sigma$  & $-\sigma$ & \xone & $+\sigma$ \\
\tableline
    0.00 &   -0.291 &   -0.092 &    0.154 &   -0.434 &   -0.214 &   -0.082  &   -0.338 &   -0.140 &    0.072 \\
    0.10 &   -0.300 &   -0.093 &    0.157 &   -0.452 &   -0.224 &   -0.081  &   -0.355 &   -0.145 &    0.071 \\
    0.20 &   -0.278 &   -0.079 &    0.146 &   -0.422 &   -0.204 &   -0.085  &   -0.310 &   -0.137 &    0.081 \\
    0.30 &   -0.258 &   -0.093 &    0.103 &   -0.396 &   -0.186 &   -0.079  &   -0.264 &   -0.117 &    0.074 \\
    0.40 &   -0.229 &   -0.067 &    0.093 &   -0.324 &   -0.155 &   -0.058  &   -0.234 &   -0.101 &    0.020 \\
    0.50 &   -0.201 &   -0.065 &    0.051 &   -0.278 &   -0.123 &   -0.061  &   -0.213 &   -0.091 &    0.009 \\
\end{tabular}
\end{center}
\end{table}

\clearpage

\begin{table}
\begin{center}
\caption{\xonp~ for two bins of stellar masses assuming the same weight for all the galaxies.
(1) Aperture radius in units of \rp.
(2)-(4) Values of \xonp~ corresponding to 15.86\%, 50\% and 84.14\% of the distribution for $\log M^{*} \leq 10.3$ galaxies.
(5)-(7) Values of \xonp~ corresponding to 15.86\%, 50\% and 84.14\% of the distribution for $\log M^{*} > 10.3$ galaxies.\label{o3n2_mass_tab}}
\begin{tabular}{ccccccc}
\tableline\tableline
$r$/\rp & \multicolumn{3}{c}{$\log M^{*} \leq 10.3$} & \multicolumn{3}{c}{$\log M^{*} > 10.3$} \\    
        & $-\sigma$ & \xonp & $+\sigma$ & $-\sigma$ & \xonp & $+\sigma$ \\
\tableline
    0.00 &   -0.445 &   -0.228 &   -0.024 &   -0.203 &   -0.055 &    0.139 \\
    0.10 &   -0.459 &   -0.237 &   -0.021 &   -0.209 &   -0.059 &    0.146 \\
    0.20 &   -0.462 &   -0.232 &   -0.022 &   -0.204 &   -0.059 &    0.144 \\
    0.30 &   -0.437 &   -0.215 &   -0.004 &   -0.199 &   -0.050 &    0.134 \\
    0.40 &   -0.403 &   -0.196 &   -0.018 &   -0.196 &   -0.057 &    0.120 \\
    0.50 &   -0.362 &   -0.188 &   -0.042 &   -0.187 &   -0.038 &    0.117 \\
\end{tabular}
\end{center}
\end{table}

\clearpage

\begin{table}
\begin{center}
\caption{\xone~ for two bins of stellar masses assuming the same weight for all the galaxies.
(1) Aperture radius in units of \re.
(2)-(4) Values of \xone~ corresponding to 15.86\%, 50\% and 84.14\% of the distribution for $\log M^{*} \leq 10.3$ galaxies.
(5)-(7) Values of \xone~ corresponding to 15.86\%, 50\% and 84.14\% of the distribution for $\log M^{*} > 10.3$ galaxies.\label{o3n2_mass_re_tab}}
\begin{tabular}{ccccccc}
\tableline\tableline
 $r$/\re & \multicolumn{3}{c}{$\log M^{*} \leq 10.3$} & \multicolumn{3}{c}{$\log M^{*} > 10.3$} \\    
         & $-\sigma$ & \xone & $+\sigma$ & $-\sigma$ & \xone & $+\sigma$ \\
\tableline
    0.00 &   -0.407 &   -0.215 &    0.005 &   -0.177 &   -0.059 &    0.171 \\
    0.10 &   -0.427 &   -0.236 &    0.008 &   -0.189 &   -0.064 &    0.178 \\
    0.20 &   -0.407 &   -0.211 &    0.004 &   -0.177 &   -0.058 &    0.176 \\
    0.30 &   -0.347 &   -0.185 &    0.009 &   -0.161 &   -0.053 &    0.121 \\
    0.40 &   -0.314 &   -0.151 &   -0.027 &   -0.173 &   -0.054 &    0.098 \\
    0.50 &   -0.276 &   -0.125 &   -0.032 &   -0.157 &   -0.050 &    0.073 \\
    0.60 &   -0.221 &   -0.102 &   -0.041 &   -0.169 &   -0.058 &    0.058 \\
    0.70 &   -0.206 &   -0.098 &   -0.025 &   -0.159 &   -0.064 &    0.027 \\
    0.80 &   -0.157 &   -0.088 &    0.002 &   -0.130 &   -0.063 &    0.004 \\
    0.90 &   -0.124 &   -0.068 &    0.014 &   -0.111 &   -0.044 &    0.004 \\
    1.00 &   -0.121 &   -0.053 &    0.004 &   -0.094 &   -0.042 &    0.012 \\
    1.10 &   -0.084 &   -0.033 &   -0.000 &   -0.084 &   -0.035 &   -0.000 \\
    1.20 &   -0.056 &   -0.024 &   -0.003 &   -0.059 &   -0.029 &   -0.002 \\
    1.30 &   -0.045 &   -0.013 &    0.004 &   -0.034 &   -0.014 &    0.005 \\
    1.40 &   -0.018 &   -0.007 &    0.009 &   -0.020 &   -0.005 &    0.006 \\
    1.50 &    0.000 &    0.000 &    0.000 &    0.000 &    0.000 &    0.000 \\
\tableline
\end{tabular}
\end{center}
\end{table}

\clearpage

\begin{table}
\begin{center}
\caption{Results of the fits of the fixed angular aperture corrections given in tables~\ref{ha_type_tab} to \ref{o3n2_mass_re_tab} to a 5th order polynomial of the type: $a_{0} + a_{1}x + a_{2}x^{2} + a_{3}x^{3} + a_{4}x^{4} + a_{5}x^{5}$.\label{poly_fits}}
\begin{tabular}{lcrrrrrr}
\tableline\tableline
 Aperture & Selection & $a_{0}$ & $a_{1}$ & $a_{2}$ & $a_{3}$ & $a_{4}$ & $a_{5}$ \\
 correction & criterion & & & & & & \\
\tableline
\xhap & All    & 0.0000 &   0.1599  &  0.5027 &  $-$0.2276 &   0.0167 &      0.0037 \\
      & Sa-Sbc & 0.0000 &   0.1588  &  0.5059 &  $-$0.2649 &   0.0459 &   $-$0.0022 \\
      & Sc-Sdm & 0.0000 &   0.1192  &  0.6799 &  $-$0.3951 &   0.0812 &  $-$0.0056 \\
      & $\log M^{*}/M_{\odot} \leq 10.3$ & 0.0000 &   0.1661 &   0.4937 &  $-$0.2044 &   0.0024 &  0.0061 \\
      & $10.3 \leq \log M^{*}/M_{\odot}$ & 0.0000 &   0.1691 &   0.4630 &  $-$0.1956 &  0.0079 &   0.0044 \\
\hline
\xhahbp & All    & 1.1253 &  0.0274 &  $-$0.3125 &   0.2959 &  $-$0.1091 &   0.0143 \\
\end{tabular}
\end{center}
\end{table}

\clearpage

\begin{table}
\begin{center}
\caption{Ratio of the H$\alpha$ flux contained in the SDSS or SAMI apertures at different redshifts to the H$\alpha$ flux within a circular aperture of 10 or 3.3~kpc diameter for the CALIFA spirals whose 10 or 3.3~kpc diameter aperture is completely covered by PMAS/PPAK.
(1) Angular diameter of SDSS/SAMI aperture$\dagger$.
(2) Linear diameter of fixed reference aperture.
(3) Redshift.
(4)-(6) Values corresponding to 15.86\%, 50\% and 84.14\% of the distribution.
Only the first values are listed. A complete version of the table, including SAMI apertures and fixed circular apertures of 3.3 kpc diameter, are be available in the electronic edition.\label{aper12_sdss_ha_tab}}
\begin{tabular}{cccccc}
\tableline\tableline
Ang. Diam. & Lin. Diam. & $z$ & $- \sigma$ & ($\alpha_{\rm Ang}$/$\alpha_{\rm Lin}$)$_{50}$ & $+ \sigma$ \\
 (``)      & (kpc)      &     &            &                                                 &            \\
\tableline
 3.0 & 10. &  0.010 & 0.030 & 0.061 & 0.207  \\
 3.0 & 10. &  0.011 & 0.030 & 0.060 & 0.206  \\
 3.0 & 10. &  0.012 & 0.030 & 0.059 & 0.205  \\
 3.0 & 10. &  0.013 & 0.029 & 0.058 & 0.204  \\
 3.0 & 10. &  0.014 & 0.029 & 0.057 & 0.204  \\
 3.0 & 10. &  0.015 & 0.029 & 0.056 & 0.203  \\
\tableline
\end{tabular}

$\dagger$Where the 3" aperture is for SDSS and the 15" aperture is for SAMI.
\end{center}
\end{table}

\begin{table}
\begin{center}
\caption{Ratio of the H$\alpha$/H$\beta$ contained in the SDSS or SAMI apertures at different redshifts to the H$\alpha$/H$\beta$ within a circular aperture of 10 or 3.3~kpc diameter for the CALIFA spirals whose 10 or 3.3~kpc diameter aperture is completely covered by PMAS/PPAK.
(1) Angular diameter of SDSS/SAMI aperture$\dagger$.
(2) Linear diameter of fixed reference aperture.
(3) Redshift.
(4)-(6) Values corresponding to 15.86\%, 50\% and 84.14\% of the distribution.
Only the first values are listed. A complete version of the table, including SAMI apertures and fixed circular apertures of 3.3 kpc diameter, are be available in the electronic edition.\label{aper12_sdss_hahb_tab}}
\begin{tabular}{cccccc}
\tableline\tableline
Ang. Diam. & Lin. Diam. & $z$ & $- \sigma$ & ($\alpha\beta_{\rm Ang}$/$\alpha\beta_{\rm Lin}$)$_{50}$ & $+ \sigma$ \\
 (``)      & (kpc)      &     &            &                                                 &            \\
\tableline
 3.0 & 10. &   0.010 &    0.985 &    1.120 &    1.287 \\
 3.0 & 10. &   0.011 &    0.985 &    1.120 &    1.288 \\
 3.0 & 10. &   0.012 &    0.985 &    1.120 &    1.288 \\
 3.0 & 10. &   0.013 &    0.984 &    1.120 &    1.289 \\
 3.0 & 10. &   0.014 &    0.984 &    1.120 &    1.289 \\
 3.0 & 10. &   0.015 &    0.984 &    1.120 &    1.290 \\
\tableline
\end{tabular}

$\dagger$Where the 3" aperture is for SDSS and the 15" aperture is for SAMI.
\end{center}
\end{table}

\clearpage

\begin{table}
\begin{center}
\caption{Difference of the N2 contained in the SDSS or SAMI apertures at different redshifts to the N2 within a circular aperture of 10 or 3.3~kpc diameter for the CALIFA spirals whose 10 or 3.3~kpc diameter aperture is completely covered by PMAS/PPAK.
(1) Angular diameter of SDSS/SAMI aperture$\dagger$.
(2) Linear diameter of fixed reference aperture.
(3) Redshift.
(4)-(6) Values corresponding to 15.86\%, 50\% and 84.14\% of the distribution.
Only the first values are listed. A complete version of the table, including SAMI apertures and fixed circular apertures of 3.3 kpc diameter, are be available in the electronic edition.\label{aper12_sdss_n2ha_tab}}
\begin{tabular}{cccccc}
\tableline\tableline
Ang. Diam. & Lin. Diam. & $z$ & $- \sigma$ & (N2$_{\rm Ang}$ $-$ N2$_{\rm Lin}$)$_{50}$ & $+ \sigma$ \\
 (``)      & (kpc)      &     &            &                                                 &            \\
\tableline
 3.0 & 10. &   0.010 &    0.012 &    0.079 &    0.234 \\
 3.0 & 10. &   0.011 &    0.011 &    0.079 &    0.234 \\
 3.0 & 10. &   0.012 &    0.011 &    0.079 &    0.234 \\
 3.0 & 10. &   0.013 &    0.011 &    0.079 &    0.234 \\
 3.0 & 10. &   0.014 &    0.011 &    0.079 &    0.234 \\
 3.0 & 10. &   0.015 &    0.011 &    0.080 &    0.233 \\
\tableline
\end{tabular}

$\dagger$Where the 3" aperture is for SDSS and the 15" aperture is for SAMI.
\end{center}
\end{table}

\begin{table}
\begin{center}
\caption{Difference of the O3N2 contained in the SDSS or SAMI apertures at different redshifts to the O3N2 within a circular aperture of 10 or 3.3~kpc diameter for the CALIFA spirals whose 10 or 3.3~kpc diameter aperture is completely covered by PMAS/PPAK.
(1) Angular diameter of SDSS/SAMI aperture$\dagger$.
(2) Linear diameter of fixed reference aperture.
(3) Redshift.
(4)-(6) Values corresponding to 15.86\%, 50\% and 84.14\% of the distribution.
Only the first values are listed. A complete version of the table, including SAMI apertures and fixed circular apertures of 3.3 kpc diameter, are be available in the electronic edition.\label{aper12_sdss_o3n2_tab}}
\begin{tabular}{cccccc}
\tableline\tableline
Ang. Diam. & Lin. Diam. & $z$ & $- \sigma$ & (O3N2$_{\rm Ang}$ $-$ O3N2$_{\rm Lin}$)$_{50}$ & $+ \sigma$ \\
 (``)      & (kpc)      &     &            &                                                 &            \\
\tableline
 3.0 & 10. &   0.010 &   -0.402 &   -0.180 &    0.060 \\
 3.0 & 10. &   0.011 &   -0.402 &   -0.180 &    0.060 \\
 3.0 & 10. &   0.012 &   -0.402 &   -0.181 &    0.060 \\
 3.0 & 10. &   0.013 &   -0.403 &   -0.181 &    0.060 \\
 3.0 & 10. &   0.014 &   -0.403 &   -0.181 &    0.060 \\
 3.0 & 10. &   0.015 &   -0.403 &   -0.181 &    0.060 \\
\tableline
\end{tabular}

$\dagger$Where the 3" aperture is for SDSS and the 15" aperture is for SAMI.
\end{center}
\end{table}

\clearpage

\begin{table}
\begin{center}
\caption{Differential (fiber-based vs. integrated) oxygen abundance of the SDSS galaxies estimated with the N2 and O3N2 indicators at three redshift bins.
(1) Stellar mass bin; 
(2) Number of galaxies; 
(3) Average over 25 random draws of the median values of $\log$O/H$_{\rm SDSS} - \log$O/H$_{\rm int}$~estimated using N2; 
(4) Average over 25 random draws of the median values of $\log$O/H$_{\rm SDSS} - \log$O/H$_{\rm int}$~estimated using O3N2.\label{abun_monte}}
\begin{tabular}{cccc}
\tableline\tableline
 & $N_{gal}$ & $\langle$ Median($\Delta \log$O/H$_{\rm N2}$) $\rangle$ & $\langle$ Median($\Delta \log$O/H$_{\rm O3N2}$) $\rangle$ \\    
 &          & (dex)                   & (dex)                 \\
\tableline
\multicolumn{4}{c}{$0.02 < z < 0.05$} \\
$8.5 \leq \log M^{*}/M_{\odot} < 9.1$    & 10640 & 0.025 & 0.034 \\
$9.1 \leq \log M^{*}/M_{\odot} < 9.7$    & 16964 & 0.026 & 0.035 \\
$9.7 \leq \log M^{*}/M_{\odot} < 10.3$   & 8817  & 0.027 & 0.037 \\
$10.3 \leq \log M^{*}/M_{\odot} < 10.9$  & 1698  & 0.045 & 0.013 \\
$10.9 \leq \log M^{*}/M_{\odot} < 11.5$  & 44    & 0.047 & 0.013 \\
\hline
\multicolumn{4}{c}{$0.05 < z < 0.10$} \\
$8.5 \leq \log M^{*}/M_{\odot} < 9.1$    & 994   & 0.015 & 0.020 \\
$9.1 \leq \log M^{*}/M_{\odot} < 9.7$    & 17560 & 0.022 & 0.029 \\
$9.7 \leq \log M^{*}/M_{\odot} < 10.3$   & 48676 & 0.023 & 0.031 \\
$10.3 \leq \log M^{*}/M_{\odot} < 10.9$  & 24456 & 0.036 & 0.012 \\
$10.9 \leq \log M^{*}/M_{\odot} < 11.5$  & 1348  & 0.042 & 0.013 \\
\hline
\multicolumn{4}{c}{$0.10 < z < 0.15$} \\
$8.5 \leq \log M^{*}/M_{\odot} < 9.1$    & 21    & 0.012 & 0.017 \\
$9.1 \leq \log M^{*}/M_{\odot} < 9.7$    & 1140  & 0.012 & 0.017 \\
$9.7 \leq \log M^{*}/M_{\odot} < 10.3$   & 12599 & 0.019 & 0.025 \\
$10.3 \leq \log M^{*}/M_{\odot} < 10.9$  & 29146 & 0.030 & 0.011 \\
$10.9 \leq \log M^{*}/M_{\odot} < 11.5$  & 3748  & 0.034 & 0.012 \\
\hline
\multicolumn{4}{c}{$0.15 < z < 0.20$} \\
$8.5 \leq \log M^{*}/M_{\odot} < 9.1$    & 0     & ---         & ---         \\
$9.1 \leq \log M^{*}/M_{\odot} < 9.7$    & 135   & 0.010 & 0.014 \\
$9.7 \leq \log M^{*}/M_{\odot} < 10.3$   & 1972  & 0.013 & 0.017 \\
$10.3 \leq \log M^{*}/M_{\odot} < 10.9$  & 7540  & 0.026 & 0.011 \\
$10.9 \leq \log M^{*}/M_{\odot} < 11.5$  & 3445  & 0.030 & 0.012 \\
\hline
\multicolumn{4}{c}{$0.20 < z < 0.25$} \\
$8.5 \leq \log M^{*}/M_{\odot} < 9.1$    & 1     & ---         & ---         \\
$9.1 \leq \log M^{*}/M_{\odot} < 9.7$    & 20    & 0.008 & 0.012 \\
$9.7 \leq \log M^{*}/M_{\odot} < 10.3$   & 494   & 0.011 & 0.015 \\
$10.3 \leq \log M^{*}/M_{\odot} < 10.9$  & 1710  & 0.018 & 0.011 \\
$10.9 \leq \log M^{*}/M_{\odot} < 11.5$  & 1698  & 0.028 & 0.011 \\
\hline
\multicolumn{4}{c}{$0.25 < z < 0.30$} \\
$8.5 \leq \log M^{*}/M_{\odot} < 9.1$    & 0     & ---         & ---         \\
$9.1 \leq \log M^{*}/M_{\odot} < 9.7$    & 1     & ---         & ---         \\
$9.7 \leq \log M^{*}/M_{\odot} < 10.3$   & 43    & 0.009 & 0.013 \\
$10.3 \leq \log M^{*}/M_{\odot} < 10.9$  & 493   & 0.014 & 0.011 \\
$10.9 \leq \log M^{*}/M_{\odot} < 11.5$  & 357   & 0.022 & 0.011 \\
\tableline
\end{tabular}
\end{center}
\end{table}

\clearpage

\end{document}